\pdfoutput=1

\documentclass{aa}  
\usepackage{natbib}
\bibpunct{(}{)}{;}{a}{}{,} 
\usepackage{multicol}

\usepackage{tabularx}
\usepackage{tabulary}
\usepackage{epsfig}
\usepackage{epstopdf}
\usepackage{graphics}
\usepackage{subcaption}
\usepackage{txfonts}
\begin{document}

\title{The \textit{GALEX} Ultraviolet Virgo Cluster Survey (GUViCS)}
\subtitle{III. The Ultraviolet Source Catalogs}
\author{Elysse N. Voyer\inst{1}, A. Boselli\inst{1}, S. Boissier\inst{1}, S. Heinis\inst{1,2}, L. Cortese\inst{3,4}, L. Ferrarese\inst{5}, P. Cote\inst{5}, J.-C. Cuillandre\inst{6}, S. D. J.Gwyn\inst{5}, E. W. Peng\inst{7,8}, H. Zhang\inst{7,8}, C. Liu\inst{9,10}}
\institute{Aix-Marseille Université, CNRS, LAM (Laboratoire d’Astrophysique de Marseille) UMR 7326, 13388, Marseille, France\\
           \email{elysse.voyer@oamp.fr, alessandro.boselli@lam.fr, samuel.boissier@lam.fr}
           \and
           Department of Astronomy, University of Maryland, College Park, MD 20742-2421, USA
           \and
           Centre for Astrophysics \& Supercomputing, Swinburne University of Technology, Mail H30, PO Box 218, Hawthorn, VIC 3122, Australia
           \and
           European Southern Observatory, Karl Schwarzschild Str. 2, 85748 Garching bei München, Germany
	   \and
	   Herzberg Institute of Astrophysics, National Research Council of Canada, Victoria, BC, V9E 2E7, Canada
	   \and
           Canada–France–Hawaii Telescope Corporation, Kamuela, HI 96743, USA
           \and
           Department of Astronomy, Peking University, Beijing 100871, China
           \and
           Kavli Institute for Astronomy and Astrophysics, Peking University, Beijing 100871, China
	   \and
	   Center for Astronomy and Astrophysics, Department of Physics and Astronomy, Shanghai Jiao Tong University, 800 Dongchuan Road, Shangai 200240, China
	   \and
	   INPAC, Department of Physics and Astronomy and Shanghai Key Lab for Particle Physics and Cosmology, Shanghai Jiao Tong University, Shanghai, 200240, China}
\date{Received 20 August 2013 / Accepted 13 May 2014}

\abstract
{In this paper we introduce the deepest and most extensive ultraviolet extragalactic source catalogs of the Virgo Cluster area to date. Archival and targeted \textit{GALEX} imaging is compiled and combined to provide the deepest possible coverage over $\sim$120 deg$^2$ in the NUV ($\lambda_{eff}$=2316\AA) and $\sim$40 deg$^2$ in the FUV ($\lambda_{eff}$=1539\AA) between 180$^{\circ}$ $\le$ R.A. $\le$ 195$^{\circ}$ and 0$^{\circ}$ $\le$ Decl. $\le$ 20$^{\circ}$. We measure the integrated photometry of 1770 extended UV sources of all galaxy types and use \textit{GALEX} pipeline photometry for 1,230,855 point-like sources in the foreground, within, and behind the cluster. Extended source magnitudes are reliable to $m_{UV}\sim22$, showing $\sim$0.01$\sigma$ difference from their asymptotic magnitudes. Point-like source magnitudes have a 1$\sigma$ standard deviation within $\sim$0.2 mag down to $m_{uv} \sim$23. The point-like source catalog is cross-matched with large optical databases and surveys including the SDSS DR9 ($>$ 1 million Virgo Cluster sources), the Next Generation Virgo Cluster Survey (NGVS; $>$13 million Virgo Cluster sources), and the NED ($\sim$30,000 sources in the Virgo Cluster). We find 69\% of the entire UV point-like source catalog has a unique optical counterpart, 11\% of which are stars and 129 are Virgo cluster members neither in the VCC nor part of the bright CGCG galaxy catalog (i.e., $m_{pg} <$ 14.5). These data are collected in four catalogs containing the UV extended sources, the UV point-like sources, and two catalogs each containing the most relevant optical parameters of UV-optically matched point-like sources for further studies from SDSS and NGVS. The GUViCS catalogs provide a unique set of data for future works on UV and multiwavelength studies in the cluster and background environments.}

\titlerunning{GUViCS III. The Ultraviolet Source Catalogs}
\authorrunning{Elysse N. Voyer et al.}
\maketitle

%
\section{Introduction}

Observational studies of the cosmic star-formation rate density of the universe over multiple wavelengths show that present day galaxies are forming stars at the lowest rate compared to their cosmic past \citep{lil96,mad96,con97,hop06,bou09}. Ongoing extragalactic astrophysical research is largely focused on the investigation of potential physical mechanisms and properties that have driven the evolution of the star-formation rate density to its current state. Several studies have found that galaxy mass and environment are the two major drivers evolution throughout time \citep{dre80,gav96,cow96,bos01,bos06,cle06,pog09,cle09}. However, questions remain as to how these two drivers influence the star-formation rate of galaxies, even within the local volume of the universe. Studies of star formation in local galaxies contribute to our understanding of an initial benchmark for observational and theoretical evolutionary studies, as well as enable detailed physical analysis of star-forming galaxies that requires high resolution. 

Young, massive, short-lived O and B stars emit the majority of their radiation in the ultraviolet (UV) spectral region $<$ 3000\AA. Thus, observations in broad UV bandpasses map the regions of the unobscured star formation within a galaxy \citep{ken98,bos09}. Over the past decade, imaging from the \textit{Galaxy Evolution Explorer} (\textit{GALEX}) satellite have fueled a plethora of studies that explore unobscured star formation in galaxies near and far. \textit{GALEX} took broadband near-UV (NUV; $\lambda_{eff}$=2316\AA; $\Delta\lambda$=1771-2831\AA) and far-UV (FUV; $\lambda_{eff}$=1539\AA; $\Delta\lambda$=1344-1786\AA) images within a 1.2$^{\circ}$ diameter field-of-view. It has performed extensive surveys at shallow ($\sim$100s), medium ($\sim$1,500s), and deep ($\sim$30,000s) exposure times, providing a rich archive of publicly available data. 

The goal of the \textit{GALEX} Ultraviolet Virgo Cluster Survey (GUViCS) \citep{bos11} is to provide the most extensive ($\sim$120 deg$^2$) UV source catalogs of the Virgo Cluster to date from \textit{GALEX} imaging data for in-depth and multiwavelength studies focused on the effects of the cluster environment on star formation in all types of local galaxies. The Virgo Cluster has been the interest of a multitude of work since it is the largest cluster of galaxies nearest to the Milky Way \citep[16.5 Mpc;][]{mei07,bla09} and still in the midst of virialization \citep{gav99,mei07}. These two key factors permit detailed analysis of galaxies that are undergoing (and have undergone) physical transformations attributable to a range of processes including tidal interactions, galaxy harassment, and ram-pressure stripping \citep{bos06,bos08a,bos08b}. The cluster is made up of galaxies of nearly every observed type in the local universe, from giant ellipticals and extended late-type galaxies to dwarf elliptical (dE) and ultra compact dwarfs (UCDs). However, it is primarily dominated by quiescent systems as shown by the morphology segregation effect \citep{dre80,bin87,san85}. 

Initial work has been performed on the GUViCS data set over the past three years. In GUViCS Paper I \citet{bos11} discuss, in detail, the science objectives of GUViCS and present the first measurement of the NUV and FUV luminosity functions in the central 12 deg$^2$ of the cluster (centered on M87) down to $M$ = -10. They find slopes similar to those measured for the UV luminosity functions in the field and in the B-band in Virgo. In GUViCS Paper II \citet{boi12} combine UV and multiwavelength observations (\ion{H}{I}, optical, H$\alpha$) to measure the level of star formation in gas that has been ram-pressure stripped from cluster galaxies. They find the amount of star formation is low compared to the amount of fuel available and provide constraints for future spectro-photometric models of galaxy evolution. GUViCS is also complementary to several recent multiwavelength surveys of the Virgo Cluster including the Next Generation Virgo Cluster Survey \citep[NGVS;][]{fer12} done in optical bands ($u^*g'r'i'z'$) with the Canada-France-Hawaii Telescope (CFHT), the \textit{Herschel} Virgo Cluster Survey \citep[HeViCS;][]{dav10} in the far-infrared (100-500$\mu$m), and the Arecibo Legacy Fast ALFA survey \citep[ALFALFA;][]{gio05} in \ion{H}{I}. Together, these data can facilitate large statistical studies on detailed properties of Virgo Cluster members as well as galaxies in the background of the cluster (A. Boselli et al., \textit{in preparation}).   

In this paper we construct and present UV catalogs of extended and point-like sources within and behind the Virgo Cluster. We use targeted and archival NUV and FUV \textit{GALEX} observations to construct the deepest possible data set in each band, and we identify extended sources from well known catalogs including the Virgo Cluster Catalog \citep[VCC;][]{bin85} and the Catalog of Galaxies and of Clusters of Galaxies \citep[CGCG;][]{zwi}. The resulting GUViCS catalogs are extensive, containing 1,230,855 point-like sources and 1770 extended sources. The point-like source catalog is matched with existing large optical databases and catalogs to obtain optical photometry, identify stars, find additional extended galaxies, and provide redshifts that can differentiate background galaxies and cluster members. We also perform integrated UV photometry on all GUViCS extended sources, and test its reliability, as well as the reliability of the \textit{GALEX} pipeline photometry for the point-like catalog. 

This full set of data is fundamental for several reasons. The Virgo Cluster covers one of the largest areas of the sky mapped from the UV to the far-infrared (500 microns) at an unprecedented depth, and is thus ideal for future cosmological studies. UV data are necessary to determine accurate photometric redshifts for distant objects and to study the star-formation history of galaxies at different redshifts, from the great wall at $\sim$7000 km/s to the distant universe. The large volume sampled allows the study of stellar populations in background ellipticals, and of their dependence on environment or redshift. For example, 452 clusters from the MaxBCG catalog of \citet{koe07} are included in the coordinate range considered here. GUViCS FUV and NUV data are available for $\sim$170 central bright cluster galaxies of these clusters below redshift $\sim$0.3, allowing for an analysis of the evolution of their FUV-NUV colors (Cucciati et al., \textit{in preparation}). This could reveal potential star formation or the evolution of the UV upturn. More generally, the pilot survey of the NGVS (central 4 square degrees) found thousands of early type galaxies in the redshift range 0.2-0.7, predicting that the final NGVS catalog will be extremely rich with these galaxies. GUViCS FUV, NUV, and NGVS $u^*$-band data will allow the study of their rest-frame UV properties. Additionally, between 500 to 1000 clusters will be identified in the NGVS data \citep{mei13} allowing us to study how these properties are affected by the environment. The GUViCS extended and point-like source catalogs will be the starting point for all future studies in the Virgo Cluster and in the background of Virgo.

This paper is organized as follows. In Section \ref{uvdata}, we introduce the \textit{GALEX} UV data used here. In Section \ref{build}, we describe the methods used for the detection of UV extended sources in the \textit{GALEX} catalogs, as well as the procedure used to reduce the \textit{GALEX} catalogs for point-like sources. In Section \ref{phot}, we measure and discuss the limitations on UV photometry for extended and point-like sources. In Section \ref{matchoptcat}, we discuss the matching of the GUViCS point-like source catalog to optical data sets. In Section \ref{analysis} we present the analysis of extended source statistics and magnitudes, and of new Virgo members and background sources from the UV point-like source catalog. Section \ref{summ} gives the summary and results of this work. Finally, Appendices \ref{reduc}-\ref{guvics_ned_app} provide further technical details of this work not discussed in the main text, and Appendices \ref{appendix_UVVC_extend_cat}-\ref{appendix_guvics-sdss_cat} present the GUViCS catalogs.
%
\section{Ultraviolet Data}
\label{uvdata}

The GUViCS survey is a combination of archival and targeted \textit{GALEX} observations over an area of $\sim$120 deg$^2$ centered on M87, covering 180$^{\circ}$ $\le$ R.A. $\le$ 195$^{\circ}$ and 0$^{\circ}$ $\le$ Decl. $\le$ 20$^{\circ}$ \citep{bos11}. Here, we use \textit{GALEX} archival observations of all available data since the 7$^{th}$ \textit{GALEX} data release (GR7; Feb. 27th, 2013) from:
\begin{itemize}
 \item the Nearby Galaxies Survey (NGS)
 \item the All-Sky Imaging Survey (AIS)
 \item the Medium Imaging Survey (MIS)
 \item the Deep Imaging Survey (DIS)
 \item all public guest investigator observations 
\end{itemize}
New observations were awarded specifically for the GUViCS project during \textit{GALEX} guest investigator cycle 6 (GI6-001, P.I. A. Boselli). These data cover 94 new NUV pointings in order to complete the \textit{GALEX} coverage of the cluster at $\sim$MIS depth ($\sim$1.5ks). The majority of new GUViCS data was collected between March-April 2010, and the observations were completed in March 2011. Due to the non-operational status of \textit{GALEX}'s FUV detector in early 2010, it was not possible to observe any GUViCS program fields in the FUV. Additionally, we include 27 NUV and FUV guest investigator fields from program GI2-125 (P.I. A. Boselli) that are part of a previous Virgo Cluster observational campaign proposed by several members of this team covering the Arecibo Galactic Environment Survey (AGES) area down to MIS depths. Further discussion of the \textit{GALEX} data reduction is in Appendix \ref{reduc}. 

Figures \ref{guvics_nuv_cov} and \ref{guvics_fuv_cov} show the combined archival and targeted GUViCS data coverage maps with respect to the depth of each pointing. This represents the initial data set that we use as a base for the UV Virgo Cluster catalogs. Table \ref{galex_tiles} summarizes the number of \textit{GALEX} tiles available in the GUViCS area from each survey for the NUV- and FUV-bands.
     \begin{table}
     \caption{\textit{GALEX} tiles in the GUViCS area.}             
     \label{galex_tiles}      
     \centering          
     \begin{tabular}{ccc}     
     \hline\hline  
      \textit{GALEX} Survey & \# NUV tiles & \# FUV tiles\\
      \hline   
       new GUViCS& 94 & 0\\  
       AGES & 27 & 0\\
       NGS & 23 & 20\\
       AIS & 305 & 295\\
       MIS & 19 & 14\\
       DIS & 13 & 12 \\
       guest investigator& 101 & 83\\
      \hline
      \end{tabular}
      \end{table}
It shows that $\sim$78\% of the \textit{GALEX} image tiles in the GUViCS region have an observation in the FUV-band, but the majority of these are very shallow AIS images. The gaps in \textit{GALEX} coverage shown in Figures \ref{guvics_nuv_cov} and \ref{guvics_fuv_cov} (i.e. black empty regions) were purposely avoided during observations due to the presence of bright stars in those areas. These figures also demonstrate how heavily the FUV observations are affected by Galactic cirrus compared to the NUV-band.

\begin{figure*}
\centering
 \resizebox{0.57\textheight}{!}{\includegraphics{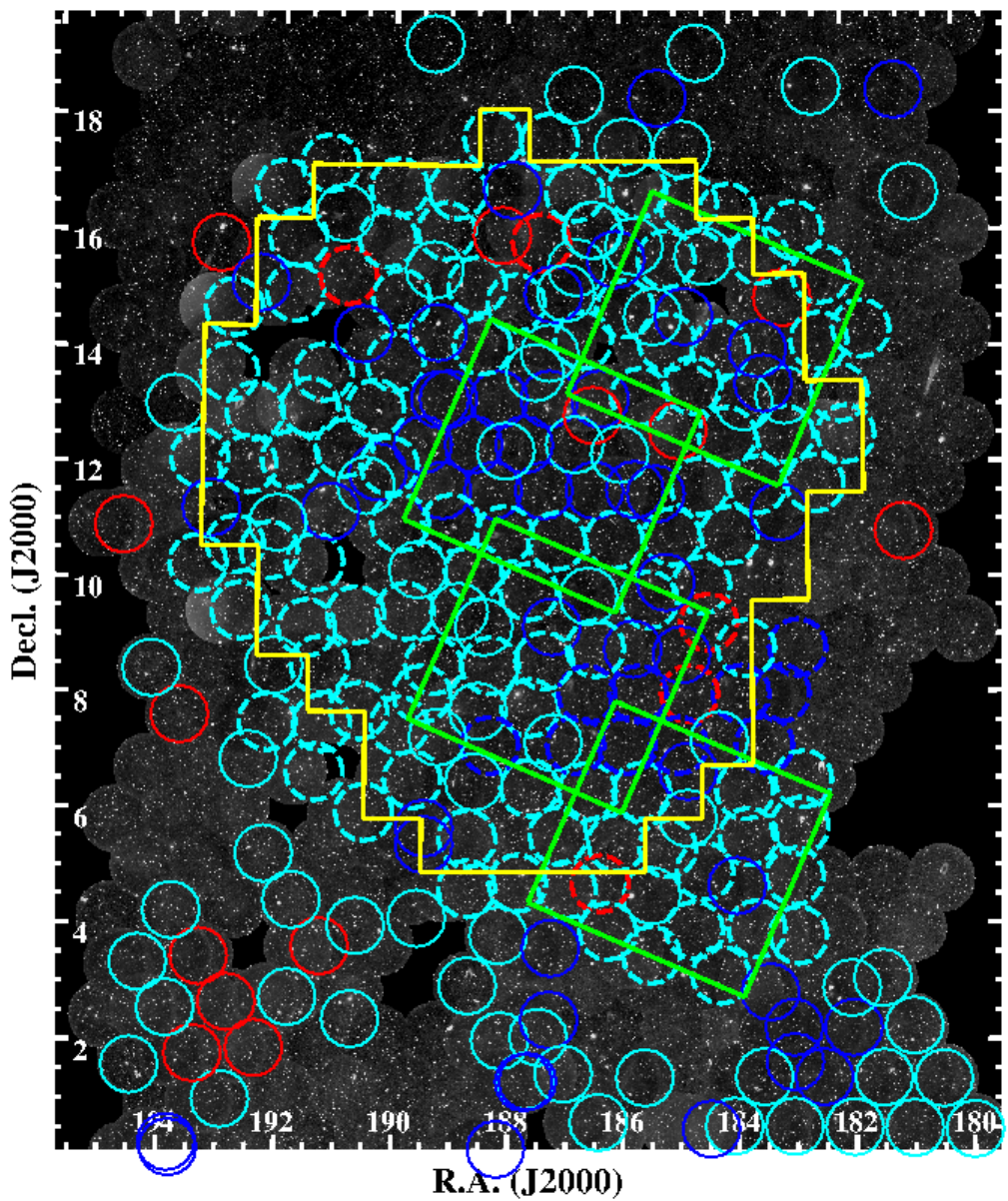}}
\caption{NUV \textit{GALEX} coverage of the GUViCS survey area. Both R.A. and Decl. are given in units of degrees. The black and white sky image mosaic is created from the AIS-like \textit{GALEX} fields (exposure time $<$ 800s) in the GUViCS area. The individual circles represent 0.6$^{\circ}$ radial \textit{GALEX} fields with deeper exposure times than AIS. Dashed circles indicate that the field is either a targeted GUViCS or AGES field (P.I. A. Boselli). Red circles are \textit{GALEX} fields with exposure times between 800s-1,500s, cyan circles are \textit{GALEX} fields with exposure times between 1,500s-30,000s, and dark blue circles are \textit{GALEX} fields with exposure times $>$ 30,000s. Gaps in \textit{GALEX} coverage (i.e. black empty regions) were purposely avoided during observations due to the presence of bright stars in those areas. The large yellow footprint represents the area covered by the NGVS in optical bands (CFHT $u^*g'r'i'z'$), the green footprints represent the area covered by the HeViCS in the infrared (100$\mu$m, 160$\mu$m, 250$\mu$m, 350$\mu$m, 500$\mu$m), and the ALFALFA survey in radio (\ion{H}{I}) covers the entire GUViCS area between 0$^{\circ}$ $<$ Decl. $<$ 20$^{\circ}$.}
\label{guvics_nuv_cov}
\end{figure*}

\begin{figure*}[!ht]
\centering
 \resizebox{0.57\textheight}{!}{\includegraphics{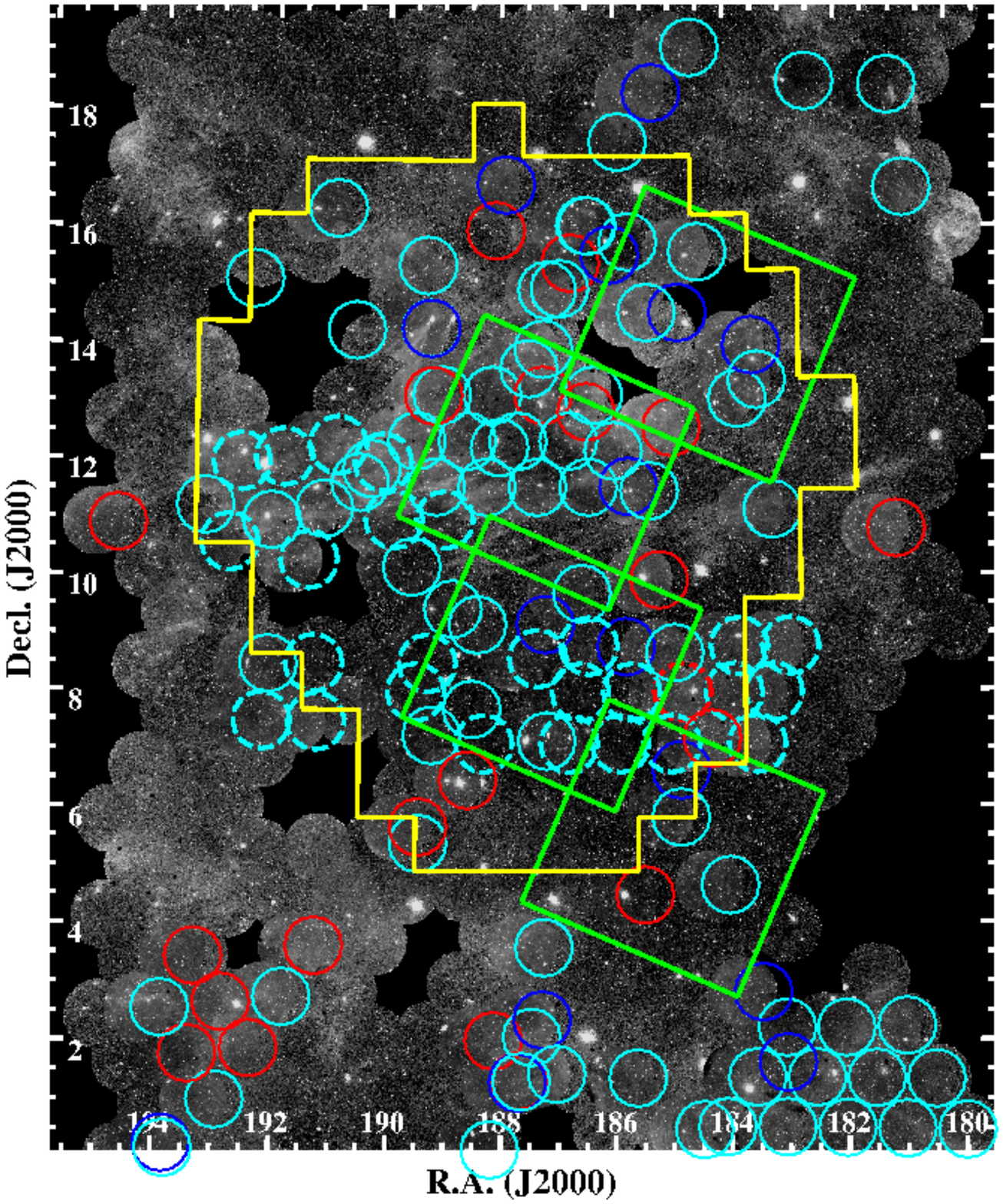}}
\caption{FUV \textit{GALEX} coverage of the GUViCS survey area. Both R.A. and Decl. are given in units of degrees. The explanation of the figure is the same as for Figure \ref{guvics_nuv_cov}. The background image demonstrates how dominant Galactic cirrus is in FUV wavelengths compared to the NUV.}
\label{guvics_fuv_cov}
\end{figure*}

  In this work we utilize the \textit{GALEX merged object catalogs} (\textit{mcats}) that are the final combined versions of the NUV and FUV catalogs output for each \textit{GALEX} tile by the automated pipeline. The \textit{mcats} are produced by matching the NUV and FUV detections in the same \textit{GALEX} tile that have signal-to-noise (S/N) $>$ 2 within a tolerance of 3$''$ according to their pair-wise match probabilities \citep{mor07}. The \textit{mcat} catalog provides a single set of coordinates for each \textit{GALEX} source as well as NUV and FUV photometric parameters in several different aperture types, where the data is available.

\section{Catalog Construction}
\label{build}
  \subsection{The GUViCS Extended Source Catalog}
  \label{build_es}
  Extended sources are treated separately from point-like sources since their large angular sizes require a separate method of source detection and photometry \citep{bos11}. In order to initially identify extended sources in the GUViCS data set we match 2096 VCC galaxies \citep{bin85}, 353 CGCG\footnote{GOLDMine contains 111 Virgo CGCG sources since the database only includes the CGCG catalog up to apparent photographic Vega magnitude m$_{pg}$ = 14.5. The 242 fainter CGCG sources are obtained from NED.} galaxies \citep{zwi}, and 95 galaxies in various other catalogs with the \textit{GALEX} \textit{mcats}. These are selected from either the GOLDMine\footnote{http://goldmine.mib.infn.it/} \citep{gav03} or NED\footnote{http://ned.ipac.caltech.edu/} databases. We obtain the following galaxy parameters from GOLDMine for sources in the VCC and CGCG: central coordinates, optical diameters (from photographic plates), apparent photographic Vega magnitude ($m_{pg}$), galaxy types, velocities (heliocentric), and membership to the various Virgo substructures. NED provides central coordinates, major diameters, minor diameters, galaxy types, and velocities (heliocentric). Position angles for 404 VCC late-type galaxies are taken from H.T. Meyer and T. Lisker (\textit{in preparation}), and were calculated based on the $r'$-band light within 2 Petrosian apertures. For CGCG sources, position angles are obtained from NED, where available.

  The deepest \textit{GALEX} image available for each extended source is used for its inspection and photometry. We perform a visual inspection of each galaxy by comparing its UV image with its optical color-combined image to determine if it is a real UV detection. We use the NGVS Graphical Search Tool\footnote{http://www2.cadc-ccda.hia-iha.nrc-cnrc.gc.ca/gsky/ndata.html} to examine deep optical images. The UV-optical comparison is particularly important for cases where small red extended sources have bluer UV background sources interloping in their optical apertures. If the optical image is not examined in these cases, then a UV background detection might be mistaken for emission from the foreground galaxy. Ultimately, we find that 1770 extended sources that are real UV detections, where 1136 are Virgo member galaxies and 634 are background galaxies. Hereafter, we will refer to this catalog as the UV Virgo Cluster extended source (UV-VES) catalog. Table \ref{ext_stats} provides a statistical breakdown of UV-VES catalog's Virgo members and background galaxies by catalog of origin.

   \begin{table}
   \caption{Extended source statistics.}             
   \label{ext_stats}      
   \centering          
   \begin{tabular}{rccc}     
   \hline\hline       
   Catalog & Total & Virgo Members\tablefootmark{a} & Background\\ 
   \hline                    
    VCC  & 1340 & 1033 & 307\\  
    CGCG &  335 &  88 & 247\\
    other (NED)  &   95 &   15 &  80\\
   \hline
   \end{tabular}
   \tablefoot{\\
   \tablefoottext{a}{Membership determinations of CGCG and other extended sources are based on velocity measurements of those sources provided in GOLDMine or NED. Membership determination of VCC sources comes from \citet{bin85,bin93}, where in some cases, it is based on galaxy velocity, while in other cases, it is based on the surface brightness versus photographic magnitude ($\sim$$m_B$ in Vega system) of the galaxy.}
    }
   \end{table}

  \subsection{The GUViCS Point-Like Source Catalog}
  \label{build_pls}
   All UV Virgo point-like sources are from the \textit{GALEX mcats}. Since there is a significant amount of \textit{GALEX} tile overlap in the GUViCS area, we initially reduce the \textit{mcats} corresponding to any overlapping tiles. Sources from the deepest observations in the regions of overlap are retained, and sources from the shallower observations are discarded. Next, all \textit{mcats} are combined into a single catalog. In this way we construct the deepest possible unified point-like source catalog from the original \textit{mcats}. Further technical details of the catalog reduction procedure are discussed in Appendix \ref{catalog_construct}. Hereafter, we will refer to this catalog as the UV Virgo Cluster point-like source (UV-VPS) catalog. 

   Initially, we clean the UV-VPS catalog of sources matched to bright stars and sources within extended VCC and CGCG galaxy apertures. We remove 12,211 bright foreground stars that are matched to SIMBAD Astronomical Database\footnote{http://simbad.u-strasbg.fr/simbad/} sources. Point-like matches to UV-detected extended sources are removed, as well as 6,702 confirmed stars and background galaxies\footnote{Object type confirmations are made with the SDSS database.} within their UV apertures. Additionally, we remove the 197 pipeline detections matched to VCC sources that are non-UV detections based on our visual inspection of VCC galaxies (see Section \ref{build_es}). We considered these pipeline detections to be spurious. If they remained in the point-like source catalog it would incorrectly include non-UV detected VCC sources that could present false matches with other surveys and catalogs.

   Next, we clean the UV-VPS catalog of spurious point-like detections (i.e. detections of noise) based on \textit{GALEX} magnitude limits in the literature \citep{mor07,ham10,gea11,hab12}. The UV-VPS catalog is not uniform in depth since it combines many different \textit{GALEX} observing campaigns with a large range of exposure times. Therefore, cuts in magnitude for spurious detections vary depending on the exposure time of each point-like source. Table \ref{spursrcs} provides the $m_{NUV}$ and $m_{FUV}$ cuts with respect to exposure time. They are $\sim$0.5 magnitudes fainter than the \citet{mor07} 5$\sigma$ limiting magnitudes for the AIS, MIS, and DIS surveys. A few fields\footnote{NUV: GI4\_012001\_PG1216p069 (33,699.85 s), GI4\_012004\_RXJ1230d8p0115 (46,087.9 s), and GI4\_012003\_3c273 (50,050.5 s); FUV: GI4\_012001\_PG1216p069 (30,170.35 s) and GI4\_012004\_RXJ1230d8p0115 (31,246.25 s).} have exposure times $>$ 30 ks in the GUViCS area. Our determination of the magnitude limit (i.e. DEEP in Table \ref{spursrcs}) after which detections are considered spurious in these images is based on investigations of detection completeness in deep \textit{GALEX} fields \citep{ham10,gea11,hab12}. Finally, random visual checks are made on samples of spurious detections in each exposure time range in Table \ref{spursrcs}. They confirm that we are indeed removing spurious detections and not real UV sources. In total, these magnitude cuts eliminate a significant fraction of sources, removing $\sim$44\% of NUV sources and $\sim$53\% of FUV sources.

     \begin{table*}
     \caption{Magnitude limits for the removal of spurious UV sources from the \textit{GALEX mcats}.}             
     \label{spursrcs}      
     \centering          
     \begin{tabular}{cccccccc}     
     \hline\hline     
  
      \textit{GALEX} Image Type & Exposure Time Range & $m_{NUV}$ limit & $m_{FUV}$ limit & $\sigma_{NUV}$\tablefootmark{a} & $\sigma_{FUV}$\tablefootmark{a} & $mean_{\Delta NUV}$\tablefootmark{a} & $mean_{\Delta FUV}$\tablefootmark{a}\\ 
                                & (s)                 &                 &                 &                                 &                                 &                                      &\\
      \hline                    
       AIS-like &   \phantom{0000} t $\le$ 381         & 21.3 & 20.4 & 0.215 & 0.301 & -0.025 & -0.026\\  
       MIS-like &         381 $<$ t $\le$ 2500        & 23.2 & 23.1 & 0.303 & 0.354 & -0.026 & -0.053\\
       DIS-like &        2500 $<$ t $\le$ 30000       & 24.9 & 25.3 & 0.342 & 0.380 & -0.037 &  0.088\\
       DEEP     &       30000 $<$ t \phantom{00000000} & 25.5 & 25.5 & 0.344 & 0.367 & -0.092 & -0.090\\
      \hline
      \end{tabular}
      \tablefoot{\\
      \tablefoottext{a}{See Figure \ref{point_phottests} for the distributions of point-source photometry from which these values are calculated.}\\
      }
      \end{table*}

      After cleaning stars, extended sources, and spurious detections from the UV-VPS catalog, 1,231,331 UV sources remain. There are 1,197,637 sources ($\sim$97\%) with NUV detections, 201,004 sources ($\sim$16\%) with FUV detections, and 33,694 sources ($\sim$2.7\%) that are only detected in the FUV-band. A discussion of the reliability and caveats of the UV-VPS can be found in Appendix \ref{rel_cavs}.

%
\section{Photometry}
\label{phot}
  As discussed in \citet{bos11}, the photometry of extended and point-like UV sources must be treated separately. We get the total magnitudes of point-like sources from the \textit{GALEX mcat} $m_{NUV}$\footnote{NUV magnitudes and errors are from the \textit{GALEX mcat} fields \textit{nuv\_mag} and \textit{nuv\_magerr}.} and $m_{FUV}$\footnote{FUV magnitudes and errors are from the \textit{GALEX mcat} fields \textit{fuv\_mag} and \textit{fuv\_magerr}.} measurements, along with their respective errors. These are \textit{GALEX} calibrated measurements of \textit{MAG\_AUTO} Kron magnitudes \citep{kro80} that are calculated by SExtractor \citep{ber96}. 

  Photometry for extended sources must be performed manually. Despite some similarities to their optical morphologies, UV observations of local galaxies tend to be clumpy in comparison since the UV picks out regions of on-going star formation and not the more homogeneous evolved stellar populations. Thus, it is quite difficult to implement a standard routine for automatic photometry (e.g. SExtractor) on a large set of nearby UV galaxies. Additionally, the large photometric apertures of extended sources are prone to contamination by foreground stars and background galaxies that must be removed on a case-by-case basis. Following the methods of \citet{bos11}, we utilize the FUNTOOLS analysis package in DS9 to measure UV extended source photometry in manually created apertures. These measurements provide aperture magnitudes that are well suited for many studies (e.g. the luminosity function of the cluster). The aperture parameters can also be used as initial input for measurements of the extrapolated UV magnitudes \citep{cor12}. Details of the UV aperture construction are given in Appendix \ref{ext_apps}.
    
\begin{figure*}
    \centering
    \resizebox{\hsize}{!}{\includegraphics{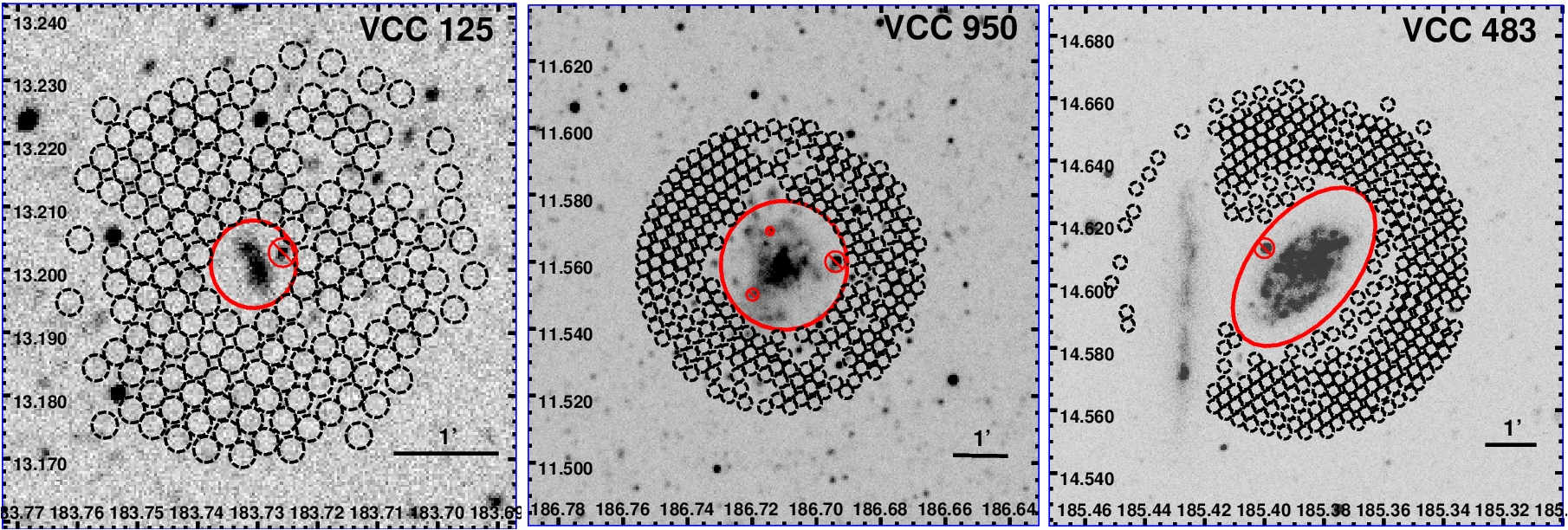}}
    \caption{Examples of NUV galaxy and background photometry apertures for extended sources VCC 125, VCC 950, and VCC 483. Horizontal axis coordinates are degrees of right ascension (J2000), and vertical axis coordinates are degrees of declination (J2000). Solid red ellipses are the UV photometry apertures, smaller red circle-backslash apertures are masked regions, and black dashed circles are the background apertures.}
    \label{ext_aps}
    \end{figure*}

  To calculate the average background flux in the vicinity of each galaxy we fill a circular annulus extending 1.5$'$ from the edge of the semi-major axis with small circular background apertures that have 7.5$''$ radii. For each galaxy, the background apertures are manually moved to a dark region of the sky if we find they contain any small UV sources. Figure \ref{ext_aps} provides examples of the final galaxy and background apertures for three VCC sources. To calculate the statistical errors on the UV magnitudes we follow the error estimation procedures described in \citet{bos03}, \citet{gil07}, and \citet{cor12}. The calculation is described in Appendix \ref{error}.
 
    \subsubsection{Testing Photometric Accuracy of Extended Sources}
\begin{figure*}
    \centering
    \resizebox{\hsize}{!}{\includegraphics{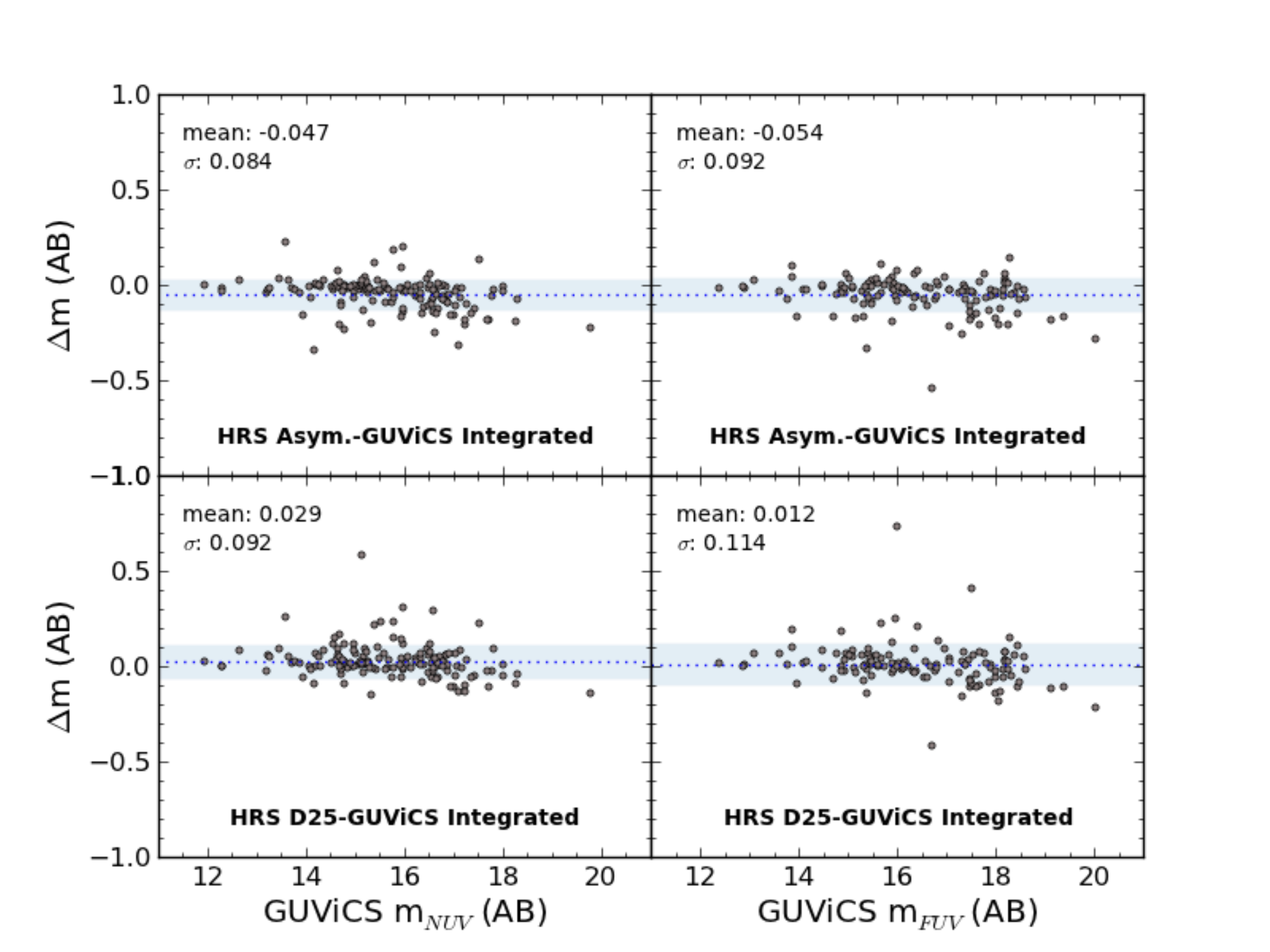}}
    \caption{Comparison of GUViCS photometry to \textit{GALEX} UV photometry for VCC sources in common with the \citet{cor12} HRS galaxy catalog. The lightly shaded rectangle spanning the length of each plot represents the 1$\sigma$ spread in the magnitude of all the differences. The dotted horizontal line marks the value of the mean of all the differences. The numerical values of these two quantities are given in the upper left corner in each plot.}
    \label{vcc_hrs}
    \end{figure*}

    We compare the results of our photometry with the catalog of \citet{cor12}, as shown in Figure \ref{vcc_hrs}. \citet{cor12} measure the UV asymptotic and $D_{25}$ magnitudes from \textit{GALEX} data of sources in the \textit{Herschel} Reference Survey \citep{bos10}. They have 143 VCC sources in common with our UV-VES catalog. We find that the standard deviation of differences between our integrated and their asymptotic magnitudes is $\sim$0.1 mag, and there is an average difference of -0.05 mag, where the asymptotic magnitudes are brighter, on average, for both the NUV- and FUV-bands. This offset is due to the fact that \citet{cor12} provide an extrapolated magnitude value that, by definition, is always larger than any possible measure within a given aperture. Additionally, we find no clear trend when comparing our integrated magnitudes to their $D_{25}$ magnitudes. 

    \begin{figure*}
    \centering
    \resizebox{\hsize}{!}{\includegraphics{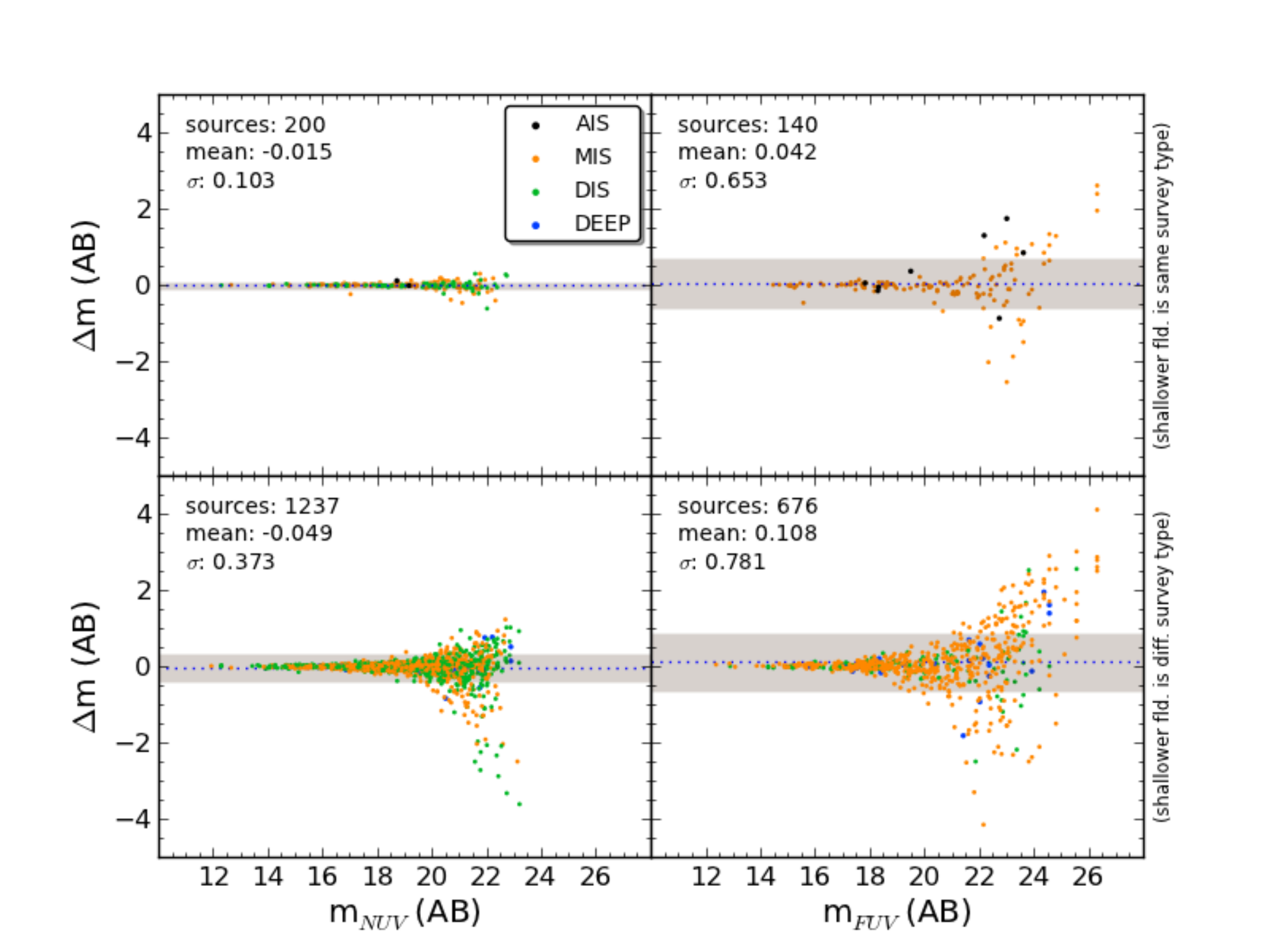}}
    \caption{Difference in UV magnitudes versus UV magnitudes of GUViCS extended galaxies with more than one independent measurement. NUV and FUV data are plotted in the left and right panels, respectively. The legend shows each point colored with respect to the \textit{GALEX} field type of the magnitude taken from the field with the deepest exposure time in the comparison (i.e. the magnitude in the GUViCS catalog). The upper set of plots show magnitude differences for photometry done in fields from the same exposure time categories (i.e. in the ranges provided in Table\ref{spursrcs}). The lower set of plots show magnitude differences for photometry done in fields from different exposure time categories (e.g. DIS-MIS). The lightly shaded rectangle spanning the length of each plot represent the 1$\sigma$ spread in the magnitude of all the differences. The dotted horizontal line marks the value of the mean of all the differences. The numerical values of these two quantities are given in the upper left corner of each plot.}
    \label{ext_phottests}
    \end{figure*}

    \begin{figure*}
    \centering
    \includegraphics[width=15cm,height=23cm,keepaspectratio]{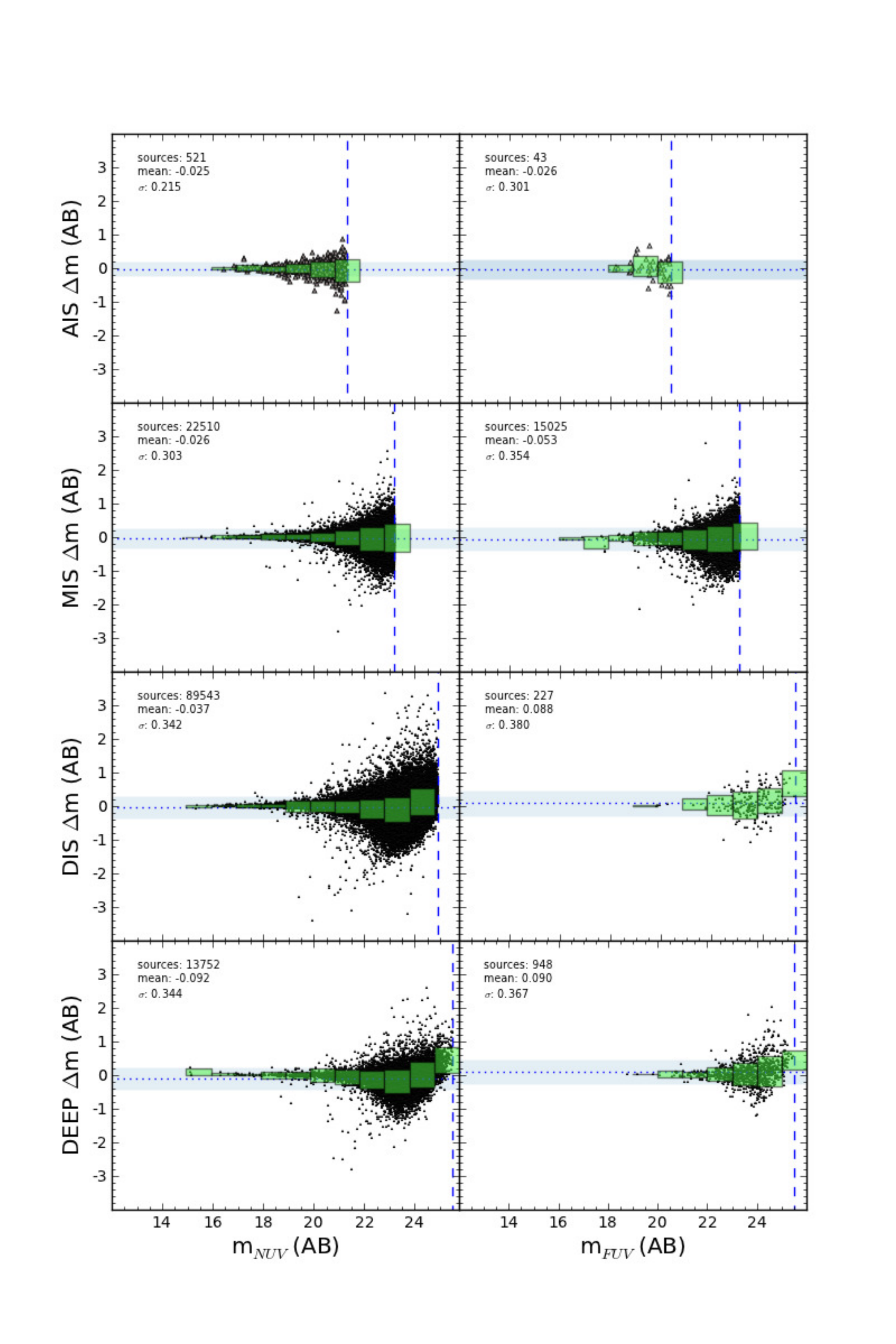}
    \caption{Difference in UV magnitudes versus UV magnitudes of GUViCS point-like sources. NUV and FUV data are plotted in the left and right panels, respectively. All plots show magnitude differences for photometry done in fields from the same exposure time categories (i.e. in the ranges provided in Table \ref{spursrcs}). The lightly shaded rectangle spanning the length of each plot represents the 1$\sigma$ spread in the magnitude of all the differences. The dotted horizontal line marks the value of the mean of all the differences. The numerical values of these two quantities are given in the upper left corner of each plot. The transparent boxes marking each magnitude bin in each distribution provide the 1$\sigma$ spread within that magnitude bin, and are centered at the mean value for that bin.}
    \label{point_phottests}
    \end{figure*}

  \subsection{Variations with Exposure Time}
  An effective way to estimate the uncertainty of a given magnitude is to determine the change in magnitude when it is measured in an image of lower exposure time. By performing this test with larger statistics in images of similar depths, it is possible to buildup a quantitative estimate of magnitude uncertainty, as well as an understanding of the faint magnitude limits. We perform these tests for both extended and point-like sources in the final GUViCS catalogs.

  Figure \ref{ext_phottests} shows the NUV and FUV change in magnitude, with respect to the magnitude in the deepest image, for all extended sources that have coverage in a shallower \textit{GALEX} field. The upper two plots show the magnitude differences between fields within the same exposure time ranges from Table \ref{spursrcs}, while the lower two plots show magnitude differences between fields in different exposure time ranges. The exposure time of the deepest field is indicated by the color of each point as shown in the legend. First, the NUV data shows much smaller differences in magnitude, on average, compared to the FUV data. This could be explained by larger NUV statistics, as well as deeper NUV coverage as a result of the early decommissioning of the \textit{GALEX} FUV instrument. However, the uncertainty in magnitude appears to increase at similar points for both sets of plots: $m_{UV} \sim$20 between images of differing exposure times, and $m_{UV} \sim$23 between images of similar exposure times. Thus, sources fainter than $m_{UV} \sim$20 can have an uncertainty of at least $\pm$0.3 magnitude (or $\pm$1 in a few cases) in an image shallower than MIS or DIS data. This uncertainty largely increases for sources fainter than $m_{UV} \sim$22.
    
  We can quantify the uncertainty in \textit{GALEX} pipeline magnitudes of point-like sources by comparing magnitudes taken in images of different depths for the same source. Figure \ref{point_phottests} shows the differences in magnitude as a function of the deepest magnitude for point-like sources with shallower overlapping \textit{GALEX} fields within similar exposure time ranges Table \ref{spursrcs}). From top to bottom are the AIS, MIS, DIS, and DEEP distributions. Their average 1$\sigma$ variation is represented by the large shaded rectangles, and the 1$\sigma$ variation within each magnitude bin is marked by the green transparent boxes centered at the median value in each bin on the vertical axis. The low statistics in the AIS plots are a result of the initial cutting a majority of spurious AIS pipeline sources from the catalog (i.e. AIS-depth sources fainter than $m_{NUV}$ = 21.3 and $m_{FUV}$ = 20.4), leaving an average of $\sim$180 NUV and $\sim$35 FUV sources in AIS fields. Thus, the chance of finding a source match between overlapping AIS images of different exposure times is highly unlikely. For most plots in Figure \ref{point_phottests}, excluding the FUV DIS and DEEP plots, the bin-by-bin uncertainty estimates remain lower than the overall 1$\sigma$ uncertainty until $m_{uv} \sim$23. Thus, we can reasonably say that pipeline photometry up to this magnitude has uncertainties within $\pm$0.2$\sigma$, on average.

%
\section{Matching with Optical Data}
\label{matchoptcat}
In the following sections we discuss the details of cross-matching the UV-VPS catalog with the SDSS 9th data release (DR9), the NGVS source catalog, and NED. We do this in order to categorize as many UV sources as possible as background objects, Virgo Cluster members, and stars, as well as to obtain optical photometry in various aperture types, $z_{phot}$, and $z_{spec}$, where available. Additionally, matching the UV-VPS catalog with the NED database locates additional extended Virgo sources that are not part of the VCC and CGCG dataset from GOLDMine.

\begin{figure}
\resizebox{\hsize}{!}{\includegraphics{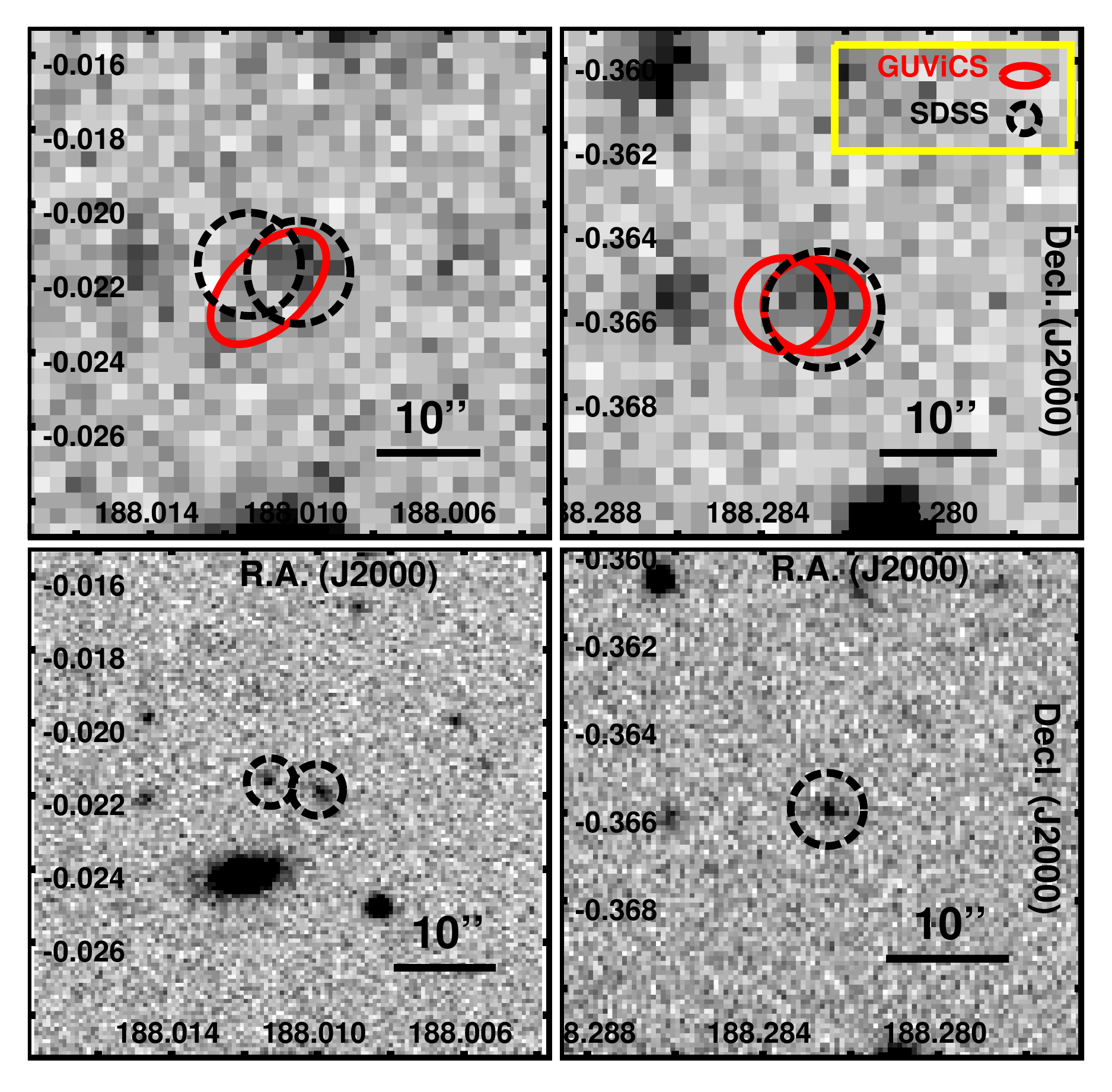}}
\caption{Examples of two multiple matching scenarios between the UV-VPS catalog and the SDSS database. The upper panel shows \textit{GALEX} NUV images, where GUViCS sources are inside the solid red Kron apertures, and SDSS sources are in the dashed black 5$''$ radii apertures. The lower panel shows SDSS $i'$ images for comparison, with SDSS sources inside the dashed black regions. \textbf{Left:} A single GUViCS source finds a match to two SDSS sources within 5$''$. \textbf{Right:} A single SDSS source has two GUViCS matches within 5$''$ of its center. These types of multiple matches are excluded from the UV-VPS catalog in Appendix \ref{appendix_UVVC_ptsrc_cat}, but can be recovered in the full SDSS-GUViCS matched catalog in Appendix \ref{appendix_guvics-sdss_cat} using the $uv\_vcc\_id$ catalog parameter.}
\label{guv_sdss_multi}
\end{figure}

We note that in the cases of matching to SDSS and NGVS, two scenarios of multiple source matching can occur. First, GUViCS sources tend to find two or more optical matches in situations where two or more individual sources are blended in the UV image. This is especially problematic for identifying the optical counterparts of chance aligned background sources, as well as background sources aligned with foreground stars or relatively small Virgo member galaxies. Secondly, there is a small percentage of multiple GUViCS sources matched to the same optical source when UV sources are shred by the \textit{GALEX} pipeline. Since all known extended sources and bright stars have already been removed from the UV-VPS catalog, this population represents cases of shredding of point-like sources only. Examples of both multiple match scenarios with SDSS data are shown in Figure \ref{guv_sdss_multi}.

Below, we discuss the details of matching with each of these optical datasets.


  \subsection{GUViCS-SDSS Matching}
  \label{sdss}
  SDSS provides the most extensive set of broadband optical data of extragalactic sources to date down to an $r'$-band point source magnitude of $m_{r'}$ = 22.2 at 95\% completeness with a median point spread function FWHM = 1.3$''$. DR9\footnote{http://www.sdss3.org/dr9/} \citep[released August 2012; ][]{ahn12} includes all data from prior SDSS data releases with improvements in the determination of stellar parameters and astrometry since DR8 \citep{aih11}.

  We match GUViCS to SDSS optical sources within a tolerance of 5$''$ of angular separation on the sky. This tolerance was selected based on the \textit{GALEX} point spread function ($\sim$5$''$); the \textit{GALEX} point spread function cannot distinguish two sources within $\sim$5$''$ of one another and thus two or more optical sources within this radius should most probably be matched with the same \textit{GALEX} source. However, we note that this match tolerance will introduce a notable number of multiply-matched sources \citep{bud09}. We find 926,147 UV sources (75\% of the UV-VPS catalog) with one or more optical counterparts in the SDSS database. About 86\% of these are one-to-one matches, 13\% are one-to-multiple optical-to-UV matches, and 1\% are one-to-multiple UV-to-optical matches. The complete statistics are provided in Table \ref{guv_sdss_mtch_stats}. 

   \begin{table}
   \caption{Statistics of GUViCS-to-SDSS UV-optical matching.}             
   \label{guv_sdss_mtch_stats}      
   \centering          
   \begin{tabular}{rrrr}     
   \hline\hline       
   Match &        & \% of Sources & \% Total \\ 
   Type  & Number &  Matched      & UV-VPS\\ 
   \hline                    
      \textbf{One-to-One} & 799486 & 86.3 & 65.0\\
      NUV & 785823 & 86.7 & 65.6\\
      FUV & 141548 & 82.6 & 70.7\\
\hline
      \textbf{Multiple}&&&\\
      \textbf{SDSS-to-GUViCS} & 119090 & 12.9 & 9.7\\
      NUV & 116634 & 12.9 & 9.7\\
      FUV & 25787 & 15.1 & 12.9\\
\hline
      \textbf{Multiple}&&&\\
      \textbf{GUViCS-to-SDSS} & 9697 & 1.0 & 0.8\\
      NUV & 5138 & 0.6 & 0.4\\
      FUV & 4897 & 2.9 & 2.4\\
   \hline
   \end{tabular}
  \tablefoot{These statistics are calculated from the final version of the UV-VPS catalog in Appendix \ref{appendix_UVVC_ptsrc_cat} that has 1,230,85 UV sources. This catalog has been cleaned of all sources within apertures of additional extended sources found in NED (see Section \ref{ned_matching}).\\
    }
   \end{table}

    \subsubsection{One-to-One Matches}
    Only the one-to-one GUViCS-SDSS matches are categorized as stars, Virgo Cluster members, and background sources. To select these sources we follow the same criteria outlined in \citet{bos11}. Stars are determined from SDSS parameters in two ways. First, we find all spectroscopically identified stars in the SDSS data\footnote{i.e. SDSS parameter \textit{class} = STAR}. Secondly, we select stars determined photometrically by SDSS with $r'$-band magnitudes $\le$ 21   \footnote{i.e Objects have SDSS parameters \textit{type} = 6 and \textit{cModelMag\_r} $\le$ 21. SDSS photometrically separates stars from galaxies by comparing their \textit{cModelMag} and \textit{psfMag} measurements. Any source that has \textit{psfMag}-\textit{cModelMag} $\le$ 0.145 is classified as a star.}. This criteria has been tested and confirmed by several authors \citep{lup01,oya08,ham10}. Next, we categorize sources with $z_{spec}$ $<$ 0.01167 ($v_{vel}$ $<$ 3,500 km/s taking into account the elongated structure of the Virgo Cluster \citep{gav99}) as Virgo Cluster members, and sources with $z_{spec}$ $\ge$ 0.01167 as background sources. In the remaining catalog we consider any sources with clean photometry, $r'$-band magnitudes\footnote{i.e. SDSS parameters \textit{clean} = 1, and \textit{cModelMag\_r} $<$ 20} $<$ 20, and $z_{phot}$ $\ge$ 0.1 as background sources \citep{oya08}.  

    All sources remaining uncategorized after the application of these conditions are considered `to be determined' (TBD) sources and are flagged as such in the final UV-VPS catalog\footnote{See the \textit{SDSS\_SRCS\_TBD} field in Table \ref{UV-VPS_cols}}. GUViCS source matches are found with 134,006 stars, 86,921 background sources, 103 Virgo members, and there are 575,500 TBD matches. We present the matched GUViCS-SDSS data in Appendix \ref{appendix_UVVC_ptsrc_cat} and \ref{appendix_guvics-sdss_cat}.

  \subsection{GUViCS-NGVS Matching}
  \label{ngvs}
  NGVS \citep{fer12} is a Large Program  carried out on the 3.6m CFHT using the 1 deg$^2$ MegaCam imager \citep{bou03}.  The survey targets 104 square degrees in Virgo, covering the cluster from its core to virial radius in 117 pointings. At the time of writing, the survey is complete in the $u^*$, $g'$, $i'$, and $z'$ filters, with only limited coverage in $r'$. The NGVS is significantly deeper than the SDSS, reaching a point source magnitude depth $m_{g'}$ = 25.7 at a S/N = 10 and a surface brightness depth of $\mu_{g'}$ = 29 mag arcsec$^{-2}$. The spatial resolution (between 0.5 and 0.9 arcsec, depending on the filter) is also superior to the SDSS, improving morphological identification of background galaxies. Details of the NGVS source catalog, used here, are provided in Appendix \ref{ngvs_cat}.

  Again, we match the UV-VPS and NGVS catalogs within a tolerance of 5$''$ in angular separation since this is approximately the minimum point spread function of \textit{GALEX}. Considering only GUViCS sources located in the NGVS footprint (706,847 UV sources, $\sim$57\% of the entire UV-VPS catalog), we find 683,888 (96\%) have at least one match in the NGVS source catalog. Of these, 31.5\% are one-to-one matches, 68\% are multiple NGVS-to-GUViCS matches, and 1\% are multiple GUViCS-to-NGVS matches. Table \ref{guv_ngvs_mtch_stats} provides the complete matching statistics. We use the NGVS catalog parameters of one-to-one GUViCS-NGVS matches to categorize UV objects as stars, Virgo members, background sources, and sources `to be determined$'$ (TBD). Background sources and Virgo members are selected solely on their $z_{spec}$, where available, and the majority of sources we find within the cluster are most likely globular clusters or ultra-compact dwarfs (UCDs). The TBD sources consist of any one-to-one matches not in the other three categories and with a good $m_{g'}$. There are 16,016 UV sources matched to stars, 521 matched to background sources, 8 matched to Virgo members, and 154,865 TBD matches.

  Currently, we do not provide NGVS photometric data for these sources in the UV-VPS catalog since the final NGVS catalog is still in preparation. However, in Appendix \ref{appendix_UVVC_ptsrc_cat}, we do provide columns that indicate which UV sources have NGVS counterparts, which are stars, and note whether these are one-to-one or multiple matches\footnote{i.e. fields \textit{NGVS\_NUM\_MTCHS}, \textit{NGVS\_STAR}, and \textit{NGVS\_multUV} in Table \ref{UV-VPS_cols}.}. Once the NGVS source catalog is published (S. Gwyn et al., \textit{in preparation}), the full data will be added to the UV-VPS catalog on the GUViCS website. 

   \begin{table}
   \caption{Statistics of GUViCS-to-NGVS UV-optical matching within NGVS footprint.}             
   \label{guv_ngvs_mtch_stats}      
   \centering          
   \begin{tabular}{rrrr}     
   \hline\hline       
   Match &        & \% of Sources & \% Total \\ 
   Type  & Number &  Matched      & UV-VPS\\
   \hline                    
   \textbf{One-to-One} & 215631 & 31.5 & 30.5\\
   NUV & 209597 & 31.5 & 30.5\\
   FUV & 42932 & 39.8 & 38.9\\
   \hline
   \textbf{Multiple}&&&\\
   \textbf{NGVS-to-GUViCS} & 466450 & 68.2 & 66.0\\
   NUV & 455947 & 68.4 & 66.3\\
   FUV & 63923 & 59.3 & 57.9\\
   \hline
   \textbf{Multiple}&&&\\
   \textbf{GUViCS-to-NGVS} & 9211 & 1.3 & 1.3\\
   NUV & 6013 & 0.9 & 0.9\\
   FUV & 3703 & 3.4 & 3.4\\

   \hline
   \end{tabular}
   \tablefoot{These statistics are calculated from the final version of the UV-VPS catalog in Appendix \ref{appendix_UVVC_ptsrc_cat} that has 1,230,85 UV sources. This catalog has been cleaned of all sources within apertures of additional extended sources found in NED (see Section \ref{ned_matching}).\\
    }
   \end{table}

    \begin{figure*}[!ht]
    \resizebox{\hsize}{!}{\includegraphics{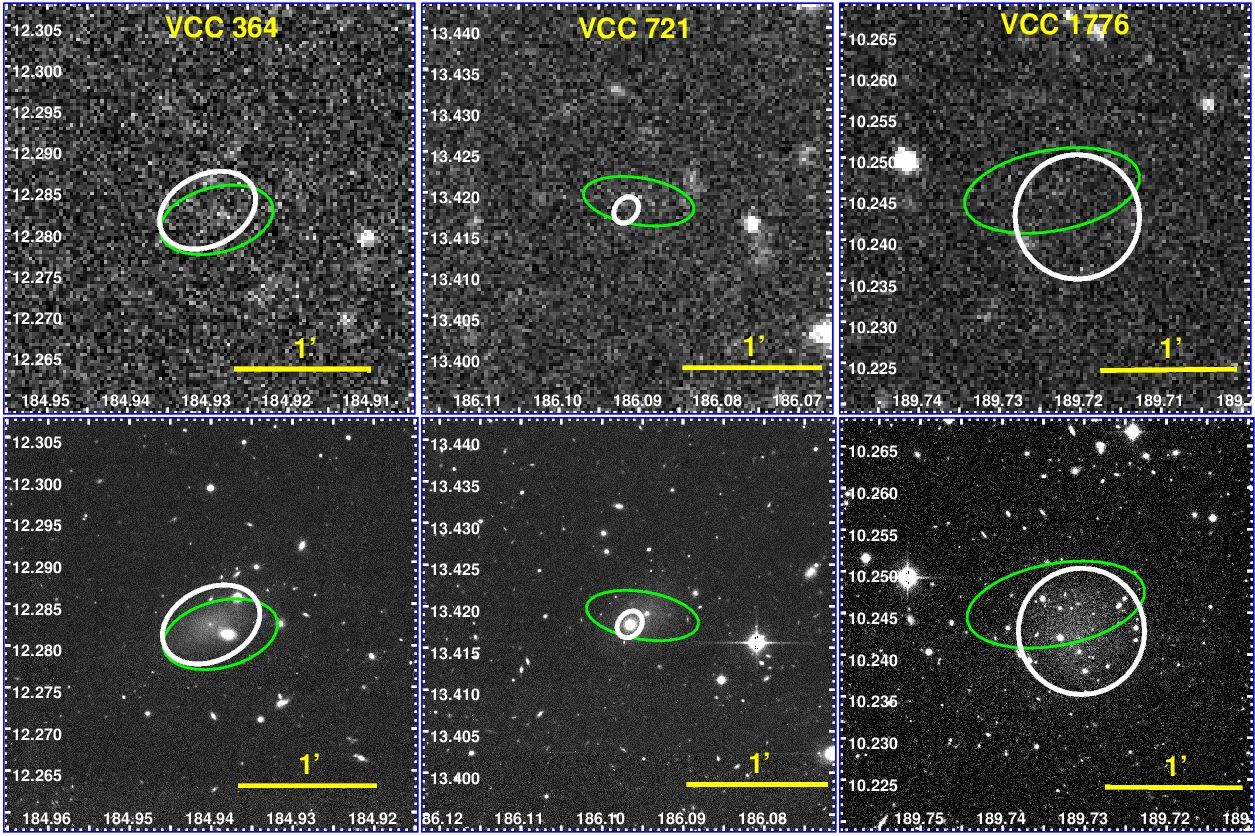}}
    \caption{Extended sources VCC 364 (type Im), VCC 721 (type dE), and VCC 1776 (type dIm/dE) are three examples where an extended source finds a match in the \textit{GALEX} pipeline catalogs, but, upon visual inspection is not detected in the UV. The horizontal axis coordinates are degrees of right ascension (J2000), and the vertical axis coordinates are degrees of declination (J2000). The upper panel shows \textit{GALEX} NUV images, and the bottom panel shows the NGVS $i'$-band images for comparison. The thick white regions are the VCC optical apertures, and the thin green regions are the \textit{GALEX} pipeline Kron elliptical apertures.}
    \label{ext_nondet}
    \end{figure*}

  \subsection{GUViCS-NED Matching}
  \label{ned_matching}
  We make a final cross-match of the UV-VPS catalog with NED in order to obtain any further redshift information on GUViCS sources, and to find any other extended sources that are not CGCG sources in the GOLDMine database. We batch query the NED database within the areas of 32 circular fields with 300$'$ radii each that together provide complete coverage of the GUViCS footprint. Since there are large overlaps between the 32 fields, we remove any duplicate sources from our query output. We retrieve NED object names, central coordinates, redshifts, and redshift quality flags (QF) for each NED source. QFs provide information on how each redshift was obtained. We use the QFs to make sure we use spectroscopically based redshifts to further categorize the UV sources. We select QFs of type SPEC (spectroscopic redshift) and 1LIN (spectroscopic redshift from a single spectral line) as reliable redshifts with the ability to sort between foreground and background sources in the cluster. All sources with redshifts associated with any other QF are flagged as TBD sources in the UV-VPS catalog. The final NED catalog contains 35,575 sources, and 26,915 (76\%) of these have $z_{spec}$, enabling the separation of Virgo Cluster members from background sources for those objects also detected in the UV data.

    \begin{figure*}[!ht]
    \resizebox{\hsize}{!}{\includegraphics{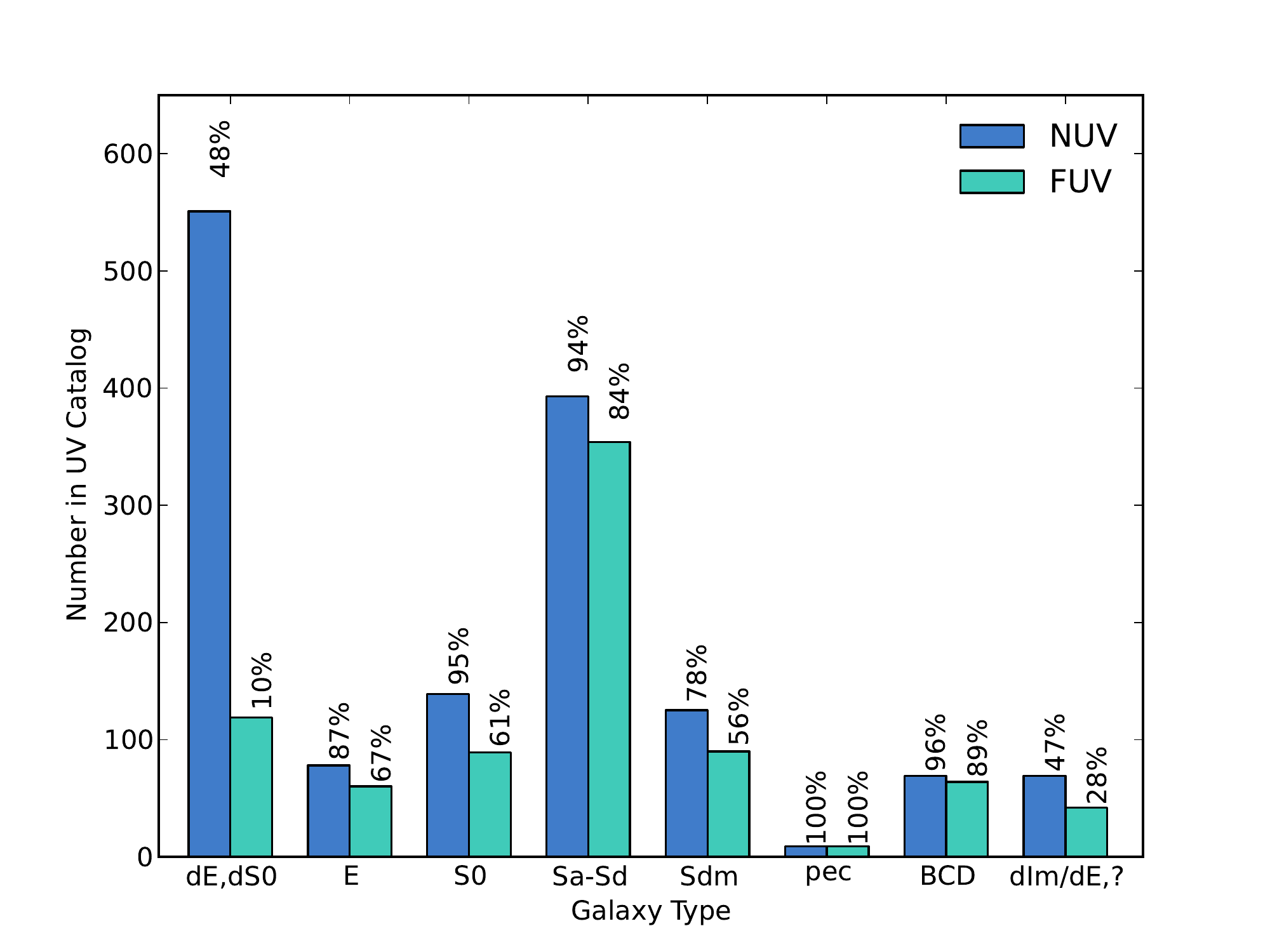}}
    \caption{Distribution of GOLDMine UV VCC and CGCG extended source galaxy types detected in the GUViCS area. Percentages are of the total number of that optical galaxy type within the GUViCS area, per band. This does not necessarily indicate the total percent detected for these types in the UV because some of these galaxies could have been categorized as non-detections due to shallow coverage in \textit{GALEX} data.}
    \label{hist_gtypes}
    \end{figure*}

  About 0.5\% of objects in the UV-VPS find a match in the NED catalog within a tolerance of 4$''$. This tolerance is 1$''$ lower than that used for matching with the SDSS and NGVS data because we know that NED sources are resolved and extended, thus the poor GALEX point spread function is less problematic in this situation. Statistics of one-to-one and multiple matches are provided in Table \ref{guv_ned_mtch_stats}. There are 332 point-like source matches of which 284 are background sources (i.e. $z_{spec}$ $\ge$ 0.01167), 27 are new Virgo members (i.e. $z_{spec}$ $<$ 0.01167), 18 are stars, 1 is a globular cluster, and 2 objects are questionable. NED redshifts and flags are listed in the UV-VPS catalog in Appendix \ref{appendix_UVVC_ptsrc_cat}. An additional 95 sources are matched to NED extended sources. Therefore, we remove these sources from the UV-VPS catalog and add them to the UV-VES catalog.

   \begin{table}
   \caption{Statistics of GUViCS-to-NED UV-optical matching.}             
   \label{guv_ned_mtch_stats}      
   \centering          
   \begin{tabular}{rrrr}     
   \hline\hline       
   Match &        & \% of Sources & \% Total \\ 
   Type  & Number &  Matched      & UV-VPS\\
   \hline                    
   \textbf{One-to-One} & 5916 & 95.9 & 0.5\\
   NUV & 5762 & 96.9 & 0.5\\
   FUV & 2935 & 96.5 & 1.5\\
   \hline
   \textbf{Multiple}&&&\\
   \textbf{NED-to-GUViCS} & 121 & 2.0 & 0.0\\
   NUV & 117 & 2.0 & 0.0\\
   FUV & 41 & 1.3 & 0.0\\
   \hline
   \textbf{Multiple}&&&\\
   \textbf{GUViCS-to-NED} & 130 & 2.1 & 0.0\\
   NUV & 65 & 1.1 & 0.0\\
   FUV & 67 & 2.2 & 0.0\\
   \hline
   \end{tabular}
   \tablefoot{These statistics are calculated from the initial version of the UV-VPS catalog that has 1,231,331 UV sources.\\
    }
   \end{table}

\section{Analysis}
\label{analysis}

  \subsection{UV Extended Sources}
  Through a visual investigation we find that 65\% (1,436) of all VCC and CGCG Virgo Cluster sources are UV detections, and 44\% of these (625) do not have any \textit{GALEX} FUV observations. There are 366 VCC and CGCG extended sources that are not detected in the UV, i.e. their signal is at the level of noise in the \textit{GALEX} image. The majority of these galaxies are dEs, irregulars, and unknown types. However, 136 of these sources only have coverage in AIS-depth images, and might be detectable at faint levels in deeper data. Figure \ref{ext_nondet} shows three examples of non-UV detections in \textit{GALEX} images deeper than AIS-depth. 

  Figure \ref{hist_gtypes} shows the distribution of extended galaxy types detected in the UV data for VCC and CGCG GOLDMine sources. The NUV detects $\gtrsim$ 90\% of all galaxy types in Virgo, except for dE, dS0, Sdm, and dIm/dE types. The majority of extended Virgo Cluster galaxies emitting in the FUV are Sa-Sd, BCDs, and peculiar galaxy types. The majority of galaxy types detected in the NUV are dE and dS0 types. However, they only represent 48\% of total number of dEs and dS0s in the entire cluster. Additionally, they are also the lowest number of detections per type in the FUV-band due to their red colors \citep{bos05}. We note that the lower percentage of FUV detected S0s compared to Es detected in the FUV is most likely due to a lack of deep FUV coverage for S0s. The morphology-density effect \citep{dre80,bin87,whi93} states that Es dominate cluster cores, and that S0s are more abundant in central parts of clusters, but are generally lacking in the core. The central part of the Virgo Cluster has deep FUV data, and thus coverage for Es, while these data are more sporadic towards the periphery, where the fraction of S0/E increases (see Figure \ref{guvics_fuv_cov}). The periphery is primarily covered by AIS FUV fields that are too shallow to detect many S0s located in the GUViCS footprint. 

  Through our visual investigation, we find that $\sim$13\% of UV detected dEs have nucleated UV cores, and we flag these objects in the UV-VES catalog. Nucleated dE are a typical class of objects in Virgo first identified by the VCC survey \citep{bin85} and discussed in \citet{bin91}. They have been extensively discussed in \citet{cot06}, and first detected in the UV by \citet{bos08a}.

    \subsubsection{Magnitude Distributions}
\begin{figure}
    \centering
    \resizebox{\hsize}{!}{\includegraphics{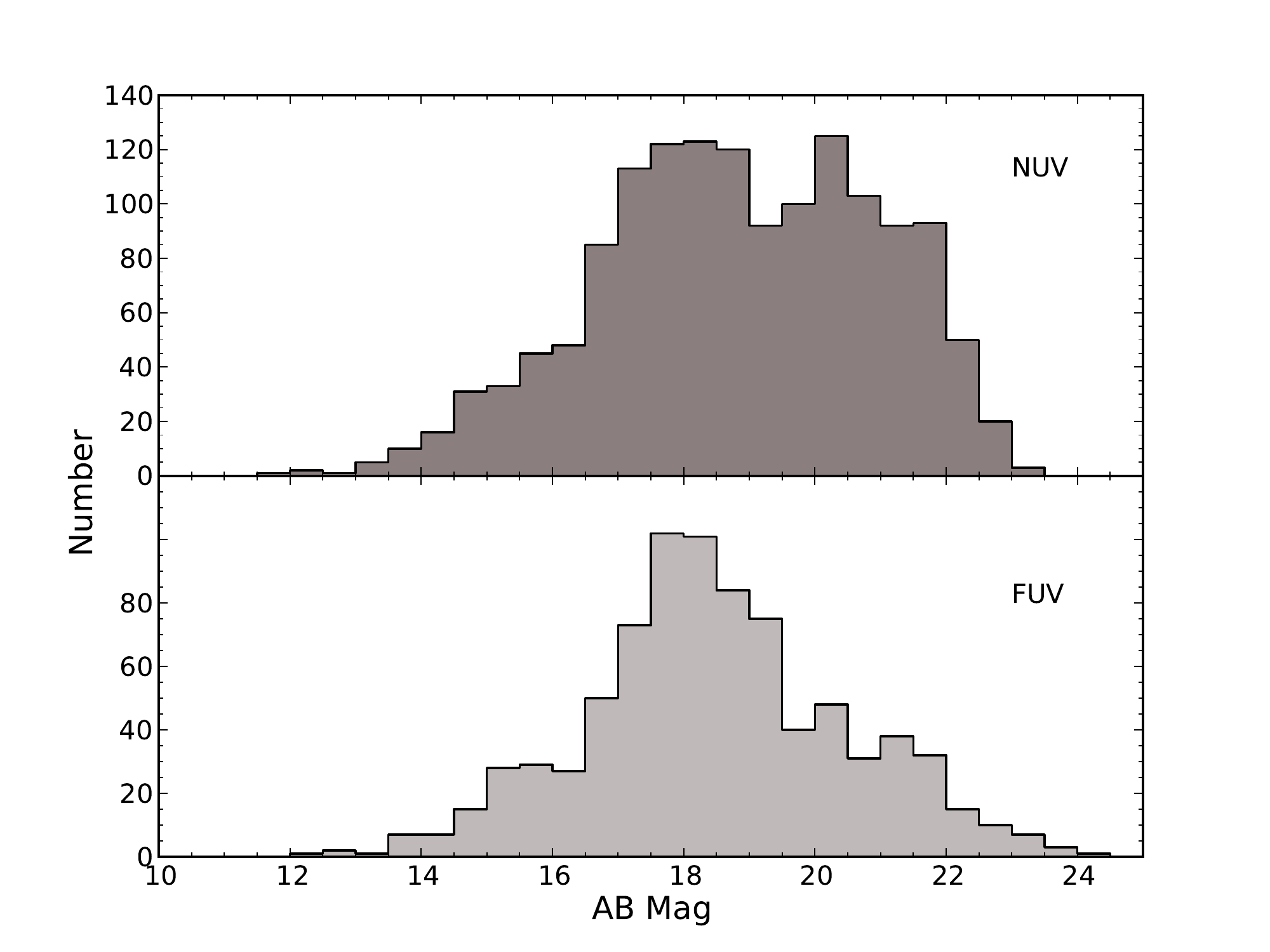}}
    \caption{Distributions of $m_{NUV}$ and $m_{FUV}$ for all extended sources in the Virgo Cluster area.}
    \label{ext_magdist}
    \end{figure}

\begin{figure*}
    \centering
    \includegraphics[width=21cm,height=24cm,keepaspectratio]{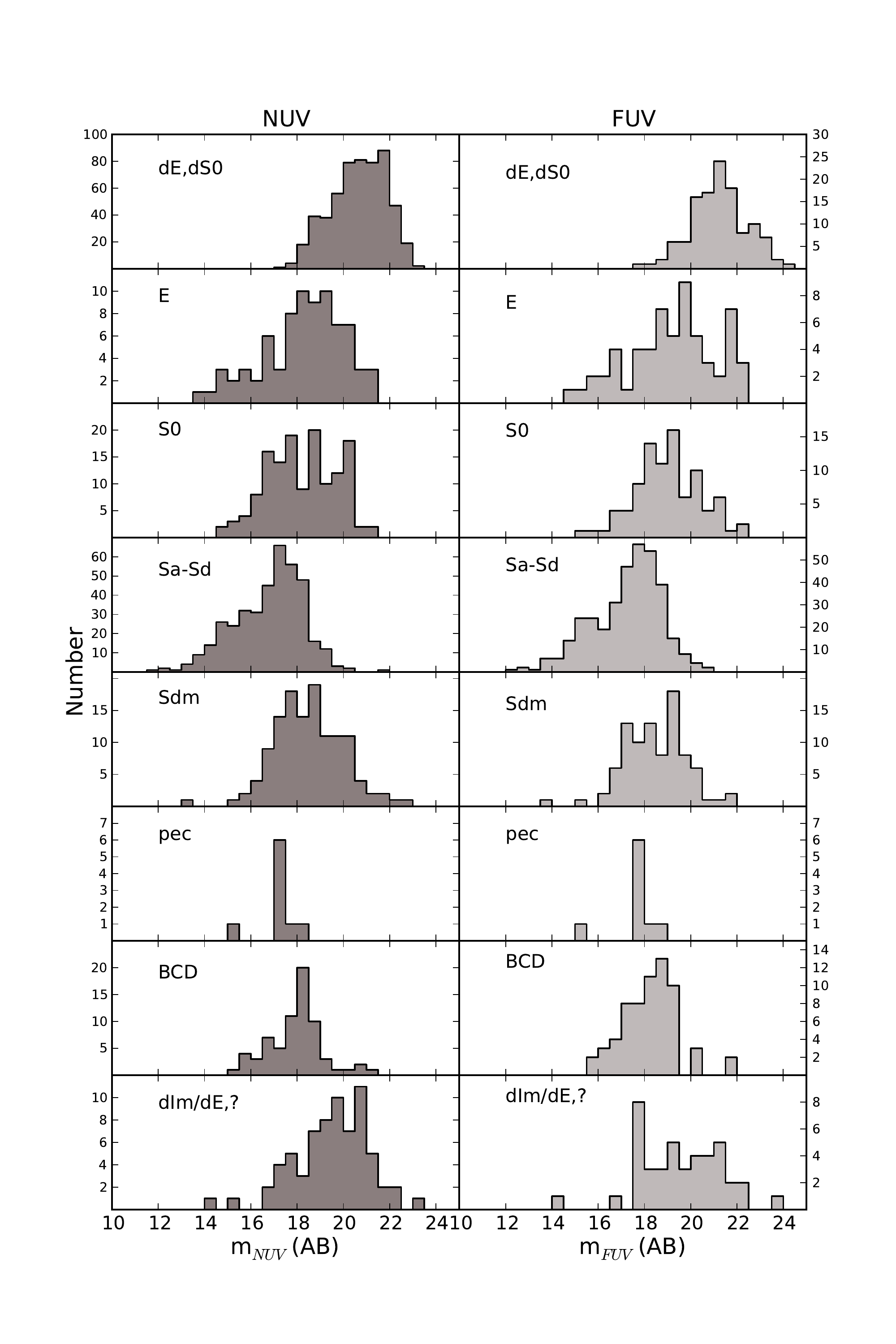}
    \caption{Distribution of NUV and FUV AB magnitudes in morphological bins for all UV extended sources in the Virgo Cluster area. Morphologies of CGCG sources are from GOLDMine and NED, and morphologies of VCC sources are from \citet{bin85,bin93}.}
    \label{ext_magdist_morph}
    \end{figure*}

    The distributions of extended source magnitudes in the NUV and FUV are shown in Figure \ref{ext_magdist}. The NUV distribution begins to drop-off after $m_{NUV}$ $\sim$22 and the FUV distribution after $m_{FUV}$ $\sim$20. Figure \ref{ext_magdist_morph} separates these plots into bins of galaxy type \citep[CGCG from GOLDMine and NED, and VCC from][]{bin85,bin93} showing that the distributions are qualitatively similar between UV bands for each galaxy type, except for the dIm/dE/? types that have a brighter peak in their FUV distribution and not in their NUV distribution.

  \subsection{UV Point-Like Cluster and Background Sources}
   \begin{table*}
   \label{allmatchtab}
   \caption{Number of UV sources categorized from one-to-one optical catalog matching.}             
   \label{optical_match_output}      
   \centering          
   \begin{tabularx}{\textwidth}{rrrrrrrrr}  
   \hline\hline       
                       &   \textbf{Total} & SDSS only & NGVS only & NED only & SDSS+NGVS & SDSS+NED & NGVS+NED & All\\ 
   \hline           
   Net Optical Matches & \textbf{847257}  & 628163    & 46634     &  971     & 166880    & 2492     & 166      & 1951\\
   Stars               & \textbf{137617}  & 121584    &  3597     &   14     &  12419    &    3     &   0      &    0\\
   Virgo Members       &    \textbf{129}  &     91    &     8     &   18     &      0    &   12     &   0      &    0\\
   Background Sources   &  \textbf{88291}  &  82715    &   238     & 1131     &    281    & 3924     &   1      &    1\\
   TBD                 & \textbf{620995}  & 466130    & 45495     &    0     & 109370    &    0     &   0      &    0\\
   \hline
   \end{tabularx}
   \end{table*}

\begin{figure}
    \centering
    \resizebox{\hsize}{!}{\includegraphics{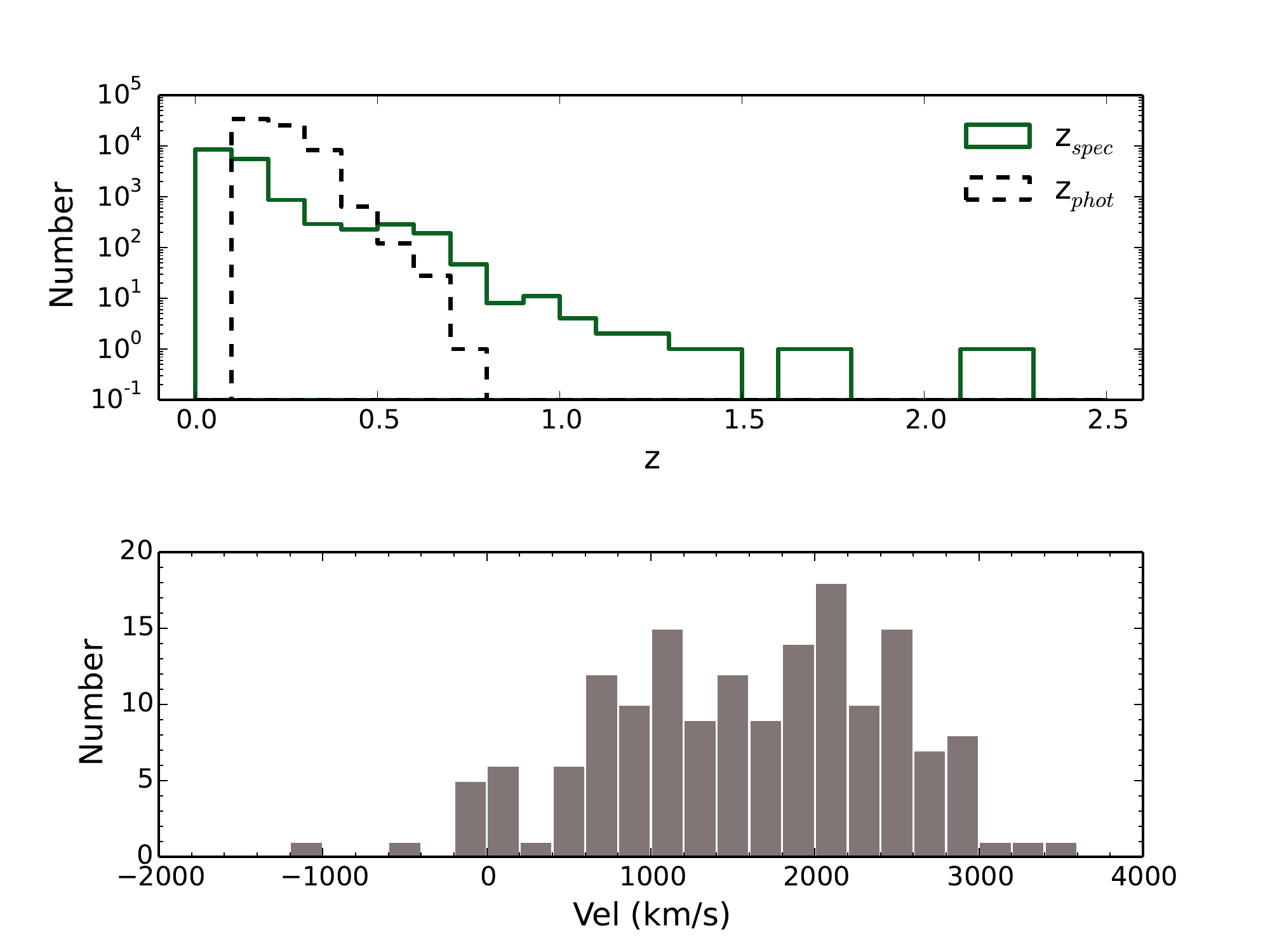}}
    \caption{Redshift distributions for Virgo Cluster members (bottom) and Virgo background galaxies (top). In the upper plot, spectroscopic redshifts are represented by the solid line histogram and photometric redshifts are represented by the dashed line histogram. All redshifts in the lower plot are spectroscopic, and $z$ $<$ 0.01167.}
    \label{point_z_dist}
    \end{figure}

  The combined statistics from matching with SDSS, NGVS, and NED are provided in Table \ref{optical_match_output}. In Appendix \ref{exam_sdss_ngvs} we provide a deeper analysis of GUViCS matches to SDSS and NGVS sources within the NGVS footprint. We find that 69\% of the entire UV-VPS catalog have a one-to-one match in at least one of these optical catalogs. However, the majority of these matches (73\%) do not have available measurements that can distinguish if they are in the foreground or background of the cluster (i.e. TBD sources). The other 27\% of one-to-one matches are either categorized as stars, Virgo members, or background sources. We find 129 new UV Virgo Cluster members that are not in the VCC or CGCG catalog, and 88,291 confirmed cluster background sources. Figure \ref{point_z_dist} shows the redshift distribution of the new Virgo members (bottom plot) and background sources (top plot). The background sources identified via $z_{phot}$ are from SDSS with $m_{r}$ $<$ 20, however, SDSS $z_{phot}$ cannot reliably identify Virgo Cluster members. 

\begin{figure*}
    \centering
     \resizebox{0.7\textheight}{!}{\includegraphics{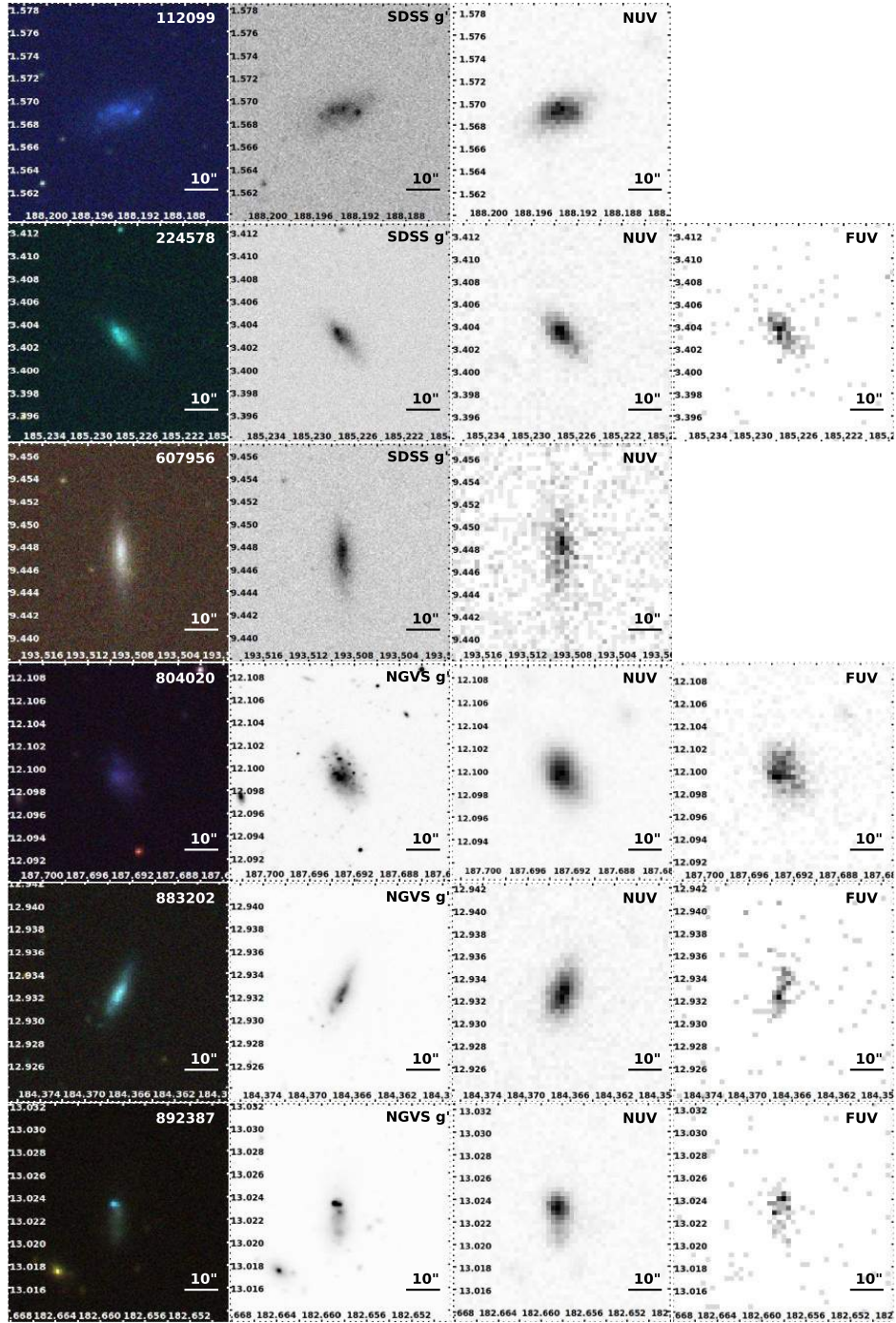}}
    \caption{Gallery of UV Virgo Cluster sources in the UV-VPS catalog that are spectroscopically confirmed Virgo Cluster members from the SDSS and NED. Shown, from left to right are the SDSS g'r'i' image, the SDSS or NGVS g' image, the \textit{GALEX} NUV image, and \textit{GALEX} FUV image, where available. Right ascension (J2000) is given on the horizontal axis in degrees and declination (J2000) on the vertical axis in degrees. The UV-VPS catalog ID is provided in the top right corner of the g'r'i' image. The full gallery is available in the online-only material for this paper.}
    \label{vm_gal}
    \end{figure*}

  The majority of the new Virgo members (121 sources) are blue dwarf-like galaxies, ranging from extremely bright to very diffuse sources in both the UV and optical images. The remaining 8 members are either globular clusters (GCs) or ultra-compact dwarfs (UCDs). Given the elongated structure of the cluster, we use the same selection for Virgo members as \citet{bos11}, i.e. everything at $z_{spec}$ $<$ 0.01167 ($vel$ $<$ 3500 km s$^{-1}$), in order to include sources from all of Virgo's substructures \citep{bin88,bin93,gav99}. A multiwavelength gallery of the new UV cluster members is presented in Figure \ref{vm_gal}. The gallery displays images from the FUV-band, NUV-band, NGVS or SDSS $g'$-band, and SDSS color combined images ($g'r'i'$). 

  The 8 GC and UCD sources are detected by matching with the NGVS catalog. Three of these sources were originally identified by \citet{han01}, and they are flagged as such in the UV-VPS catalog. The remaining 5 NGVS-matched new Virgo members are identified in E. Peng et al. (\textit{in preparation}) and H. Zhang et al. (\textit{in preparation}). They have made further observations of NGVS GCs and UCD candidates with the Hectospec instrument on the 6.5m MMT and the Anglo-Australian Telescope. Based on their initial results, we have identified these 5 as GC/UCD cluster members from Hectospec data. The 5 GCs/UCDs are classified in the NGVS catalog as stars based on their photometry, and they all have $m_{NUV}$ $>$ 21. Additionally, this team has made spectroscopic observations of 519 of the UV background sources with matches in the NGVS catalog as UV sources. Upon the publication of  E. Peng et al. (\textit{in preparation}) and H. Zhang et al. (\textit{in preparation}), we will update the GUViCS catalog with these data.

  In addition to these 8 compact Virgo members identified in GUViCS-NGVS matching, we find 15 more UCDs in the UV-VPS catalog. Six were first identified as GUViCS UV sources in \citet{bos11}, and 9 are UCDs from \citet{bro11}. Ultimately, these 23 total compact sources identified in the UV-VPS catalog are excellent candidates for further studies of the UV-optical properties, and stellar populations, of compact objects in Virgo.

\begin{figure*}
    \centering
    \resizebox{\hsize}{!}{\includegraphics{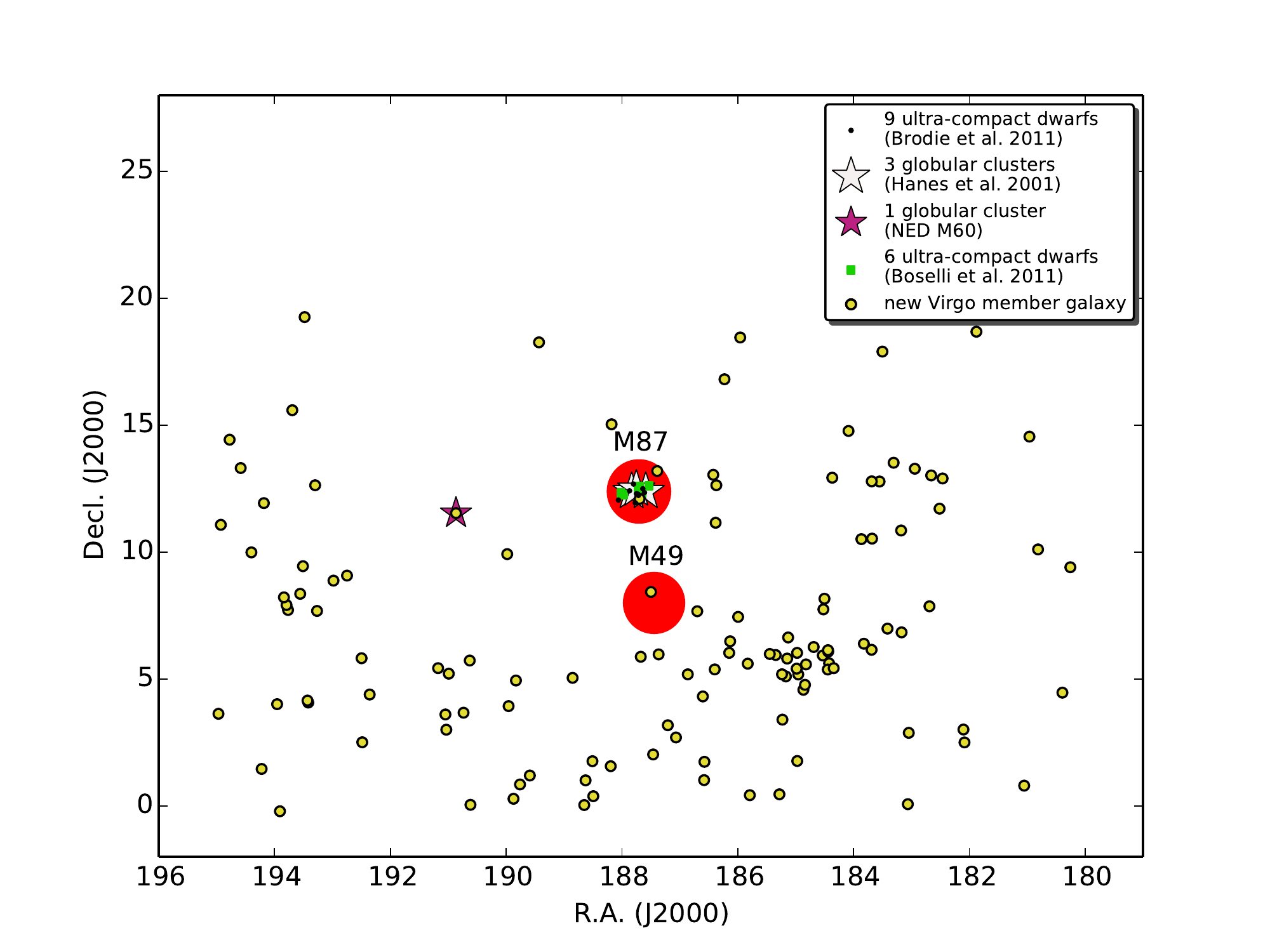}}
    \caption{Spatial mapping of Virgo Cluster members that are neither part of the VCC, nor the CGCG, nor an additional extended source from NED. The positions of M49 and M87 are shown for reference. Three globular clusters are identified from the NGVS and one from NED. The 6 UCDs from \citet{bos11} and 9 UCDs from \citet{bro11} are also shown. Although we currently find low statistics for GCs and UCDs in Virgo from the literature, ongoing spectroscopic studies (e.g. Peng et al., \textit{in preparation}) are expected to find larger samples of these sources. Thus, here we point out that the brightest GCs and UCDs are detectable in the UV, and future studies can be performed with upcoming larger samples.}
    \label{vm_distr}
    \end{figure*}

  Figure \ref{vm_distr} shows the 2-dimensional spatial distribution of the 121 new UV Virgo galaxies and the 23 UV-detected compact objects in the cluster. The majority of compact sources are located around M87, except for one GC around M60. The new blue dwarf-like galaxies avoid the central regions of the cluster and extend to the cluster outskirts, following the well studied morphology-density relation \citep{dre80,whi93} that shows redder early-type sources mainly populate the central regions of a cluster, and the highest density of bluer, less evolved, sources are typically at larger cluster radii. \citet{bos08a,bos08b} show that dwarf galaxies falling into the cluster are affected by ram-pressure stripping that quenches their star formation. They also show that dwarf galaxies may also be the remnants of low luminosity disk galaxies affected by ram-pressure stripping as they fall into the cluster. Figure \ref{vm_distr} demonstrates that star-forming dwarf galaxies are rare in the center of the cluster, lending evidence to the idea that star-formation is quenched by the ram-pressure stripping scenario in the cluster outskirts. This sample provides a good data set for further studies on the fate of star-forming dwarf galaxies in Virgo and their contributions to the overall cluster evolution (A. Boselli et al., \textit{in preparation}).     


  We investigate the remaining UV point-sources within the NGVS footprint that were not matched to any optical counterpart in SDSS, NGVS, and NED. If we consider all these sources to be spurious UV detections, they would contaminate the catalog at extremely low levels. Down to completeness limits \citep{mor07} for AIS-depth images, the contamination would be $<$ 1\% in NUV and 3.5\% in FUV. For MIS-depth and DIS-depth it would be $<$ 1\% in NUV and FUV. Therefore, we can estimate that statistical studies making use of the entire UV-VPS catalog can expect very low levels of spurious source contamination, $\sim$1\% on average, in both the NUV and FUV. However, we further investigate these unmatched UV sources to determine if they are actually all spurious detections, or if other situations occur.

  The NUV and FUV S/N and magnitude distributions for the remaining sources with no optical counterparts are shown in Figure \ref{no_match_sn_uvmag_dist}. The majority of these sources have UV magnitudes beyond the reliability estimates for the UV-VPS catalog, as discussed in Appendix \ref{complete}. The S/N distributions in both UV bands peak near zero, and we find (via visual checks) that the majority of these unmatched sources are indeed spurious detections. We have also investigated the 22 NUV sources with S/N > 100 and find the majority of these sources are stars not identified in the NGVS or the SDSS, a few are spurious detections in ghosts of stars, and at least one is a real background source that is not in the SDSS catalogs, or in the current version of the NGVS catalog. These sources may also be found in the distributions at lower S/N. There are 273 NUV and 96 FUV sources that have $m_{UV}$ $<$ 21. We randomly select and investigate several of these sources and find them to be one of the following: 
  \begin{itemize}
    \item spurious detection in the ghosts of a stars\footnote{i.e. \textit{[FUV][NUV]\_GOOD\_PHOTOM\_FLAG}=1}
    \item source within the halo of a bright star that has been masked in SDSS and NGVS
    \item a real, bright, UV-optical source that is not in the current version of the NGVS source catalog, nor in SDSS 
    \item small background source that is blended by the GALEX pipeline into a single source with a center $>$ 5$''$ from the optical center of any of the individual sources (thus, not matched to any SDSS or NGVS source)
  \end{itemize}
  We expect that the real, bright, UV-optical sources will find one-to-one matches in the final NGVS catalog. Also, we reiterate that the optical sources blended in the UV are a symptom of the \textit{GALEX} pipeline catalogs. In a brief investigation of GUViCS sources with $m_{UV}$ $\ge$ 21, we find a combination of the situations listed above, and additional spurious UV detections. 

\section{Summary and Results}
\label{summ}
The GUViCS catalogs provide the most extensive UV dataset of the Virgo Cluster to date, covering $\sim$120 deg$^2$ in the NUV-band and $\sim$40 deg$^2$ in the FUV band between 180$^{\circ}$ $\le$ R.A. $\le$ 195$^{\circ}$ and 0$^{\circ}$ $\le$ Decl. $\le$ 20$^{\circ}$. These catalogs are available at the Strasbourg Astronomical Data Center\footnote{http://cds.u-strasbg.fr/} (CDS), and updated versions will be kept on the GUViCS website\footnote{http://galex.oamp.fr/guvics/index.html} starting shortly after the publication of this paper. 

There are a total of 1,232,626 UV source detections, of which 1,771 are classified as extended galaxies, and the remainder are classified as point-like sources. About 69\% (847,257 sources) of the point-like sources are either new Virgo members (129 sources), background galaxies (88,291 sources), or foreground stars (137,617 sources). We find UV emission in the majority of all extended galaxies and BCDs in Virgo, suggesting some level of star formation not hidden by dust, or the presence of a UV-upturn for ellipticals. However, UV emission is detected in just less than half of all dwarf galaxies (dE,dS0,dIm) in Virgo. Additionally, we find that $\sim$13\% of UV detected dE galaxies have nucleated UV cores \citep{bin91,cot06,lis06, bos08a}. The majority of newly detected Virgo galaxies are extremely blue objects, primarily with diffuse or compact dwarf-like galaxy types. Their spatial distribution in Virgo shows that these galaxies extend to the outskirts of the cluster and sparsely populate the cluster core. This evidence suggests that the environment plays a role in shutting down star formation, and is consistent with the picture proposed by \citet{bos08a,bos08b} where ram-pressure stripping acts to this effect. These newly cataloged dwarf-like Virgo members might also be low-luminosity disks that have been affected by ram-pressure stripping during infall. In the future, it would be compelling to study these objects with respect to the known dE, dS0, BCD, and dIm populations detected in the UV-VES catalog.  

Finally, several interesting compact objects are identified among the UV point-like sources. There are 4 confirmed globular clusters \citep[NED,][]{han01}, 9 ultra-compact dwarfs identified from \citet{bro11}, and 6 ultra-compact dwarfs originally identified in GUViCS by \citet{bos11}. We also find 5 additional globular cluster/ultra-compact dwarf candidates from preliminary matching to the catalogs of E. Peng et al. (\textit{in preparation}) and H. Zhang et al. (\textit{in preparation}) from Hectospec observations of NGVS compact object candidates. We hope that the UV detection of these sources will motivate future studies on the UV properties of compact objects using larger statistical samples that will be available soon. 


%
\begin{acknowledgements}
\\We would like to thank Hagen Meyer and Thorsten Lisker for sharing their position angles of VCC galaxies. We would like to thank Alan McConnachie for his work on the NGVS point-source separation statistics (APS) and for instructive disucussions on its application to the GUViCS point-like source catalogs. We would also like to thank Ted Wyder for useful discussions on the \textit{GALEX} pipeline data and catalogs. This work is supported by the French Agence Nationale de la Recherche (ANR) 
    \begin{figure}
    \centering
    \resizebox{\hsize}{!}{\includegraphics{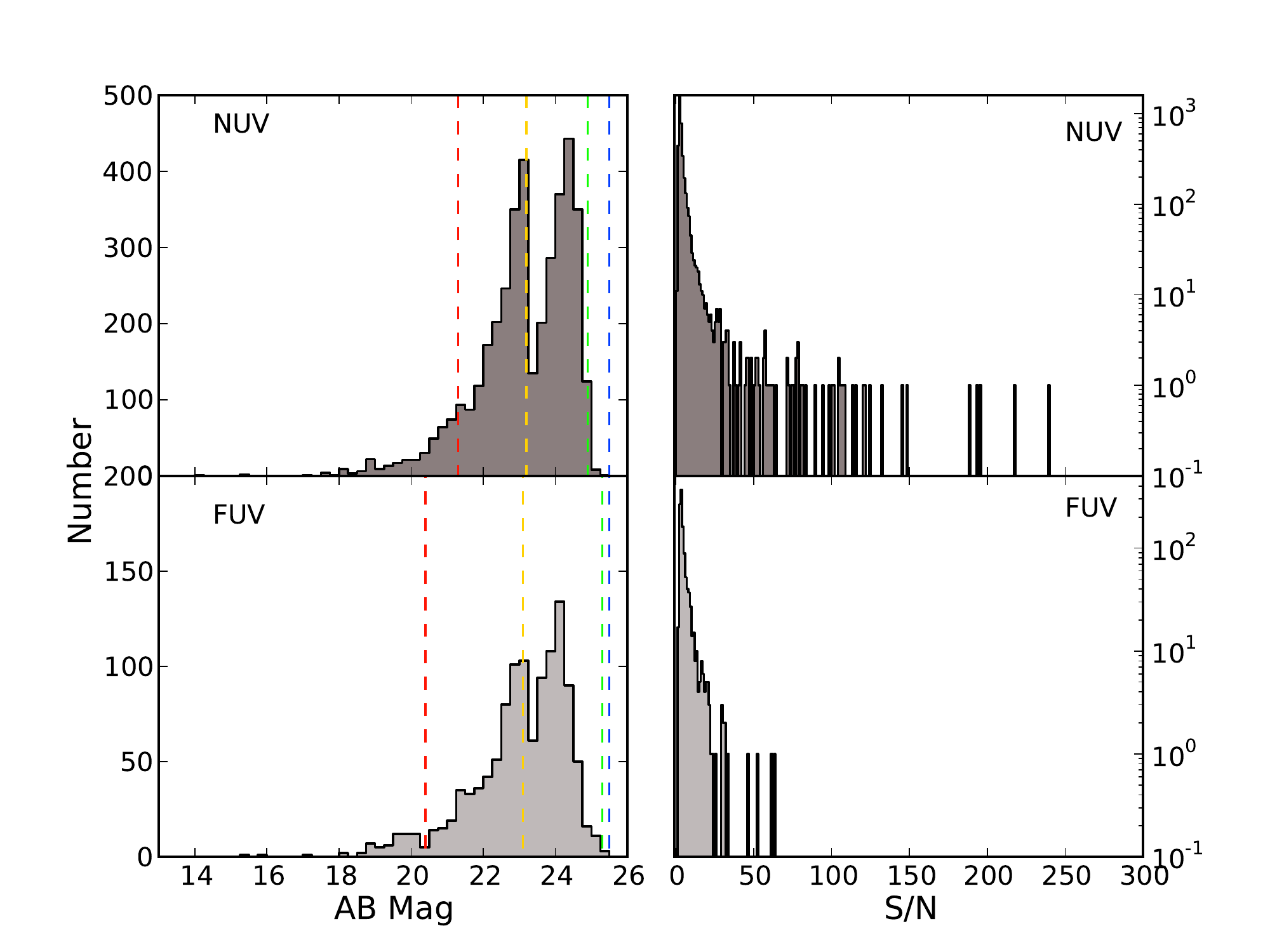}}
    \caption{Distributions of NUV (3,948 sources) and FUV (1,163 sources) signal-to-noise and magnitudes for the UV-VPS catalog that did not find an optical counterpart in the SDSS, NGVS, or NED within the NGVS footprint. From left to right, the vertical lines in the left-hand plots indicate the spurious magnitude limits for the AIS (red), MIS (yellow), DIS (green), and DEEP (blue) \textit{GALEX} images (Table \ref{spursrcs}).}
    \label{no_match_sn_uvmag_dist}
    \end{figure}
Grant Program Blanc VIRAGE (ANR10-BLANC-0506-01). We wish to thank the \textit{GALEX} Time Allocation Committee for the generous allocation of time devoted to GUViCS, and the Canada- France-Hawaii Telescope direction and staff for the support that enabled the NGVS. Finally, we thank the anonymous referee for helping to improve this manuscript.

\\
\\GALEX is a NASA Small Explorer, launched in 2003 April, operated for NASA by the California Institute of Technology under NASA contract NAS5-98034. We gratefully acknowledge NASA’s support for the construction, operation and science analysis for the \textit{GALEX} mission, developed in cooperation with the Centre National d’Etudes Spatiales of France and the Korean Ministry of Science and Technology. This work uses data from the SDSS-III database, the NASA/IPAC Extragalactic Database (NED), the GOLD Mine Database, and the SIMBAD database. Funding for SDSS-III has been provided by the Alfred P. Sloan Foundation, the Participating Institutions, the National Science Foundation, and the U.S. Department of Energy Office of Science. The SDSS-III web site is http://www.sdss3.org/. SDSS-III is managed by the Astrophysical Research Consortium for the Participating Institutions of the SDSS-III Collaboration including the University of Arizona, the Brazilian Participation Group, Brookhaven National Laboratory, University of Cambridge, Carnegie Mellon University, University of Florida, the French Participation Group, the German Participation Group, Harvard University, the Instituto de Astrofisica de Canarias, the Michigan State/Notre Dame/JINA Participation Group, Johns Hopkins University, Lawrence Berkeley National Laboratory, Max Planck Institute for Astrophysics, Max Planck Institute for Extraterrestrial Physics, New Mexico State University, New York University, Ohio State University, Pennsylvania State University, University of Portsmouth, Princeton University, the Spanish Participation Group, University of Tokyo, University of Utah, Vanderbilt University, University of Virginia, University of Washington, and Yale University. NED is operated by the Jet Propulsion Laboratory, California Institute of Technology, under contract with the National Aeronautics and Space Administration. The SIMBAD database is operated at CDS, Strasbourg, France. This research used the facilities of the Canadian Astronomy Data Centre operated by the National Research Council of Canada with the support of the Canadian Space Agency.
\end{acknowledgements}

%
\bibliographystyle{aa}
\bibliography{guvics_ev}

\begin{appendix}
\label{appendix}
  \section{\textit{GALEX} Pipeline Data Reduction}
  \label{reduc}
  The details of the \textit{GALEX} image reduction are described in \citet{mor07}. All \textit{GALEX} images used here were reduced via various versions of the \textit{GALEX} $``$Ops7'' pipeline. The source detection and UV photometry portion of the \textit{GALEX} pipeline utilizes the Source Extractor (SExtractor) software \citep{ber96} in combination with a program developed by the \textit{GALEX} team, \textit{poissonbg}. Both the FUV and NUV data generally have low, non-Gaussian, background count rates of $\sim$10$^{-3}$ and $\sim$10$^{-4}$ counts, respectively. Since SExtractor assumes Gaussian statistics to generate its background maps, it is necessary for the \textit{GALEX} pipeline to create its own background and detection threshold maps via \textit{poissonbg} using a modified sigma clipping procedure sampling the entire Poisson distribution. Further details of the \textit{poissonbg} procedure are provided in \citet{mor07}. To produce the final NUV and FUV source catalogs, SExtractor and \textit{poissonbg} are run twice. During the first run the background map and detection threshold map, made with \textit{poissonbg}, are inputs to SExtractor along with the background-subtracted intensity map. This run produces a segmentation map that is used to mask bright sources in a second run of \textit{poissonbg} in order to produce the final background and threshold maps. These maps are then used as input for the final run of SExtractor. In both runs the detection threshold maps are used to detect sources while the background-subtracted intensity map is used for source photometry.

  \section{GUViCS Point-Like Source Catalog Construction}
  \label{catalog_construct}
\begin{figure}
    \resizebox{\hsize}{!}{\includegraphics{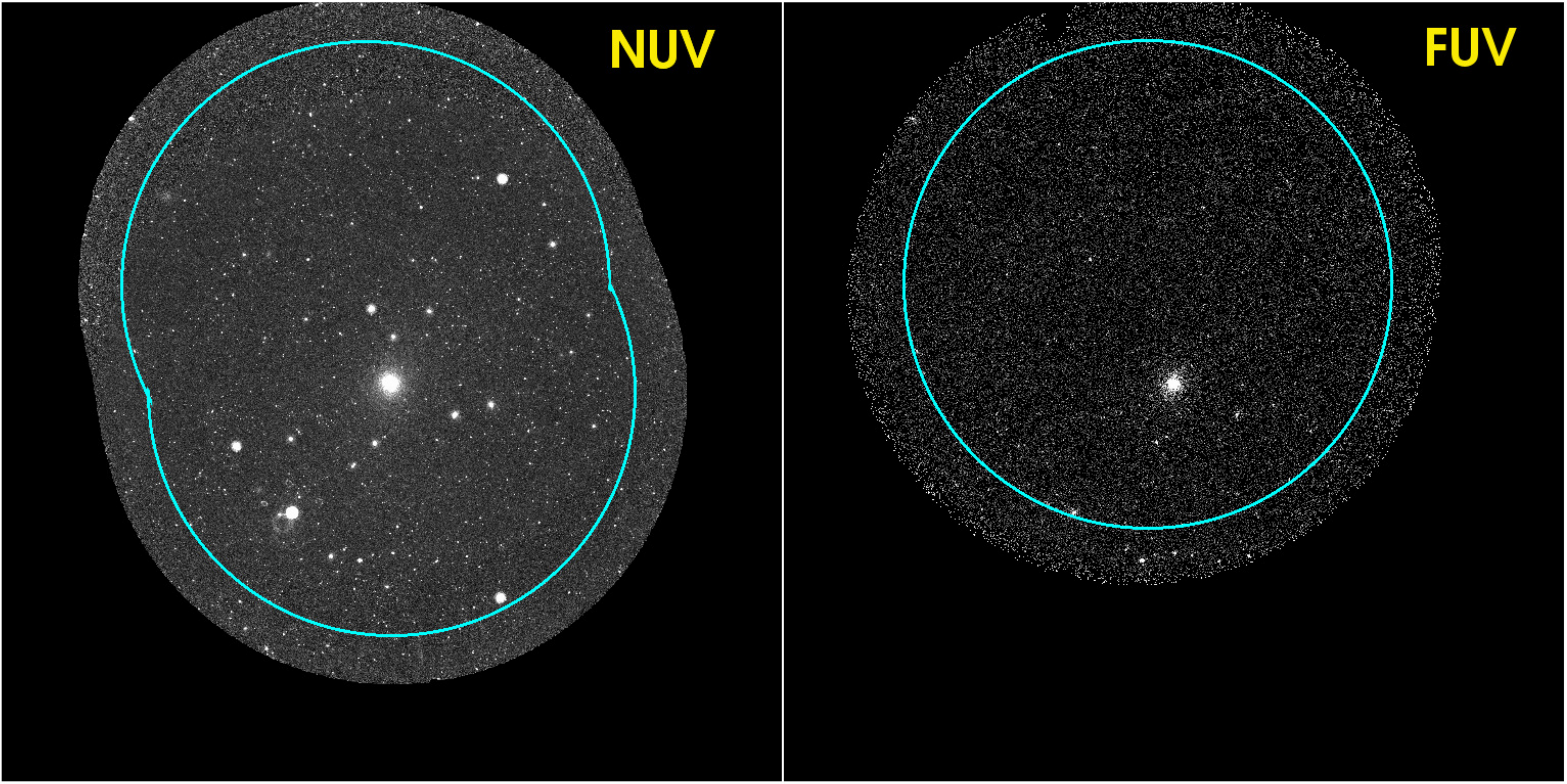}}
    \caption{The NUV and FUV images of \textit{GALEX} tile NGA\_Virgo\_MOS01 is an example of the irregular NUV image geometry also occuring for NGA\_Virgo\_MOS07 and NGA\_Virgo\_MOS11 tiles. Sources are only considered within the cyan region, i.e. inward of 0.1$^{\circ}$ from the image edge.}
    \label{mos}
    \end{figure}

  We make an initial reduction of all \textit{GALEX} merged catalogs (\textit{mcats}) by removing sources beyond a radius of 0.5$^{\circ}$ from the field centers. Beyond this radius \textit{GALEX} photometry is unreliable for several reasons: uncertainties in detector sensitivity, distortions in the point spread function, high probability of reflections from adjacent bright stars, and frequent concentrated areas of high detector background. There are three \textit{GALEX} tiles\footnote{NGA\_Virgo\_MOS01, NGA\_Virgo\_MOS07, and NGA\_Virgo\_MOS11} that have irregular NUV image geometries due to the telescope drifting during observations. An example is shown in Figure \ref{mos}. For these fields, we include all \textit{mcat} sources inwards of 0.1$^{\circ}$ from the field edge, excluding the small upper portion that is cut off.
    
\begin{figure}[!ht]
      \resizebox{\hsize}{!}{\includegraphics{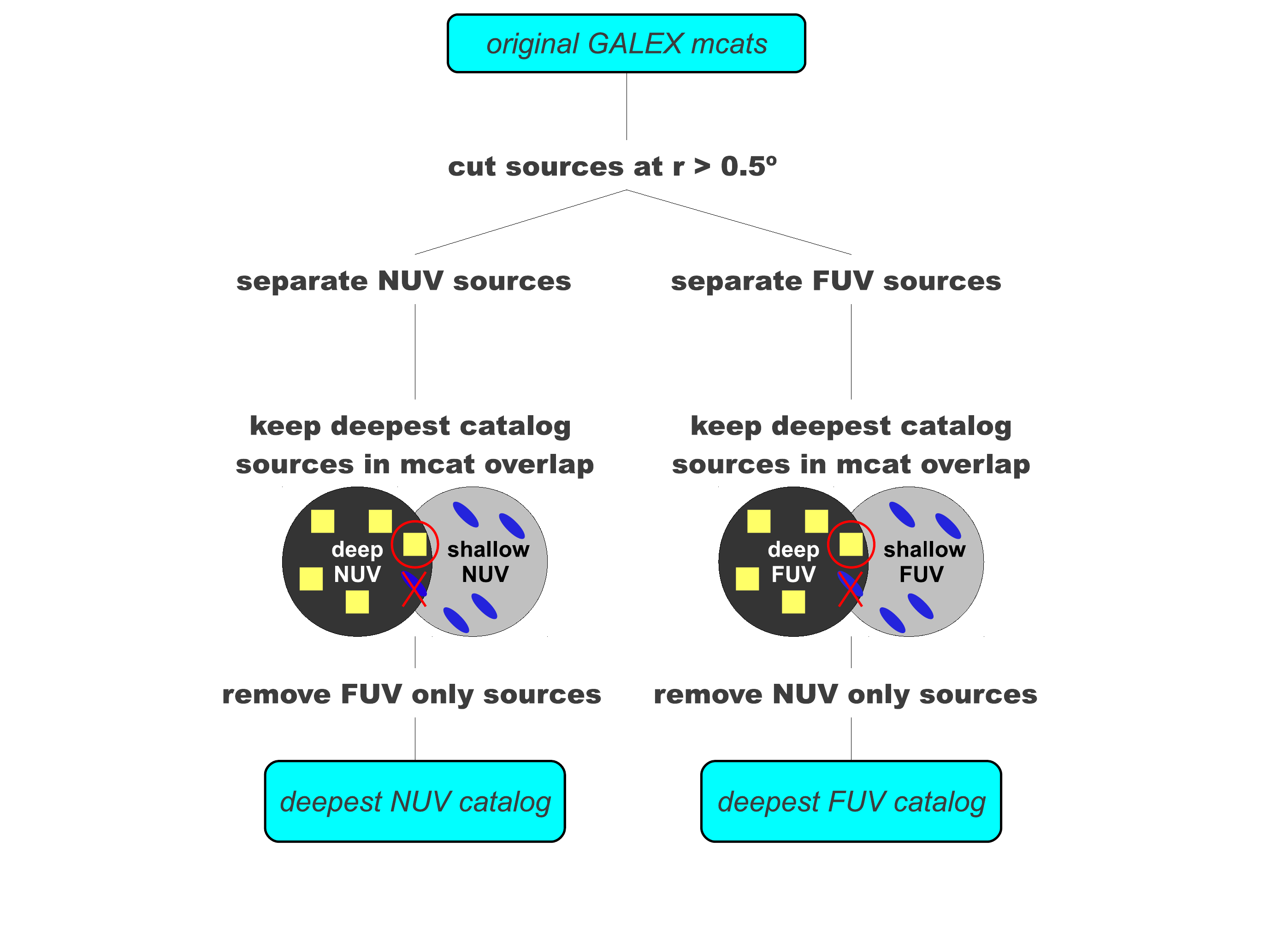}}
      \caption{Flow chart summarizing the steps taken to produce the deepest NUV and FUV point-like source catalogs from the original \textit{GALEX} \textit{mcats}.}
      \label{flow1}
      \end{figure}

  Following this, we determine the areas where individual \textit{GALEX} tiles overlap in the GUViCS field, and retain only those sources from the deepest overlapping \textit{GALEX} \textit{mcat} (i.e. highest exposure time) in each case. This procedure is carried out separately for NUV and FUV detected sources, producing separate deepest NUV and deepest FUV GUViCS catalogs at this stage of reduction. This is necessary since there are cases where the deepest image for the NUV and FUV bands in a configuration of overlapping \textit{GALEX} tiles do not always come from the same \textit{GALEX} tile because the exposure time for the observation of a tile in the NUV does not necessarily directly scale with its exposure time in the FUV. This could be due to technical issues while carrying out an observation, or to the varying initial objectives of the different \textit{GALEX} data sets. Finally, the separated NUV and FUV catalogs are cleaned of sources without detections in the respective bands, i.e. sources from the original \textit{mcats} that are only detected in the opposite band. The flow chart in Figure \ref{flow1} summarizes all of the above steps. 

  At this point, we asses the reliability of the \textit{GALEX} pipeline photometry from each source's artifact flag that was originally set during the automated SExtractor source detection. The artifact flag is the bitwise combination of all values within a source-centered 3x3 pixel box in the artifact pipeline FITS image. Here, we define bit values of 2, 4, and 512 to indicate unreliable photometry for a given source. The bit value of 2 warns that a source's flux could be affected by an interloping detector window reflection in the NUV only. The bit value of 4 warns that a flux measurement could be affected by a reflection of the instrument's dichroic beam splitter. The bit value of 512 only applies to \textit{GALEX} observations taken after May 2010, and warns of the potential inclusion of the ghost image of a bright object within a source's flux measurement. To indicate whether the \textit{GALEX} pipeline photometry is considered reliable for each source, we add a flag to the NUV and FUV catalogs\footnote{i.e. \textit{NUV/FUV\_GOOD\_PHOTOM\_FLAG}, see Appendix \ref{appendix_UVVC_ptsrc_cat} for details}. We set this flag to 0 if the photometry is considered reliable, and 1 if it is considered unreliable. We find that $\sim$3\% of the NUV detections and $\sim$0.1\% of FUV detections have poor photometry based on these artifact flags. However, sources with unreliable photometry remain in the final point-like source catalogs since these artifacts primarily affect a sources flux measurement and not necessarily the detection of that source. We urge users to proceed cautiously when selecting these objects with questionable photometry for further study. 

  Next, the NUV and FUV catalogs are recombined into a single band-merged catalog via a two step cross-matching procedure. First, the NUV and FUV catalogs are cross-matched by \textit{GALEX} tile. The \textit{GALEX} \textit{mcat} ID is used to rematch sources detected in the the NUV and FUV images from the same tile. These sources were originally matched in the \textit{GALEX} reduction pipeline as described in Appendix \ref{reduc}. Sources originally detected in only the NUV or FUV image for each tile are added as is to the tile's re-merged catalog. Finally, sources originally having detections in both bands for a \textit{GALEX} tile, but now only appearing in a single band's deepest reduced catalog should have an equivalent detection in another deeper \textit{GALEX} tile in the opposite band (i.e. this detection was removed from the opposite band's reduced catalog for the same field due to deeper field overlap in the area where that source is located). These objects are flagged accordingly. 

\begin{figure}[!ht]
      \resizebox{\hsize}{!}{\includegraphics{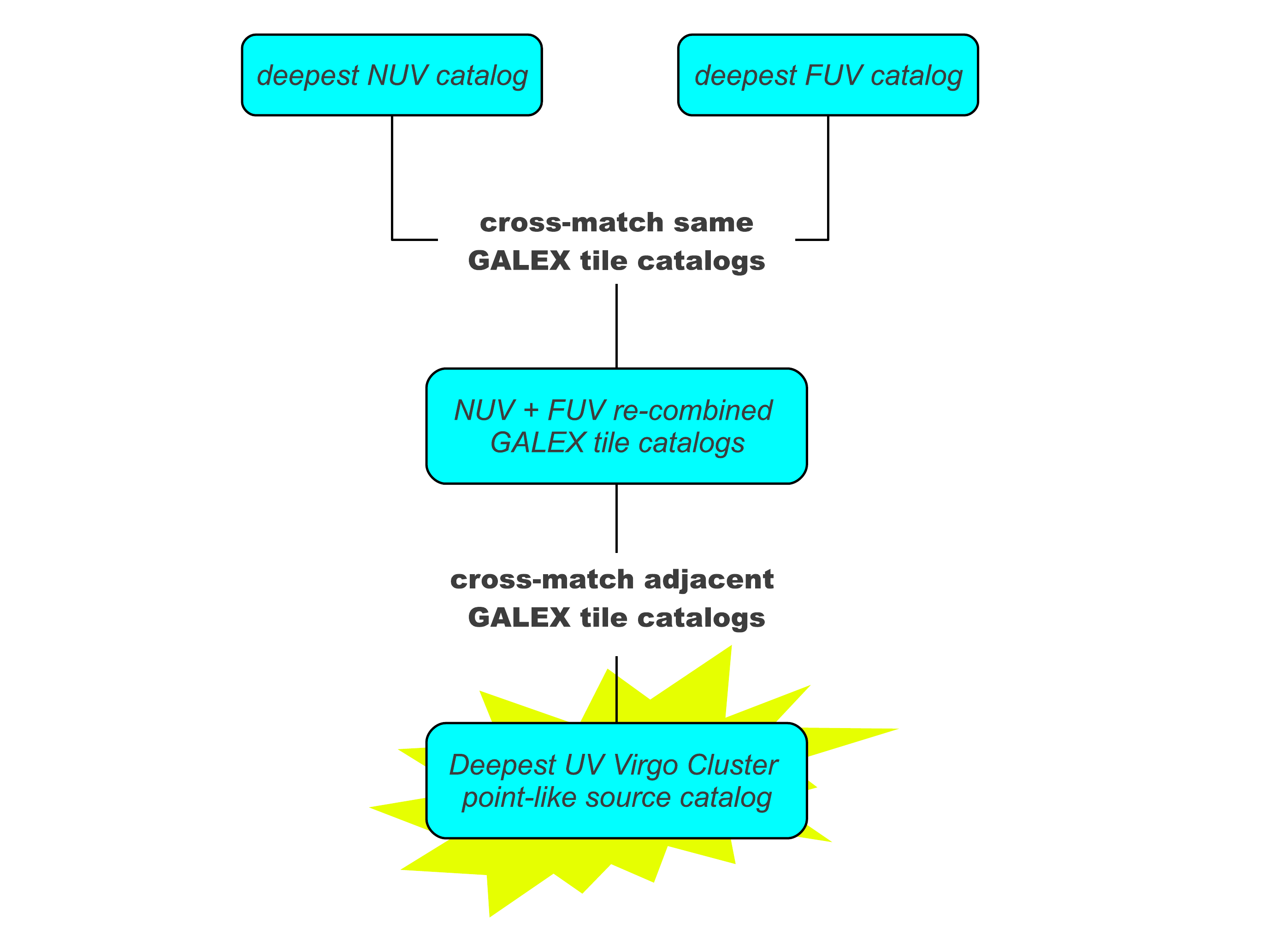}}
      \caption{Flow chart summarizing the final steps to produce the deepest UV Virgo Cluster point-like source (UV-VPS) catalog by cross-matching the deepest NUV and FUV catalogs.}
      \label{flow2}
      \end{figure}

  The second cross-matching is performed between all the recombined band-merged \textit{GALEX} tile catalogs. In this way we recover dual-band pipeline photometry of individual sources located in areas where the deepest NUV and FUV \textit{GALEX} images are from two different overlapping tiles. The same criteria used for the \textit{GALEX} \textit{mcat} band-merging is used here; match tolerance of 3$''$ and S/N >$2$. This procedure will recover sources that always had an NUV-to-FUV match. It will also find new NUV-to-FUV matches where the detection in the opposite band may have been too faint in the original \textit{GALEX} tile, but is a definite detection in a deeper overlapping tile, or, if there never was an FUV image taken for a given NUV only tile, but an older FUV image is found to serendipitously cover the same area. The flow chart in Figure \ref{flow2} summarizes this final set of steps taken to create the deepest UV Virgo Cluster point-like source (UV-VPS) catalog. We keep track of the rematching origins for dual-band photometry per source by adding a flag to the UV-VPS catalog\footnote{\textit{[NUV][FUV]\_SRCrem\_FLAG}}. The possible values of this flag are listed and explained in Table \ref{srcrem_flag}.

   \begin{table}
   \caption{Description of \textit{NUV/FUV\_SRCrem\_FLAG}.}             
   \label{srcrem_flag}      
   \centering          
   \begin{tabular}{cl}     
   \hline\hline       
   Flag Value & Description\\ 
   \hline                    
    -1                  & no image taken in this band\\  
     0                  & source is detected in this band\\
     1\tablefootmark{a} & source never detected in this band\\
     2\tablefootmark{b} & source detected in this band but deeper\\ 
                        & \textit{GALEX} image overlaps source location\\
     3\tablefootmark{c} & spurious source detection in this band\\ 
   \hline
   \end{tabular}
   \tablefoot{\\
   \tablefoottext{a}{Null values for this band in the catalog are from original \textit{GALEX} pipeline, and set to -99.0 and -999.0.}\\
   \tablefoottext{b}{Null values for this band in the catalog are set to -77.0 and -777.000.}\\
   \tablefoottext{c}{Section \ref{build_pls} describes the exposure time dependent spurious detection limits applied based on \citet{mor07}. Null values for this band in the catalog are set to -77.0 and -777.000.}\\
   }
   \end{table}

  \section{Reliability and Caveats of the GUViCS Point-Like Source Catalog}
  \label{rel_cavs}
    \subsection{Reliability}
    \label{complete}
    Several studies have already tested the reliability and completeness of \textit{GALEX} data at various depths. The 5$\sigma$ limiting magnitudes of \textit{GALEX} data with AIS, MIS, and DIS depths have been calculated by \citet{mor07} to be $m_{NUV}$ = 20.8, 22.7, 24.4 and $m_{FUV}$ =19.9, 22.6, 24.8, respectively. \citet{xu05} calculate the UV number counts in a combination of \textit{GALEX} MIS and DIS data down to magnitudes of 23.6 and 23.8 in NUV and FUV, respectively, selecting \textit{GALEX} sources with a modified version of the pipeline SExtractor input parameters to improve the deblending of faint UV point sources. They find the data is 80\% complete at depths of $m_{NUV}$ and $m_{FUV}$ = 24.5 \citep{arn05}. \citet{ham10} calculate the NUV and FUV number counts in the background of the Coma Cluster. They work with $\sim$26 ks (i.e. DIS-like) \textit{GALEX} imaging and use pipeline catalog photometry for sources up to $m_{NUV}$ and $m_{FUV}$ = 21, and Bayesian deblending photometry at fainter magnitudes. They calculate the completeness of the \textit{GALEX} pipeline catalogs for their data and find these data to be $\sim$90\% complete at $m_{NUV}$ $\sim$22 and $m_{FUV}$ $\sim$23, and only show marginal differences with completeness magnitudes calculated using the modified SExtractor run parameters from \citet{xu05}. The primary cause of incompleteness in the \textit{GALEX} data is source confusion, i.e. when sources, and possibly the background of an image, experience distortions resulting in underestimated fluxes. \citet{xu05} estimate that their \textit{GALEX} pipeline data become limited by source confusion after $m_{NUV}$ $\sim$24 and $m_{FUV}$ $\sim$25.3, while \citet{ham10} estimate brighter limits between $m_{NUV}$ = 23-23.5 for an upper limit of $m_{FUV}$ = 24. 

    Since the \textit{GALEX} coverage of the GUViCS field has large variations in depth, the reliability and completeness of the UV Virgo Cluster point-like source (UV-VPS) catalog also varies from region to region. In order to provide a qualitative assessment of the reliability of the UV-VPS catalog, i.e. where source confusion and spurious pipeline detections begin to dominate the data, we visually investigate 100 randomly selected 10$'$x10$'$ regions in each of the exposure time regimes outlined for AIS, MIS, DIS, and DEEP integrations in Table \ref{spursrcs} with respect to the S/N ratios of objects within these areas. The S/N of sources are investigated in 5 different bins in the NUV- and FUV-bands listed in Table \ref{s_to_n_bins}. We find the NUV pipeline point-like source detections to be trustworthy until the range  3$<$ S/N $<$ 3.5 in AIS images, 3.5 $<$ S/N $<$ 5 in MIS and DIS images, and no less than S/N = 5 in DEEP images. For the FUV GUViCS fields, we find AIS pipeline detections to be trustworthy up to a range of 3.5 $<$ S/N $<$ 5  for AIS and DEEP images and up to a range of 3 $<$ S/N $<$ 3.5 for MIS and DIS images. In the FUV, the AIS and DEEP images appear to have similar detection limitations due to the low signal of the AIS data providing very low statistics in the FUV, while the DEEP images suffer much less from low-level noise. Thus, we recommend that lower-signal data should be used with caution when selecting sources from the UV-VPS catalog, and to always investigate potential biases that may result from a given data selection.

   \begin{table}
   \caption{Bins of signal-to-noise for investigation of reliability of the UV-VPS catalog.}             
   \label{s_to_n_bins}      
   \centering          
   \begin{tabular}{ccc}     
   \hline\hline       
   Bin \# & NUV Range & FUV Range \\ 
   \hline                    
   1 & 2.0 $<$ S/N $<$ 2.5 & 3.0 $<$ S/N $<$ 3.5\\
   2 & 2.5 $<$ S/N $<$ 3.0 & 3.5 $<$ S/N $<$ 4.0\\
   3 & 3.0 $<$ S/N $<$ 3.5 & 4.0 $<$ S/N $<$ 4.5\\ 
   4 & 3.5 $<$ S/N $<$ 5.0 & 4.5 $<$ S/N $<$ 5.0\\
   5 & 5.0 $<$ S/N         & 5.0 $<$ S/N\\ 
   \hline
   \end{tabular}
   \end{table}

    \subsection{Caveats}
    \label{caveats}
    The \textit{GALEX} pipeline has well documented issues with source detection in particular situations. In UV observations galaxies are often clumpy and irregular compared to their optical morphologies, causing single sources to be identified as multiple detections (i.e. shredding). This can occur for large nearby galaxies, as well as for distant objects. For example, we find cases of large Virgo Cluster galaxies that have extended UV emission far enough away from the main source to be detected as separate sources by the \textit{GALEX} pipeline. A good case is VCC1217 which is studied in \citet{fum11}. They investigate the properties of 12 $`$fireball$`$ sources associated with this dwarf irregular galaxy that extend up to 3.5$'$ south-east of the galaxy. Several of these sources are independently detected in the \textit{GALEX} pipeline catalogs, and could be mistaken as background sources. However, such cases are rare compared to the statistical size of the catalog. Additionally, \textit{GALEX} data is also susceptible to blending effects where two or more sources are detected as a single source due to the large point spread function of $\sim$5$''$. Source blending is more problematic in the background of the cluster than for foreground Virgo members, mainly affecting photometry for higher redshift sources. Several studies have used Bayesian deblending methods to obtain photometry of distant sources \citep{gui06,ham10}, however we do not apply this technique here with positive results.

\begin{figure}
    \resizebox{\hsize}{!}{\includegraphics{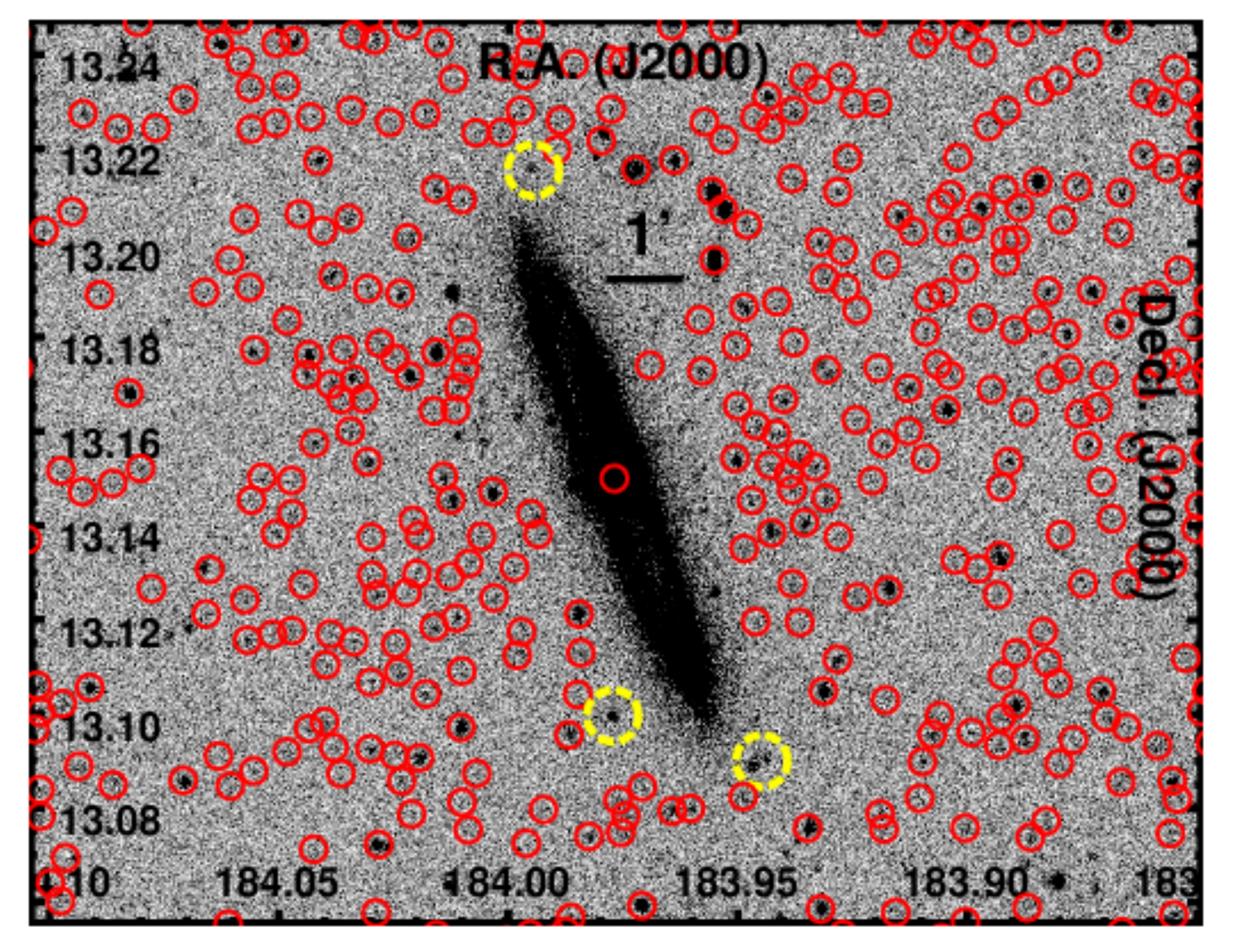}}
    \caption{\textit{GALEX} NUV image centered on VCC 167 (NGC 4216). The yellow dashed regions indicate the locations of known background sources from the SDSS, while the red regions are the \textit{GALEX} pipeline catalog sources detected in this NUV image. This clearly demonstrates how the \textit{GALEX} pipeline excludes source detections in proximity to extended sources $>$ 1$'$ in diameter.}
    \label{ext_nondet_back}
    \end{figure}

    It is also known that the \textit{GALEX} pipeline excludes detections of smaller sources in proximity to larger UV extended sources $<$ 1' in diameter (see Section \ref{build_es}). We inspect several of these cases in the UV-VPS catalog, selecting the small UV-bright sources near extended galaxies in deep NGVS optical images. We find that these are obvious background galaxies, not HII regions associated with the extended source. Figure \ref{ext_nondet_back} shows examples of several such sources around NGC 4216 in its NUV image. The cyan circles are the SExtracted detections from the original \textit{GALEX} merged catalog (\textit{mcat}) for this tile, and the dashed yellow circles mark known background objects near the galaxy that were not detected by the pipeline. Additionally, this explains why there are sources detected in FUV images that are not detected in the respective NUV images. In such cases, the FUV profile diameter of a given extended source is more compact than in the NUV, causing the automatic \textit{GALEX} algorithms to detect nearby background sources in the FUV and not the NUV.

    During our investigation of the UV-VPS catalog we notice some additional issues to be cautious of specifically concerning \textit{GALEX} pipeline detection and photometry of point-like background sources in the cluster area. First, pipeline-derived Kron apertures \citep{kro80} can be too large compared to actual galaxy size in cases where background sources are larger and less point-like. In some cases this is a result of the pipeline blending visually close background sources; the Kron aperture will stretch to encompass adjacent background sources. Secondly, during an investigation of known brightest (background) cluster galaxies in the Virgo area, we find that several obvious UV brightest cluster galaxies are not detected by the \textit{GALEX} pipeline. We believe this could be a side effect of the double image processing procedure applied to all \textit{GALEX} data in the pipeline (see Appendix \ref{reduc}). Finally, in very crowded areas of the cluster background, we have noticed individual sources that are included in the photometric apertures of multiple neighboring background sources. The \textit{GALEX} pipeline photometry is unreliable in these cases since the flux of a single source is included in the flux measurements of multiple adjacent sources. 

    One must be mindful of all the above caveats when using the UV-VPS catalog. Potential source selections, and their corresponding catalog data, should not be used without an initial investigation to see if the issues discussed here are present in the data set, and to what degree.
  
  \section{UV Aperture Construction for Extended Sources}
  \label{ext_apps}
    Extended source UV apertures are produced using optical semi-major and semi-minor radii and position angles, where available. First, we view these apertures in the deepest \textit{GALEX} NUV and FUV image of each source to determine if the galaxy is a real UV detection in both bands. Many sources are only detected in NUV observations, and in these cases their FUV data entries in the UV-VES catalog are set to null. Next, the apertures are adjusted from their optical shapes to their NUV shapes. When a galaxy's aperture is increased from the optical size to fit the full UV profile, the aperture edges are checked against their location in the deep NGVS optical imaging, or in the SDSS imaging when the NGVS is not available. We do this in order to make sure that the aperture was not extended too far, and that it does not include a nearby source separate from the primary galaxy. Finally, masks are applied to distinct, interloping, background sources and foreground stars within the NUV apertures. We confirm the presence of background sources within apertures by, once again, checking the NGVS optical images. Stars are confirmed by identification in the SDSS database. The final NUV apertures are also used for FUV photometry.

  \section{Extended Source Magnitude Error Calculation}
  \label{error}
    To calculate the error on extended source aperture magnitudes we follow the procedures from \citet{bos03}, \citet{gil07}, and \citet{cor12}. The components of the error are the error on the sky level ($err_{sky}$), the instrumental error ($err_{inst}$), the shot noise ($err_{Poiss}$), and the \textit{GALEX} calibration error in each band ($err_{GALEX}$). The $err_{GALEX}$ is a constant value essentially representing the zero point errors of 0.03 and 0.05 in the NUV and FUV filters, respectively \citep{mor07}. The $err_{Poiss}$ is calculated from the net pixels in the galaxy:

    \begin{equation}
    err_{Poiss} = \sqrt{N_{e^{-}}}\frac{1}{Gt}
    \end{equation}

    where $N_{e^-}$ is the number of electrons, $G$ is the image gain in e$^-$ ADU$^{-1}$, and $t$ is exposure time in seconds. The $err_{sky}$ and $err_{inst}$ are calculated from the pixels in the galaxy aperture and the pixels in $\sim$10 square sky regions of randomly selected sizes surrounding the galaxy. These errors account for variations in flux measurements resulting from Galactic cirrus and flat-field variations across the galaxy aperture. They are defined as:

    \begin{equation}
    err_{sky} = N_{pix}STD[<b_1>... <b_n>]
    \end{equation}
    \begin{equation}
    err_{inst} = \sqrt{N_{pix}}<STD[b_1]...STD[b_n]>
    \end{equation}

    where $N_{pix}$ is the number of pixels in the galaxy aperture, $STD[b_n]$ is the standard deviation of the values of all pixels in sky box $n$, and $<b_n>$ is the average of all pixels in sky box $n$. Thus, the total error on a UV magnitude of an extended source is:

    \begin{equation}
        err_{Tot} = \sqrt{ (  \frac{2.5}{ln(10)} \sqrt{err_{sky}^{2}+err_{inst}^{2}+err_{Poiss}^{2}} )^2+err_{GALEX}^{2}}
    \end{equation}

    where the factor of 2.5/$ln$(10) converts the total statistical error into AB magnitude units. The uncertainty is then $err_{Tot}$ over the magnitude.

  \section{The NGVS Source Catalog}
  \label{ngvs_cat}
  The NGVS source catalog was created using SExtractor on 117 individual field tiles covering the survey area (S. Gwyn et al., \textit{in preparation}). SExtractor uses the deep $g'$-band data for source detection, creating separate photometric catalogs in the 5 bands that are combined into the final source catalog of 24,934,007 objects, used here. We note that this catalog is tailored to the detection of point sources (stars and globular clusters) and relatively compact, high surface brightness galaxies; it is not designed to detect faint, low surface brightness dwarf galaxies, the brightest VCC galaxies, or sources that fall in high surface brightness areas near the centers of bright VCC galaxies. The catalog provides a star/galaxy classification. It uses the $g'$- and $i'$-band data to calculate a statistic quantifying the distance from a stellar locus, defined by the difference in flux between two apertures, taking into account the measured width of the locus and the photometric error on the magnitude. The $`$pointiness' of a source increases with higher absolute values of this statistic and it is reliable up to $m_{g'}$ and $m_{i'}$ = 24; more positive values identify image noise while more negative values select increasingly extended galaxies. Stars are identified as having probabilities $>$ 1.5 (i.e. within 1.5$\sigma$ of the growth curve).

  The NGVS catalog is supplemented by a compilation of spectroscopic redshifts from SDSS DR7 \citep{aba09}, NGVS follow up observations with the Hectospec instrument on the 6.5m MMT (E. Peng et al., \textit{in preparation}), the VCC, and the database of globular cluster from \citet{han01} and \citet{str11}. From these data, redshifts are further selected based on reasonable errors, confidence intervals, and warnings, resulting in 6,108 spectroscopic redshifts. Here, we exclude SDSS DR7 redshifts from use in the categorization of GUViCS point-like sources since the UV catalog has already been matched with SDSS DR9.

  \section{Examining UV-Optical Matches Between the SDSS and the NGVS}
  \label{exam_sdss_ngvs}
  The NGVS catalog reaches point source depths at S/N=10 that are $\sim$3.5 magnitudes deeper than the 95\% point source completeness limits of SDSS in the $i'$- and $g'$-bands. Within the area of GUViCS covered by the NGVS footprint (see Figure \ref{guvics_nuv_cov}) we compare UV-optical matches between the SDSS and the NGVS. There are 40,270 GUViCS-to-NGVS 1-to-1 source matches that do not have a SDSS counterpart at magnitudes fainter than the $g'$-band 95\% completeness level of the SDSS. This shows how the NGVS detects a large number of additional sources at faint optical magnitudes, thus demonstrating the importance of deep optical imaging for studying UV sources. We remind the reader that the current NGVS source catalog is an ongoing work and currently lacks certain types of sources, such as new faint Virgo dwarf galaxies and high-surface brightness objects, like ultra-compact dwarfs or globular clusters, that happen to lie close to bright galaxies. 

  Since the NGVS is incredibly deep, it is reasonable to assume that all sources with magnitudes brighter than the $g'$-band completeness of the SDSS (i.e. $<$ 22.2) would be detected in both surveys. In order to examine this assumption, we calculate the statistics for non-stellar GUViCS-SDSS and GUViCS-NGVS one-to-one matches. There are 198,569 SDSS-GUViCS and 108,884 NGVS-GUViCS matched sources with $g'$-band magnitudes $<$ 22.2. Of the SDSS sources, 43.8\% correspond to a 1-to-1 GUViCS-NGVS match, 55.4\% correspond to a GUViCS-to-multiple-NGVS match, and 0.8\% have no GUViCS-NGVS counterpart. Of the NGVS sources, 95.4\% correspond to a 1-to-1 GUViCS-SDSS match, only 3.5\% correspond to a GUViCS-to-multiple-SDSS match, and 1.1\% have no SDSS counterpart. 

  The small percentage of sources with SDSS matches and no NGVS counterparts could be reduced when the NGVS catalog is complete. They could also be sources that were removed from the edges of NGVS fields where the nominal exposure time drops by one half. In the opposite respect, the small percentage of sources with NGVS matches brighter than the SDSS completeness limit, and no SDSS matches, have two possible explanations. Either the SDSS source is not in the DR9 database because it is too close to an SDSS image edge and thus removed, or the SDSS field itself was not included in DR9 due to timeout issues\footnote{Timeout issues affect $\sim$1\% of DR8 fields that are then included in DR9 and the issue occurs when extended galaxies or bright stars cause larger processing and deblending times in the newer methods implemented for SDSS sky-subtraction.} during image processing. 

  At $g'$-band magnitudes fainter than the SDSS completeness limit of 22.2, there is no longer an expected correspondence between the SDSS and NGVS GUViCS matches. However, there are still many GUViCS-to-SDSS matches towards fainter magnitudes. Interesting candidates for future work could be the bright UV sources that are very blue with extremely faint optical counterparts, observable only at NGVS depths. These sources would have magnitudes fainter than the completeness limit of the SDSS and brighter than the completeness limit of the NGVS.

    \begin{figure}
    \centering
    \resizebox{\hsize}{!}{\includegraphics{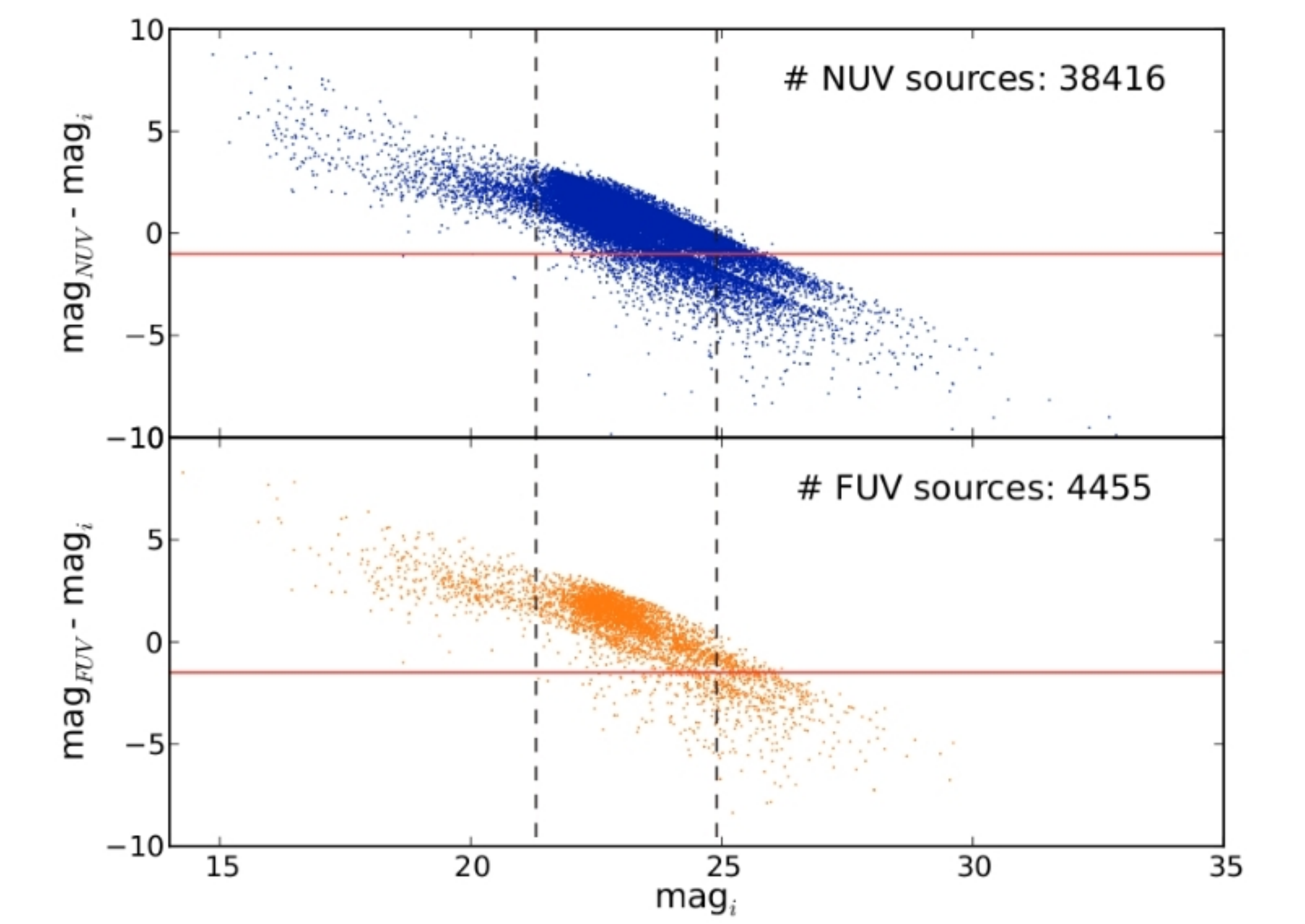}}
    \caption{Color-magnitude diagrams for all NUV (top) and FUV (bottom) UV-VPS catalog sources with $i'$-band coverage in NGVS. The dashed line at $m_{i'}$=21.3 marks the SDSS 95\% $i'$-band completeness limit, and the dashed line at $m_{i'}$=24.9 marks the NGVS point-source depth at S/N=10. The red horizontal lines at $m_{NUV}$-$m_{i'}$=-1 and $m_{FUV}$-$m_{i'}$=-1.5 represent the physical color limits for the typical colors of instantaneous bursts of star formation observed at the peak of activity.}
    \label{color_mag_all}
    \end{figure}
  Figure \ref{color_mag_all} shows the UV-optical color magnitude diagrams for all NUV and FUV detected sources in the UV Virgo Cluster point-like source (UV-VPS) catalog. The sharp upper edges of these distributions are due to the detection limit of \textit{GALEX}. Generally, we detect very blue galaxies with the NGVS, however, colors below -1 and -1.5 are physically unrealistic for the NUV-band and FUV-band, respectively. The extremely blue colors are unrealistic because these limits are the typical colors of instantaneous bursts of star formation observed at the peak of activity. We believe the overly blue colors could result from unrealistic Kron apertures output by the \textit{GALEX} pipeline. Source blending effects, due to the poor resolution of \textit{GALEX}, might result in unphysical colors if several optical sources are associated with one UV source (see Appendix \ref{caveats}). This would cause SExtractor to calculate an unrealistic aperture size for the UV source, resulting in very blue UV-i colors. This is one potential cause of the extremely blue colors for sources with $i'$-band magnitudes brighter than the NGVS completeness limit. 

\begin{figure}
    \centering
    \resizebox{\hsize}{!}{\includegraphics{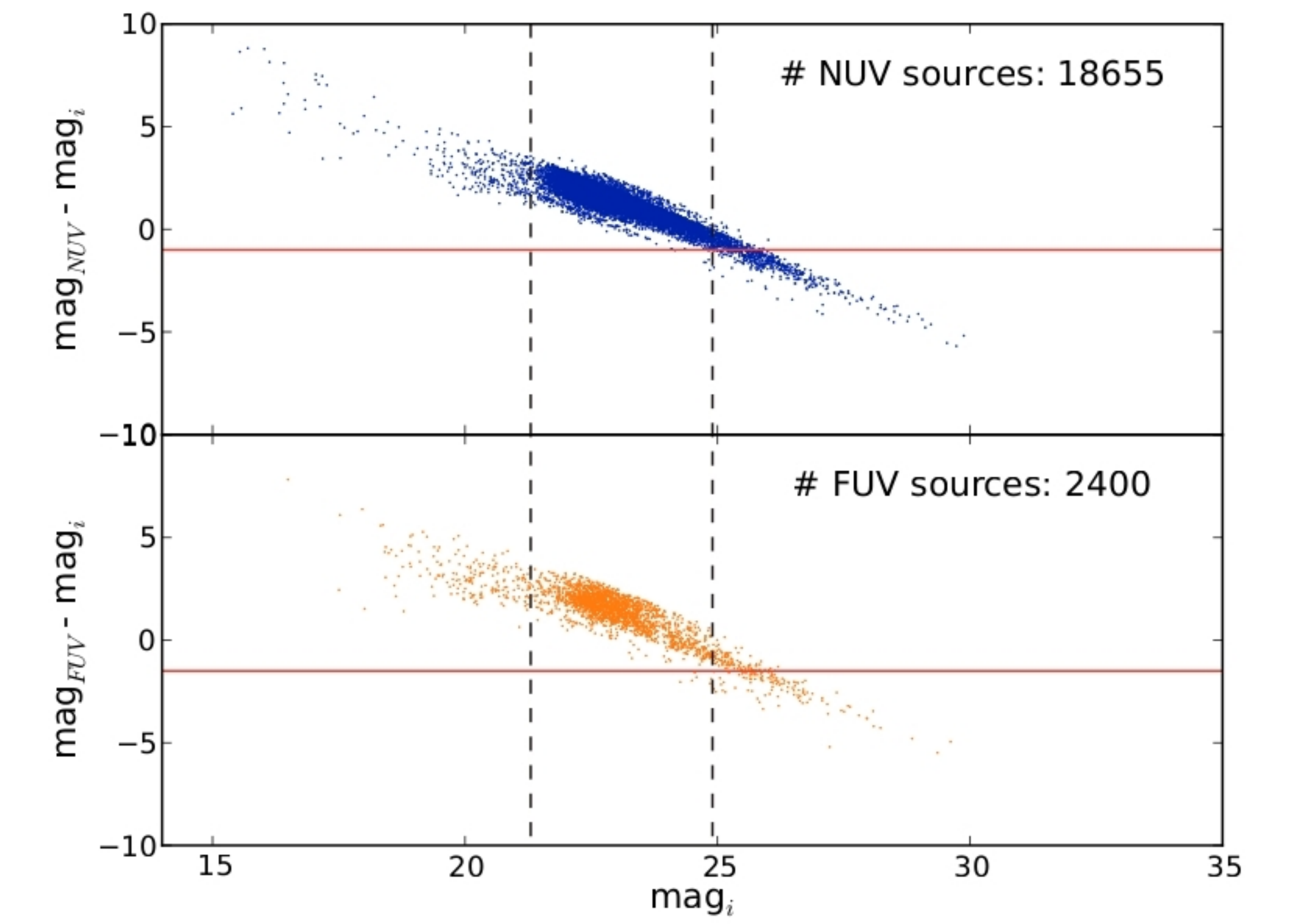}}
    \caption{Color-magnitude diagrams for NUV (top) and FUV (bottom) UV-VPS catalog sources with $i'$-band coverage in NGVS that have UV Kron apertures with semi-major axis $<$ 10$''$. The dashed line at $m_{i'}$=21.3 marks the SDSS 95\% $i'$-band completeness limit, and the dashed line at $m_{i'}$=24.9 marks the NGVS point-source depth at S/N=10. The red horizontal lines at $m_{NUV}$-$m_{i'}$=-1 and $m_{FUV}$-$m_{i'}$=-1.5 represent the physical color limits for the typical colors of instantaneous bursts of star formation observed at the peak of activity.}
    \label{color_mag_10arc}
    \end{figure}

  We test this hypothesis by constructing the same color-magnitude diagrams for only those sources with compact Kron apertures, i.e. semi-major axis sizes $<$ 10$''$, shown in Figure \ref{color_mag_10arc}. 
  At magnitudes brighter than the NGVS completeness limits the color-magnitude diagrams now behave physically. Therefore, it is likely that incorrect \textit{GALEX} pipeline apertures are the cause of the unrealistic colors in Figure \ref{color_mag_all}. However, the unphysical result is still found at magnitudes fainter than the completeness limit in Figure \ref{color_mag_10arc}. Two plausible explanations may be that these sources are spurious UV and optical detections, or, the sources with $m_{i'}$ $>$ 25 have weak UV magnitudes. Clearly, the error on the magnitude becomes very important here. This analysis also warns that \textit{GALEX} pipeline Kron magnitudes should not be trusted automatically when taken from the GUViCS catalogs, particularly for use in color analysis. 

\begin{figure}
    \centering
    \resizebox{\hsize}{!}{\includegraphics{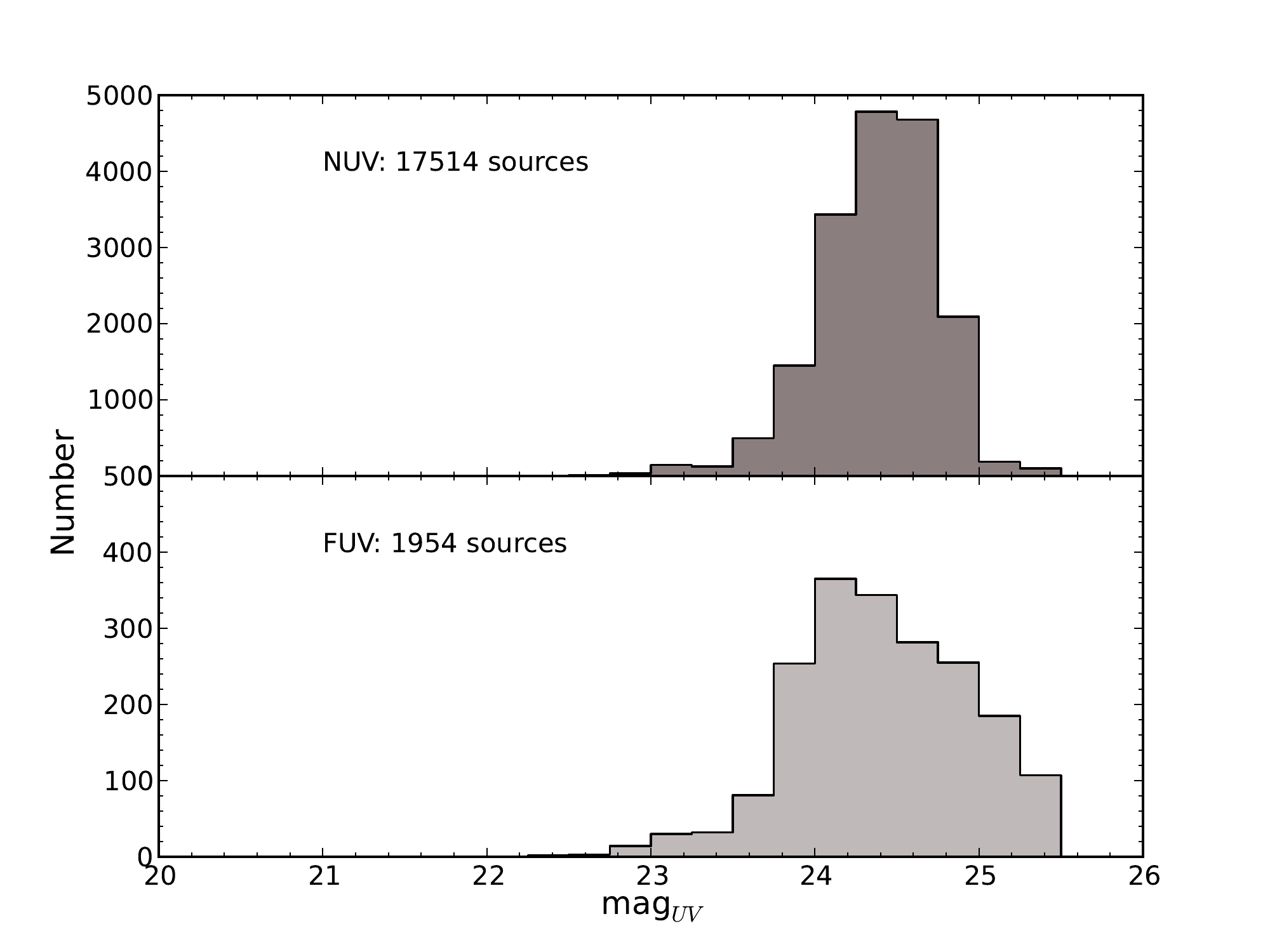}}
    \caption{NUV and FUV magnitude distributions for UV-VPS catalog sources between the SDSS and NGVS $i'$-band limits in Figure \ref{color_mag_10arc}, i.e. sources with UV Kron aperture semi-major axis $<$ 10$''$.}
    \label{mag_dist_uv_opt}
    \end{figure}

\begin{figure*}
    \centering
     \resizebox{0.6\textheight}{!}{\includegraphics{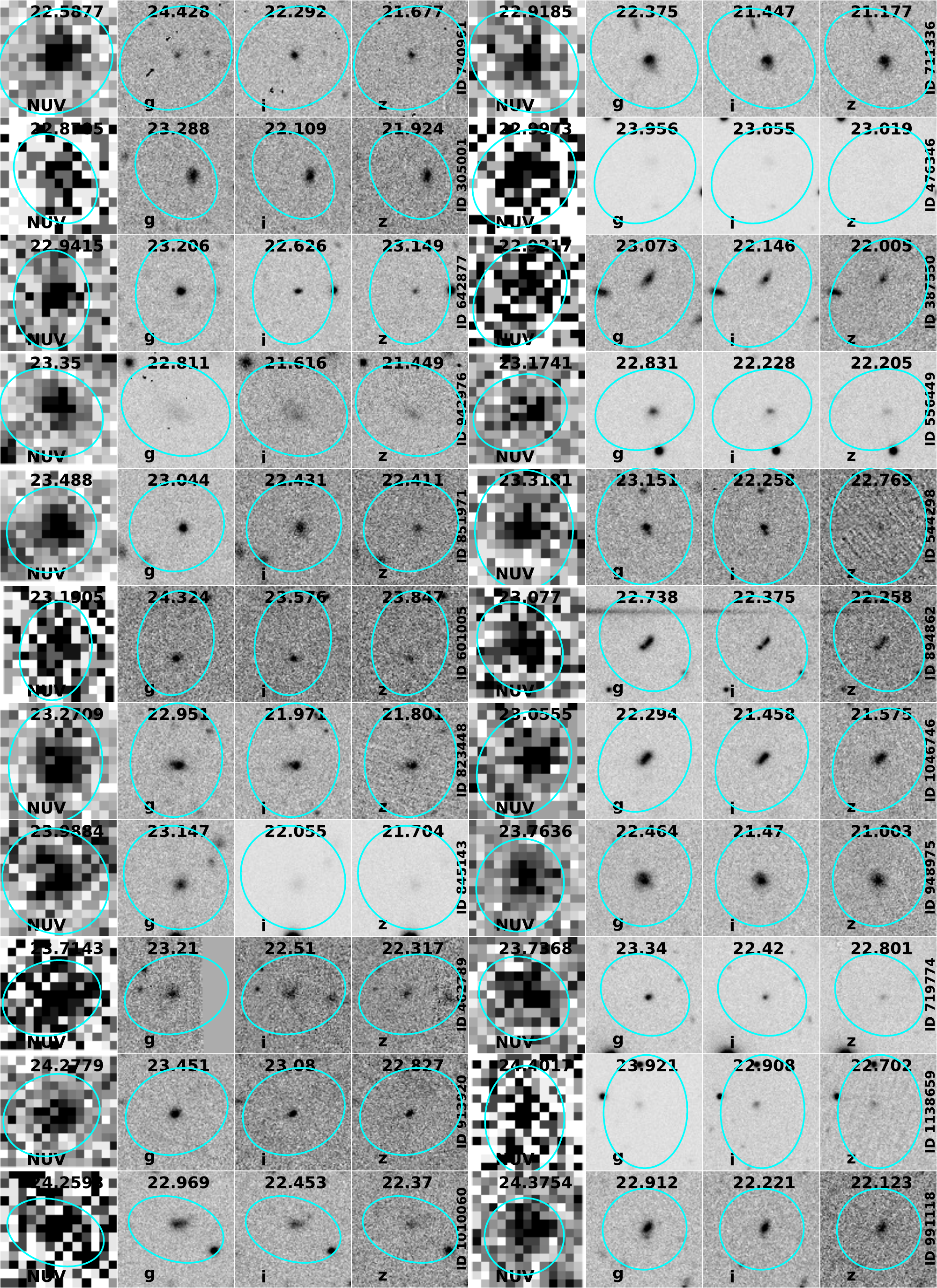}}
    \caption{Image gallery of a selection of true UV source detections between the SDSS and NGVS $i'$-band limits in Figure \ref{color_mag_10arc}, i.e. sources with UV Kron aperture semi-major axis $<$ 10$''$. Each line of the two columns shows the \textit{GALEX} NUV, NGVS $g'$-band, NGVS $i'$-band, and NGVS $z'$-band images. The magnitudes in each band are listed at the top of each stamp, and the \textit{UV\_VPS\_ID} is listed vertically along the right edge of the $z'$-band image. The cyan apertures are the NUV Kron apertures automatically calculated in the \textit{GALEX} pipeline.}
    \label{uv_opt_gallery_good}
    \end{figure*}

\begin{figure*}
    \centering
     \resizebox{0.6\textheight}{!}{\includegraphics{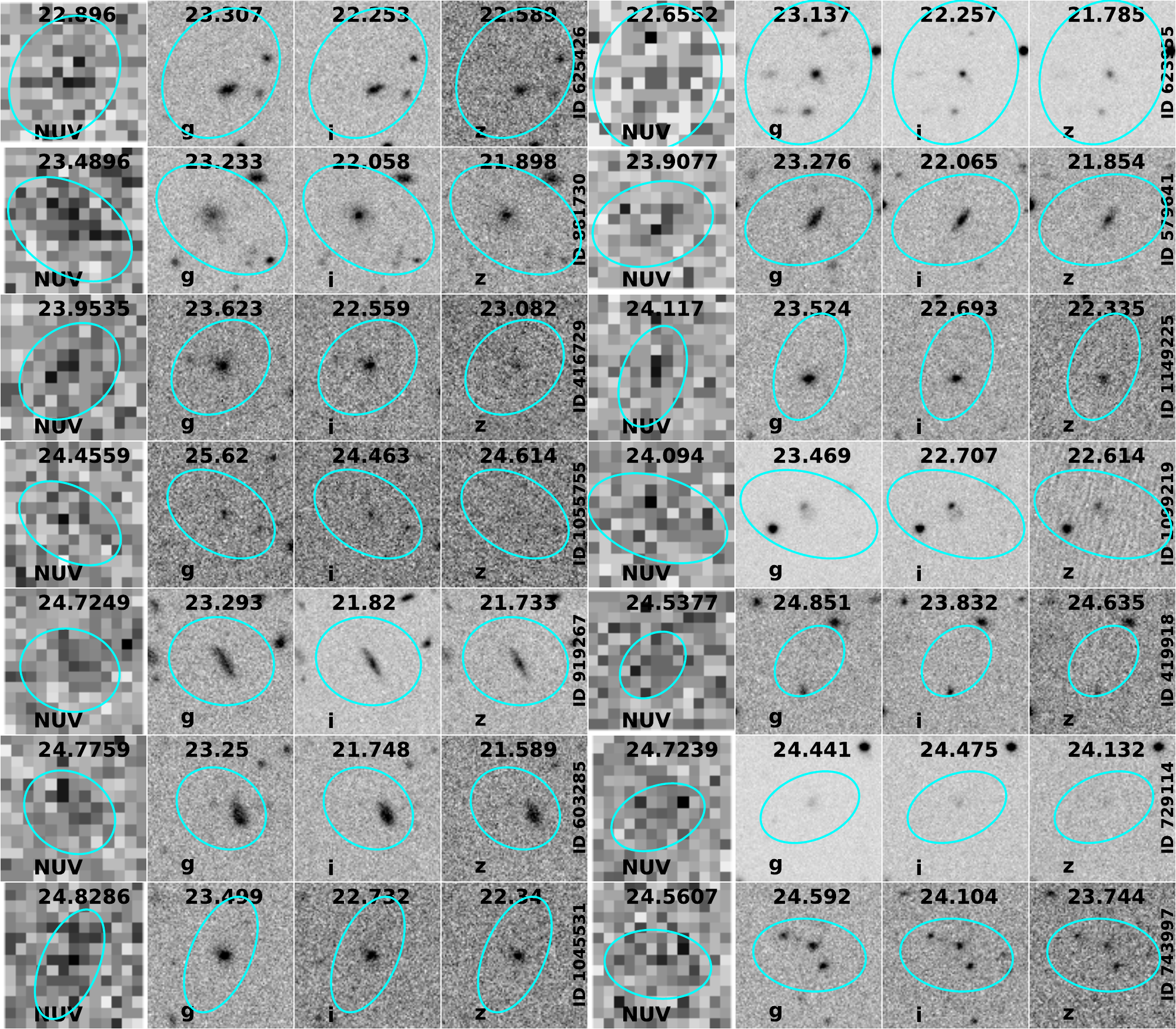}}
    \caption{Image gallery of a selection of spurious UV source detections between the SDSS and NGVS $i'$-band limits in Figure \ref{color_mag_10arc}, i.e. sources with UV Kron aperture semi-major axis $<$ 10$''$. Each line of the two columns shows the \textit{GALEX} NUV, NGVS $g'$-band, NGVS $i'$-band, and NGVS $z'$-band images. The magnitudes in each band are listed at the top of each stamp, and the \textit{UV\_VPS\_ID} is listed vertically along the right edge of the $z'$-band image. The cyan apertures are the NUV Kron apertures automatically calculated in the \textit{GALEX} pipeline.}
    \label{uv_opt_gallery_spur}
    \end{figure*}

  The UV and optical magnitude distributions for sources between the SDSS and the NGVS $i$'-band limits of the color-magnitude diagram in Figure \ref{color_mag_10arc} are shown in Figure \ref{mag_dist_uv_opt}. UV-optical images of several of these sources are provided in the galleries in Figures \ref{uv_opt_gallery_good} and \ref{uv_opt_gallery_spur}.
  All UV sources have $m_{NUV}$ and $m_{FUV}$ $>$ 22 with the distributions peaking $\sim$24.2 in both bands. The first gallery, in Figure \ref{uv_opt_gallery_good}, shows examples of good UV detections, while the second gallery, in Figure \ref{uv_opt_gallery_spur}, shows examples of spurious UV detections that remain in the UV-VPS catalog after cleaning. For two objects in Figure \ref{uv_opt_gallery_good} (ID 387550 and ID 462789) the UV image is clearly blending multiple optical sources within the \textit{GALEX} Kron aperture, while the majority of other objects have one optical-to-faint-UV detection. These sources should be tested further to determine whether they are background or cluster members, and thus are good candidates for the application of improved \textit{GALEX} photometry techniques focused on point-like or higher redshift sources (e.g. Bayesian deblending and photometry). Overall, the analysis in this appendix demonstrates the great advantage of having deep multiwavelength data in the Virgo Cluster.

%
\clearpage
\onecolumn
\section{The UV Virgo Cluster Extended Source Catalog}
\label{appendix_UVVC_extend_cat}

The UV Virgo Cluster Extended Source (UV-VES) catalog provides the deepest and most extensive UV photometric data of extended galaxies in Virgo. The catalog contains 1,771 sources and 23 fields of data for each source. If certain data is not available for a given source then a null value is entered. For example, an extended source may not have \textit{GALEX} FUV imaging available, so all FUV data entries are be entered as -999.000. Here, Table \ref{UV-VES_cols} provides a full description of all columns in the UV-VES catalog, and Table \ref{UV-VES_cat} is a preview of the catalog that highlights some of the most relevant columns. The full catalog is available online at the CDS.

\begin{longtab}
\begin{longtable}{lllll} 
\caption{\label{UV-VES_cols}Description of columns in the UV-VES catalog.}\\
\hline\hline  
Field & Format & Units & Label & Explanation \\
\hline
\endhead
\hline
\endfoot                        
      VCCID        &     I4 &    - &   id & ID number from \citet{bin85,bin93} \\
      NGCID        &     I4 &    - &    id & NGC ID number\\
      OTHER\_ID    &    I10 &    - &    id & Identification from another catalog\\
      RA           &   F9.5 &   deg & coord & Right Ascension\\
      Decl         &   F9.5 &   deg & coord & Declination\\
      GFLD\_NUV    &    A30 &     - &  name & Deepest \textit{GALEX} field available for NUV photometry\\
      GFLD\_FUV    &    A30 &     - &  name & Deepest \textit{GALEX} field available for FUV photometry\\
      EXPT\_NUV    &   F8.3 &     s &  time & Exposure time of NUV \textit{GALEX} image\\
      EXPT\_FUV    &   F8.3 &     s &  time & Exposure time of FUV \textit{GALEX} image\\
      m$_{B}$         &   F5.2 &  Vega &   mag & B-band magnitude from \citet{bin85,bin93}, where available\\ 
      Vel          &     I6 &  km/s &   vel & Galaxy velocity from \citet{bin85,bin93}\\
      SemiMaj      &   F7.2 &arcsec &  radi & Semi-major axis of the object determined in the NUV\\
      SemiMin      &   F7.2 &arcsec &  radi & Semi-minor axis of the object determined in the NUV\\
      PA           &   F4.1 &   deg & angle & Position angle\\
      Type         &     I2 &     - &  type & Galaxy type from GOLDMine, where available \citep{bin85,bin93}\\
      flux\_NUV    &   F8.3 & cnt/s &  flux & NUV integrated flux\\
      fluxerr\_NUV &   F8.3 & cnt/s &   uncert & Error on NUV integrated flux\\
      flux\_FUV     &   F8.3 & cnt/s &  flux & FUV integrated flux\\
      fluxerr\_FUV &   F8.3 & cnt/s &   uncert & Error on FUV integrated flux\\
      MAG\_NUV      &   F8.3 &    AB &   mag & NUV integrated magnitude\\
      MAGERR\_NUV  &   F8.3 &    AB &   uncert & Error on NUV integrated magnitude\\
      MAG\_FUV      &   F8.3 &    AB &   mag & FUV integrated magnitude\\
      MAGERR\_FUV  &   F8.3 &    AB &   uncert & Error on FUV integrated magnitude\\
     \hline
     \end{longtable}
\end{longtab}

\begin{sidewaystable*}
      \clearpage
       \caption{Sub-set of the UV-VES catalog.}             
      \label{UV-VES_cat}      
      \centering  
      \tiny    
      \begin{tabular}{cccccccccccccccc}     
      \hline\hline     
       VCCID & NGCID & RA & Decl & GFLD\_NUV& GFLD\_FUV&Vel&Type&m$_{NUV}$&mErr$_{NUV}$&m$_{FUV}$&mErr$_{FUV}$\\
       \hline                    
1 & 0 & 182.08458 & 13.68390 & GI6\_001002\_GUVICS002 & - & 2240 & 17 & 18.707 & 0.034 & -999.000 & -999.000 \\
3 & 0 & 182.11208 & 13.52390 & GI6\_001002\_GUVICS002 & AIS\_227\_sg44 & 6801 & 20 & 18.211 & 0.032 & 18.527 & 0.057 \\
4 & 0 & 182.13042 & 15.09470 & GI6\_001003\_GUVICS003\_0001 & AIS\_227\_sg32 & 587 & 12 & 18.494 & 0.046 & 18.842 & 0.086 \\
5 & 0 & 182.14084 & 15.11860 & GI6\_001003\_GUVICS003\_0001 & - & 0 & -1 & 21.998 & 0.173 & -999.000 & -999.000 \\
6 & 0 & 182.21500 & 9.13170 & AIS\_227\_sg70 & AIS\_227\_sg70 & 8175 & 3 & 17.549 & 0.041 & 17.796 & 0.056 \\
7 & 0 & 182.32709 & 11.43030 & GI6\_001005\_GUVICS005 & AIS\_227\_sg56 & 18887 & 7 & 17.920 & 0.034 & 18.487 & 0.071 \\
9 & 0 & 182.34250 & 13.99250 & GI6\_001004\_GUVICS004\_0001 & - & 1716 & -1 & 17.805 & 0.069 & -999.000 & -999.000 \\
10 & 0 & 182.35374 & 13.57440 & GI6\_001002\_GUVICS002 & AIS\_227\_sg44 & 1980 & 17 & 17.797 & 0.031 & 18.389 & 0.065 \\
11 & 0 & 182.39917 & 6.74330 & GI6\_001006\_GUVICS006\_0001 & - & 0 & -1 & 20.731 & 0.161 & -999.000 & -999.000 \\
12 & 0 & 182.43459 & 12.12580 & GI6\_001005\_GUVICS005 & AIS\_227\_sg56 & 8660 & 3 & 18.475 & 0.038 & 18.656 & 0.080 \\
14 & 0 & 182.46333 & 11.25670 & GI6\_001005\_GUVICS005 & AIS\_227\_sg56 & 17918 & 17 & 18.452 & 0.031 & 18.864 & 0.058 \\
15 & 0 & 182.47749 & 13.05000 & GI6\_001002\_GUVICS002 & - & 2535 & 3 & 17.927 & 0.057 & -999.000 & -999.000 \\
16 & 0 & 182.50458 & 14.61560 & GI6\_001004\_GUVICS004\_0001 & - & 6770 & 9 & 18.886 & 0.035 & -999.000 & -999.000 \\
17 & 0 & 182.50708 & 14.36690 & GI6\_001004\_GUVICS004\_0001 & AIS\_227\_sg33 & 817 & 12 & 17.154 & 0.159 & 17.451 & 0.066 \\
19 & 0 & 182.55750 & 13.18810 & GI6\_001002\_GUVICS002 & - & 6948 & 17 & 20.470 & 0.047 & -999.000 & -999.000 \\
21 & 0 & 182.59625 & 10.18860 & AIS\_227\_sg69 & AIS\_227\_sg69 & 485 & -3 & 18.021 & 0.046 & 19.446 & 0.103 \\
22 & 0 & 182.60167 & 13.17060 & GI6\_001002\_GUVICS002 & AIS\_227\_sg44 & 1699 & 17 & 18.569 & 0.032 & 19.244 & 0.069 \\
24 & 0 & 182.64874 & 11.76080 & GI6\_001005\_GUVICS005 & AIS\_227\_sg56 & 1296 & 17 & 17.601 & 0.031 & 18.050 & 0.056 \\
25 & 4152 & 182.65625 & 16.03310 & GI6\_001007\_GUVICS007\_0001 & AIS\_227\_sg31 & 2169 & 7 & 14.266 & 0.030 & 14.638 & 0.051 \\
26 & 0 & 182.67041 & 14.64670 & NGA\_NGC4192 & - & 2466 & 12 & 19.385 & 0.037 & -999.000 & -999.000 \\
27 & 0 & 182.67416 & 13.33140 & NGA\_NGC4168 & AIS\_227\_sg44 & 6816 & 7 & 16.701 & 0.033 & 17.082 & 0.058 \\
28 & 0 & 182.68959 & 15.86530 & GI6\_001007\_GUVICS007\_0001 & AIS\_227\_sg31 & 6325 & 7 & 17.654 & 0.032 & 17.992 & 0.060 \\
29 & 0 & 182.70416 & 15.25250 & GI6\_001003\_GUVICS003\_0001 & - & 462 & -1 & 21.198 & 0.089 & -999.000 & -999.000 \\
30 & 0 & 182.72708 & 15.94640 & GI6\_001007\_GUVICS007\_0001 & - & 2084 & 20 & 19.441 & 0.040 & -999.000 & -999.000 \\
31 & 0 & 182.73792 & 9.21940 & GI2\_125022\_AGESstrip1\_22 & GI2\_125022\_AGESstrip1\_22 & 2240 & 20 & 17.569 & 0.030 & 17.919 & 0.050 \\
32 & 0 & 182.76125 & 12.10390 & GI6\_001005\_GUVICS005 & - & 1894 & 0 & 18.161 & 0.037 & -999.000 & -999.000 \\
33 & 0 & 182.78209 & 14.27470 & GI6\_001004\_GUVICS004\_0001 & - & 1186 & -1 & 19.122 & 0.038 & -999.000 & -999.000 \\
34 & 0 & 182.79124 & 13.58750 & AIS\_227\_sg44 & AIS\_227\_sg44 & 265 & 7 & 16.952 & 0.036 & 17.339 & 0.054 \\
35 & 0 & 182.83116 & 11.91006 & GI6\_001005\_GUVICS005 & - & 0 & 20 & 21.061 & 0.071 & -999.000 & -999.000 \\
38 & 0 & 182.94833 & 12.14360 & GI6\_001011\_GUVICS011\_0001 & AIS\_227\_sg55 & 4017 & 7 & 16.479 & 0.032 & 16.909 & 0.057 \\
39 & 0 & 182.99709 & 15.40140 & AIS\_227\_sg32 & - & 7094 & 6 & 18.674 & 0.137 & -999.000 & -999.000 \\
40 & 0 & 183.01334 & 14.90440 & NGA\_NGC4192 & AIS\_227\_sg32 & 6949 & 6 & 18.018 & 0.040 & 18.351 & 0.072 \\
45 & 0 & 183.03166 & 15.10940 & NGA\_NGC4192 & AIS\_227\_sg32 & 15246 & 17 & 18.692 & 0.031 & 19.430 & 0.066 \\
46 & 0 & 183.04625 & 12.89310 & NGA\_NGC4168 & - & 1437 & -1 & 20.133 & 0.054 & -999.000 & -999.000 \\
47 & 4165 & 183.04916 & 13.24640 & GI1\_079003\_NGC4189 & GI1\_079003\_NGC4189 & 1862 & 3 & 17.021 & 0.048 & 17.531 & 0.054 \\
48 & 0 & 183.06291 & 12.48830 & GI6\_001011\_GUVICS011\_0001 & AIS\_227\_sg55 & -8 & 10 & 16.364 & 0.031 & 16.717 & 0.053 \\
49 & 4168 & 183.07208 & 13.20530 & GI1\_079003\_NGC4189 & GI1\_079003\_NGC4189 & 2307 & 0 & 16.912 & 0.047 & 18.658 & 0.127 \\
50 & 0 & 183.07916 & 15.48310 & GI6\_001012\_GUVICS012\_0001 & - & 1230 & -1 & 19.904 & 0.060 & -999.000 & -999.000 \\
51 & 0 & 183.08542 & 9.98640 & GI6\_001013\_GUVICS013 & AIS\_227\_sg79 & 9651 & 5 & 18.004 & 0.032 & 18.454 & 0.061 \\
52 & 0 & 183.09459 & 6.97940 & GI2\_125020\_AGESstrip1\_20 & GI2\_125020\_AGESstrip1\_20 & 2088 & 12 & 19.733 & 0.037 & 19.951 & 0.056 \\
53 & 0 & 183.09416 & 8.77670 & GI2\_125022\_AGESstrip1\_22 & GI2\_125022\_AGESstrip1\_22 & 10329 & 3 & 18.218 & 0.032 & 18.549 & 0.052 \\
54 & 0 & 183.10542 & 15.39580 & GI6\_001012\_GUVICS012\_0001 & - & 7183 & 1 & 20.305 & 0.079 & -999.000 & -999.000 \\
56 & 0 & 183.13000 & 15.27030 & NGA\_NGC4192 & - & 7089 & 5 & 18.205 & 0.046 & -999.000 & -999.000 \\
57 & 0 & 183.13583 & 11.35280 & GI1\_079002\_NGC4178 & GI1\_079002\_NGC4178 & 20241 & 5 & 17.879 & 0.032 & 18.318 & 0.052 \\
58 & 0 & 183.13458 & 12.12390 & GI6\_001011\_GUVICS011\_0001 & AIS\_227\_sg55 & 2209 & 5 & 15.423 & 0.030 & 15.950 & 0.053 \\
59 & 0 & 183.13583 & 12.31000 & GI6\_001011\_GUVICS011\_0001 & AIS\_227\_sg55 & 6995 & 7 & 17.405 & 0.031 & 17.873 & 0.057 \\
60 & 0 & 183.14375 & 11.07500 & GI1\_079002\_NGC4178 & GI1\_079002\_NGC4178 & 4528 & 11 & 17.413 & 0.031 & 17.748 & 0.051 \\
       \hline
      \end{tabular}
       \tablefoot{This is sub-section of this table that shows a variety of columns as a preview of this catalog. Please see the online version of this paper for access to the entire catalog content.}
\end{sidewaystable*}

\clearpage
\onecolumn
\section{The UV Virgo Cluster Point Source Catalog}
\label{appendix_UVVC_ptsrc_cat}

The primary GUViCS UV Virgo Cluster Point-Like Source (UV-VPS) catalog is the largest UV based compilation of archival and targeted observations in the Virgo Cluster to date. The catalog contains 1,230,855 sources and 99 data fields for each source. If certain data is not available for a given source a null value is entered. This catalog provides the most useful \textit{GALEX} pipeline NUV and FUV photometric parameters, and categorizes sources as stars, Virgo members, and background sources, when possible. It also provides identifiers for optical matches in the SDSS and NED, and indicates if a match exists in the NGVS, only if GUViCS-optical matches are one-to-one. NED $z_{spec}$ are also listed for GUViCS-NED one-to-one matches. 

Additionally, the catalog is useful for quick access to optical data on one-to-one GUViCS-SDSS matches. It contains the most relevant SDSS parameters for each UV matched source including: SDSS ID, object type based on photometry, $u'g'r'i'z'$ Petrosian magnitudes, $r'$-band composite model magnitudes (i.e. a linear combination of best fit exponential and de Vaucouleurs models), clean photometry flag, photometric redshift, spectroscopic survey (if applicable), spectroscopic warning flag, and spectroscopic redshift. 

The only parameter available in the catalog for UV sources that have multiple SDSS matches is the total number of multiple matches, i.e. \textit{SDSS\_NUM\_MTCHS}. Multiple GUViCS sources matched to the same SDSS source are also flagged as 'mUV' in the \textit{SDSS\_mUV} field, and given a total number of matches, \textit{SDSS\_NUM\_MTCHS}, of 1. All other fields for multiple matches are set to a null value of -9999. In order to obtain full optical SDSS data for multiply matched UV sources in both scenarios, the user can cross-correlate the GUViCS ID of the sources of interest with the full GUViCS-SDSS matched catalog in Appendix \ref{appendix_guvics-sdss_cat}. 

Here, Table \ref{UV-VPS_cols} provides a full description of all fields in the UV-VPS catalog, and Table \ref{UV-VPS_cat} is a preview of the catalog that highlights some of the most relevant data fields. The full catalog is available online at the CDS.

\begin{longtab}
\tiny
\begin{longtable}{lllll} 
\caption{\label{UV-VPS_cols}Description of columns in the UV-VPS catalog.}\\
\hline\hline  
Field & Format & Units & Label & Explanation \\
\hline
\endhead
\hline
\endfoot                
UV\_VPS\_ID&I4&-&id&ID number from this paper\\
UV\_RA&F8&deg&coord&Right ascension at UV center\\
UV\_DEC&F8&deg&coord&Declination at UV center\\
E\_BV&F8&AB&mag&Galactic reddening expressed as E(B-V) from the Galactic extinctions maps \\&&&&\citet{sch98}\\
FOV\_RADIUS\_NUV&F8&deg&distance&Distance in degrees from the center of FOV for NUV source\\
FOV\_RADIUS\_FUV&F8&deg&distance&Distance in degrees from the center of FOV for FUV source\\
NUV\_FLUX&F4&micro-Janskys&flux&Calibrated NUV flux in micro-Janskys from SExtractor AUTO aperture.\\ &&&&Value -999 means no NUV detection\\
NUV\_FLUXERR&F4&micro-Janskys&flux&Error on NUV\_FLUX\\
NUV\_MAG&F4&AB&mag&Calibrated NUV AB magnitude from the SExtractor AUTO magnitude\\
NUV\_MAGERR&F4&AB&mag&Error on NUV\_MAG\\
FUV\_FLUX&F4&micro-Janskys&flux&Calibrated FUV flux in micro-Janskys from SExtractor AUTO aperture.\\ &&&&Value -999 means no FUV detection\\
FUV\_FLUXERR&F4&micro-Janskys&flux&Error on FUV\_FLUX\\
FUV\_MAG&F4&AB&mag&Calibrated NUV AB magnitude from the SExtractor AUTO magnitude\\
FUV\_MAGERR&F4&AB&mag&Error on FUV\_MAG\\
NUV\_S2N&F4&-&ratio&NUV signal-to-noise from NUV\_FLUX value\\
FUV\_S2N&F4&-&ratio&FUV signal-to-noise from FUV\_FLUX value\\
FUV\_NCAT\_FLUX&F4&micro-Janskys&flux&The FUV flux derived using the source position given in the NUV catalog\\
FUV\_NCAT\_FLUXERR&F4&micro-Janskys&flux&Error on FUV\_NCAT\_FLUX\\
FUV\_NCAT\_MAG&F4&AB&mag&FUV calibrated AB magnitude derived from FUV\_NCAT\_FLUX\\
FUV\_NCAT\_MAGERR&F4&AB&mag&Error on FUV\_NCAT\_MAG\\
FUV\_NCAT\_S2N&F4&-&ratio&Signal to noise for fuv\_ncat\_flux\\
NUV\_SKYBG&F4&cts sec$-1$ arcsec$^2$&flux&Background value in photons per second per square arcsecond at the\\ &&&&source position taken in \textit{GALEX} pipeline -nd-skybg.fits file image\\
FUV\_SKYBG&F4&cts sec$-1$ arcsec$^2$&flux&Background value in photons per second per square arcsecond at the\\ &&&&source position taken in \textit{GALEX} pipeline -fd-skybg.fits file image\\
NUV\_ARTIFACT&I4&binary&number&Logical OR of artifact flags for pixels within a 3x3 pixel box in \textit{GALEX}\\ &&&&pipeline -nd-flags.fits image\\
FUV\_ARTIFACT&I4&binary&number&Logical OR of artifact flags for pixels within a 3x3 pixel box in \textit{GALEX}\\ &&&&pipeline -fd-flags.fits image\\
NUV\_NUMBER&I4&-&number&Running object number from original \textit{GALEX} pipeline -xd-mcat.fits\\
NUV\_FLUX\_RADIUS\_2&F4&pix&-&Half-light radius of NUV source from SExtractor AUTO aperture\\
NUV\_FLUX\_RADIUS\_4&F4&pix&-&90\% radius of NUV source from SExtractor AUTO aperture\\
NUV\_KRON\_RADIUS&F4&-&ratio&NUV Kron apertures in units of A or B\\
NUV\_MU\_MAX&F4&-&number&Peak NUV surface brightness above background\\
NUV\_A\_IMAGE&F4&pix&-&Profile RMS along NUV major axis\\
NUV\_B\_IMAGE&F4&pix&-&Profile RMS along NUV minor axis\\
NUV\_THETA\_IMAGE&F4&deg&-&Position angle\\
NUV\_ERRA\_IMAGE&F4&pix&-&RMS position error along NUV major axis\\
NUV\_ERRB\_IMAGE&F4&pix&-&RMS position error along NUV minor axis\\
NUV\_ERRTHETA\_IMAGE&F4&deg&-&Error ellipse NUV position angle\\
NUV\_FWHM\_WORLD&F4&deg&-&NUV FWHM assuming a gaussian core\\
NUV\_FLAGS&I2&value&flag&NUV extraction flags\\
NUV\_CLASS\_STAR&F4&value&flag&SExtractor star/galaxy classifier\\
FUV\_NUMBER&I4&-&number&Running object number from original \textit{GALEX} pipeline -xd-mcat.fits\\
FUV\_FLUX\_RADIUS\_2&F4&pix&-&Half-light radius of FUV source from SExtractor AUTO aperture\\
FUV\_FLUX\_RADIUS\_4&F4&pix&-&90\% radius of NUV source from SExtractor AUTO aperture\\
FUV\_KRON\_RADIUS&F4&-&ratio&FUV Kron apertures in units of A or B\\
FUV\_MU\_MAX&F4&-&number&Peak FUV surface brightness above background\\
FUV\_A\_IMAGE&F4&pix&-&Profile RMS along FUV major axis\\
FUV\_B\_IMAGE&F4&pix&-&Profile RMS along FUV minor axis\\
FUV\_THETA\_IMAGE&F4&-&deg&Position angle\\
FUV\_ERRA\_IMAGE&F4&pix&-&RMS position error along FUV major axis\\
FUV\_ERRB\_IMAGE&F4&pix&-&RMS position error along FUV minor axis\\
FUV\_ERRTHETA\_IMAGE&F4&deg&-&Error ellipse FUV position angle\\
FUV\_FWHM\_WORLD&F4&deg&-&FUV FWHM assuming a gaussian core\\
FUV\_FLAGS&I2&value&flag&FUV extraction flags\\
FUV\_CLASS\_STAR&F4&value&flag&SExtractor star/galaxy classifier\\
NUV\_ORG\_PIPELINE\_FLD&A35&-&name&Deepest NUV \textit{GALEX} field where source located\\
FUV\_ORG\_PIPELINE\_FLD&A35&-&name&Deepest FUV \textit{GALEX} field where source located\\
NUV\_EXPTIME&F8&sec&time&Exposure time in seconds of NUV\_ORG\_PIPELINE\_FLD\\
FUV\_EXPTIME&F8&sec&time&Exposure time in seconds of FUV\_ORG\_PIPELINE\_FLD\\
NUV\_GUVICS\_FIELD\_OVERLAPS&I4&-&number&Number of shallower NUV fields overlapping source location\\
FUV\_GUVICS\_FIELD\_OVERLAPS&I4&-&number&Number of shallower FUV fields overlapping source location\\
NUV\_GOOD\_PHOTOM\_FLAG&I4&value&flag&Based on if NUV\_ARTIFACT contains 2,4,512; 0=good and 1=bad\\
FUV\_GOOD\_PHOTOM\_FLAG&I4&value&flag&Based on if FUV\_ARTIFACT contains 2,4,512; 0=good and 1=bad\\
NUV\_SRCREM\_FLAG&I4&value&flag&See Table \ref{srcrem_flag} of paper for description of values\\
FUV\_SRCREM\_FLAG&I4&value&flag&See Table \ref{srcrem_flag} of paper for description of values\\
SDSS\_NUM\_MTCHS&I8&-&number&Number of optical SDSS sources matched to this UV source\\
SDSS\_multUV&A10&-&flag&Flag is 'mUV' if this UV sources shares a SDSS match with one or\\ &&&&more other UV sources\\
SDSS\_ObjID&I8&-&id&SDSS object identifier\\
SDSS\_RA&F8&deg&coord&Right ascension in degrees at center of SDSS source in the $r'$-band\\
SDSS\_DEC&F8&deg&coord&Declination in degrees at center of SDSS source in the $r'$-band\\
SDSS\_MATCH\_SEP&F8&deg&-&Angular separation between UV source and SDSS optical match\\
SDSS\_phot\_Clean&I1&value&flag&SDSS flag indicates if photometry is clean; clean=1 and not clean=0\\
SDSS\_phot\_ObjType&I1&-&value&Type classification of object: 0=unknown; 1=cosmic ray; 2=defect;\\ &&&&3=galaxy; 4=ghost; 5=knownobj; 6=star; 7=trail; 8=sky; 9=notatype\\
SDSS\_survey&A30&-&-&SDSS survey in which object is detected\\
SDSS\_specz&F8&$z$&redshift&SDSS spectroscopic redshift\\
SDSS\_specz\_Err&F8&$z$&redshift&Error on SDSS\_specz\\
SDSS\_photzNN&F8&$z$&redshift&SDSS photometric redshift from 'neural network'\\ &&&&algorithm\\
SDSS\_photzNN\_Err&F8&$z$&redshift&Error on SDSS\_photzNN\\
SDSS\_photzRF&F8&$z$&redshift&SDSS photometric redshift from 'random forest'\\ &&&&algorithm\\
SDSS\_photzRF\_Err&F8&$z$&redshift&Error on SDSS\_photzRF\\
SDSS\_STAR&A5&-&flag&Flag value is '*' if source is a star by any definition\\
SDSS\_STAR\_CLASS&A5&-&flag&Flag value is 'PHOT' if star determined via photo point spread function,\\ &&&&'SPEC' if star found via spectra, or\\ &&&&'BOTH' if both methods apply\\
SDSS\_BCKD\_SRCS&A10&-&flag&Flag value is 'BK' if object is behind the Virgo Cluster\\ &&&&based on its SDSS photometric or spectroscopic redshift\\
SDSS\_BCKD\_SRCS\_CLASS&A10&-&flag&Flag value is 'PHOT' if background source determined\\ &&&&from $z_{phot}$, 'SPEC' if background source determined from $z_{spec}$, or\\ &&&&'BOTH' if both methods apply\\
SDSS\_BCKD\_SRCS\_CLASS\_BREAKDOWN&A10&-&flag&Flag value various combinations of 'PNN','PRF',\\ &&&&and 'SPEC' to show the details of which $z_{phot}$ spec and if $z_{spec}$ was\\ &&&&used to classify the background source\\
SDSS\_SRCS\_TBD&A5&-&flag&Flag value is 'TBD' if SDSS source is uncategorizable\\ &&&&based on optical parameters, otherwise value is '-'\\
SDSS\_VM&A5&-&flag&Flag value is 'VM' if source is a member of the Virgo\\ &&&&Cluster based on $z_{phot}$ or $z_{spec}$, otherwise value is '-'\\
NGVS\_NUM\_MTCHS&I8&-&number&Number of optical NGVS sources matched to this UV\\ &&&&source\\
NGVS\_multUV&A5&-&flag&Flag is 'mUV' if this UV sources shares a NGVS match\\ &&&&with one or more other UV sources\\
NGVS\_STAR&A1&-&flag&Flag is '*' if based on NGVS star/galaxy classification\\ &&&&parameters\\
NED\_NUM\_MTCHS&I8&-&number&Number of optical NED sources matched to this UV\\ &&&&source\\
NED\_multUV&A7&-&flag&Flag is 'mUV' if this UV soruces hares a NED match\\ &&&&with on or more other UV sources\\
NED\_ObjID&A30&-&id&Primary Object identification from the NED database\\
NED\_RA&F8&deg&-&Right ascension in degrees of NED optical source\\
NED\_DEC&F8&deg&-&Declinationa in degrees of the NED optical source\\
NED\_MATCH\_SEP&F8&deg&-&Angular separation between UV source and NED\\ &&&&optical match\\
NED\_z&F8&$z$&redshift&Spectroscopic redshift of the NED source\\
NED\_z\_QF&A8&-&flag&The NED quality flag on NED\_z\\
NED\_BCKD\_SRCS&A2&-&flag&Flag is 'BK' if object is behind the Virgo Cluster based\\ &&&&on its NED spectroscopic redshift\\
NED\_VM&A2&-&flag&Flag is 'VM' if object is a Virgo Cluster member based\\ &&&&on its NED spectroscopic redshift\\
NED\_STAR&A1&-&flag&Flag is '*' if object is a star based on visual investigation\\
     \hline
\multicolumn{5}{l}{
      \tablefoot{Many of the \textit{GALEX} column descriptions are taken directly from the $mcat$ catalog column descriptions here: http://www.galex.caltech.edu/wiki/Public:Documentation/Appendix\_A.1. }}
      \end{longtable}
\end{longtab}

\begin{sidewaystable*}
      \clearpage
       \caption{Sub-set of the UV-VPS catalog.}             
      \label{UV-VPS_cat}      
      \centering          
      \tiny
      \begin{tabular}{ccccccccccc}     
      \hline\hline     
       UV\_VPS\_ID & GFLD\_NUV & NUV\_EXPTIME & $m_{NUV}$ & $mErr_{NUV}$ & SDSS\_ID& NED\_ID &SDSS\_TBD &NGVS\_STARS& NGVS\_BCKD\_SRCS&NED\_VM\\
       \hline                    
1&AIS\_229\_sg42&308.149993896&20.5348&0.122755&1237648720700244214&2QZ-J125816.9-002810&?&-&-&-\\
2&AIS\_229\_sg42&308.149993896&20.3257&0.10154&1237648720700178676&-&-&-&-&-\\
3&AIS\_229\_sg42&308.149993896&20.6323&0.126446&1237648720700244003&-&-&-&-&-\\
4&AIS\_229\_sg42&308.149993896&21.2726&0.17099&1237648720700244035&-&-&-&-&-\\
5&AIS\_229\_sg42&308.149993896&20.0739&0.0855449&1237648720700309536&-&-&-&-&-\\
6&AIS\_229\_sg42&308.149993896&20.7085&0.137504&1237648720700244142&-&-&-&-&-\\
7&AIS\_229\_sg42&308.149993896&20.729&0.150416&1237648720700309542&-&-&-&-&-\\
8&AIS\_229\_sg42&308.149993896&20.2624&0.0946499&1237648720700309513&-&-&-&-&-\\
9&AIS\_229\_sg42&308.149993896&21.2977&0.223296&1237648720700178852&-&-&-&-&-\\
10&AIS\_229\_sg42&308.149993896&20.7242&0.12678&1237648720700113080&-&-&-&-&-\\
11&AIS\_229\_sg42&308.149993896&20.4946&0.113604&-9999&-&-&-&-&-\\
12&AIS\_229\_sg42&308.149993896&20.4678&0.114593&-9999&-&-&-&-&-\\
13&AIS\_229\_sg42&308.149993896&19.8451&0.0737518&1237648704045187093&-&-&-&-&-\\
14&AIS\_229\_sg42&308.149993896&20.2797&0.112614&1237648704044990631&2dFGRS-N326Z076&?&-&-&-\\
15&AIS\_229\_sg42&308.149993896&20.8952&0.144493&1237648704045056121&-&-&-&-&-\\
16&GI1\_033005\_NGC4517&6402.60009766&22.2375&0.0924708&1237648704042304049&-&?&-&-&-\\
17&GI1\_033005\_NGC4517&6402.60009766&24.4108&0.386695&-9999&-&-&-&-&-\\
18&GI1\_033005\_NGC4517&6402.60009766&23.1221&0.166083&1237648704042303707&-&-&-&-&-\\
19&AIS\_229\_sg42&308.149993896&19.6255&0.0666697&1237648704044990536&-&-&-&-&-\\
20&GI1\_033005\_NGC4517&6402.60009766&22.4737&0.228617&1237648704042304117&-&?&-&-&-\\
21&GI1\_033005\_NGC4517&6402.60009766&24.1593&0.341123&1237648704042304099&-&?&-&-&-\\
22&GI1\_033005\_NGC4517&6402.60009766&24.8536&0.771642&-9999&-&-&-&-&-\\
23&GI1\_033005\_NGC4517&6402.60009766&24.43&0.421629&1237648704042369790&-&?&-&-&-\\
24&GI1\_033005\_NGC4517&6402.60009766&21.441&0.0574551&1237648704042303675&-&-&-&-&-\\
25&GI1\_033005\_NGC4517&6402.60009766&23.8077&0.340966&1237648704042303898&-&-&-&-&-\\
26&GI1\_033005\_NGC4517&6402.60009766&22.1606&0.0836092&1237648704042303826&-&-&-&-&-\\
27&GI1\_033005\_NGC4517&6402.60009766&24.0378&0.330953&1237648704042370066&-&?&-&-&-\\
28&AIS\_229\_sg32&201.0&19.0389&0.055305&1237648704044597280&-&-&-&-&-\\
29&GI1\_033005\_NGC4517&6402.60009766&24.059&0.451261&1237648704042304454&-&?&-&-&-\\
30&GI1\_033005\_NGC4517&6402.60009766&22.4869&0.120524&1237648704042303950&-&?&-&-&-\\
31&GI1\_033005\_NGC4517&6402.60009766&24.4723&0.37787&1237648704042369450&-&?&-&-&-\\
32&GI1\_033005\_NGC4517&6402.60009766&23.5099&0.218884&1237648704042303784&-&?&-&-&-\\
33&GI1\_033005\_NGC4517&6402.60009766&24.4337&0.409638&-9999&-&-&-&-&-\\
34&GI1\_033005\_NGC4517&6402.60009766&23.1899&0.253893&-9999&-&-&-&-&-\\
35&AIS\_229\_sg42&308.149993896&20.9902&0.152329&1237648704045056031&-&-&-&-&-\\
36&GI1\_033005\_NGC4517&6402.60009766&21.3041&0.117376&-9999&-&-&-&-&-\\
37&GI1\_033005\_NGC4517&6402.60009766&23.3221&0.289265&-9999&-&-&-&-&-\\
38&GI1\_033005\_NGC4517&6402.60009766&24.5103&0.517109&-9999&-&-&-&-&-\\
39&GI1\_033005\_NGC4517&6402.60009766&22.3735&0.16679&1237648704042238492&-&?&-&-&-\\
40&GI1\_033005\_NGC4517&6402.60009766&24.6084&0.444212&-9999&-&-&-&-&-\\
41&GI1\_033005\_NGC4517&6402.60009766&24.5515&0.451711&1237648704042303639&-&-&-&-&-\\
42&GI1\_033005\_NGC4517&6402.60009766&24.3917&0.478081&-9999&-&-&-&-&-\\
43&AIS\_229\_sg42&308.149993896&20.5733&0.127777&1237648704045056132&-&-&-&-&-\\
44&GI1\_033005\_NGC4517&6402.60009766&23.9409&0.299737&1237648704042238731&-&?&-&-&-\\
45&GI1\_033005\_NGC4517&6402.60009766&22.7844&0.155426&1237648704042304295&-&?&-&-&-\\
46&GI1\_033005\_NGC4517&6402.60009766&24.7583&0.450381&-9999&-&-&-&-&-\\
47&GI1\_033005\_NGC4517&6402.60009766&21.4907&0.0512307&1237648704042369167&-&-&-&-&-\\
48&GI1\_033005\_NGC4517&6402.60009766&24.7559&0.464035&1237648704042304170&-&?&-&-&-\\
49&-&-99.0&-999.0&-999.0&-9999&-&-&-&-&-\\
50&AIS\_229\_sg32&201.0&17.9924&0.0330155&1237648704044597315&-&-&-&-&-\\
       \hline
      \end{tabular}
       \tablefoot{This is a sub-sample of the entire table that shows a variety of data fields (for NUV only) as a preview of this catalog. Please see the online version of this paper for access to the entire catalog content.\\}
\end{sidewaystable*}

\clearpage
\onecolumn
\section{The GUViCS-SDSS Matched Catalog}
\label{appendix_guvics-sdss_cat}

The GUViCS-SDSS matched catalog provides the most relevant SDSS data on all GUViCS-SDSS matches, including one-to-one matches and multiply matched sources. The catalog gives full SDSS identification information, complete SDSS photometric measurements in multiple aperture types, and complete redshift information (photometric and spectroscopic). It is ideal for large statistical studies of galaxy populations at multiple wavelengths in the background of the Virgo Cluster. The catalog can also be used as a starting point to study and search for previously unknown UV-bright point-like objects within the Virgo Cluster. The catalog contains 1,056,793 sources and 85 data fields for each source. If certain data is not available for a given source that field is given a null value. Here, Table \ref{all_sdss_cols} provides a full description of all fields in the GUViCS-SDSS matched catalog, and Table \ref{all_sdss_cat} is a preview of the catalog that highlights some of the most relevant data fields. The full catalog is available online at the CDS.

\begin{longtab}
\tiny
\begin{longtable}{lllll} 
\caption{\label{all_sdss_cols}Description of columns in the GUViCS-SDSS matched catalog.}\\
\hline\hline  
Field & Format & Units & Label & Explanation \\
\hline
\endhead
\hline
\endfoot         
UV\_VPS\_ID&I4&-&id&ID number from this paper\\       
SDSS\_ObjID&I8&-&id&SDSS object identifier\\
SDSS\_SpecObjID&I8&-&id&SDSS identifier pointing to spectrum of object if it exists\\
UV\_RA&F8&deg&coord&Right ascension at UV center\\
UV\_DEC&F8&deg&coord&Declination at UV center\\
SDSS\_MATCH\_SEP&F8&deg&-&Angular separation between UV source and SDSS optical match\\
SDSS\_RA&F8&deg&coord&Right ascension in degrees at center of SDSS source in the $r'$-band\\
SDSS\_DEC&F8&deg&coord&Declination in degrees at center of SDSS source in the $r'$-band\\
SDSS\_phot\_Clean&I1&value&flag&SDSS flag indicates if photometry is clean; clean=1 and not clean=0\\
SDSS\_phot\_ObjType&I1&value&-&Type classification of object: 0=unknown; 1=cosmic ray; 2=defect; 3=galaxy; 4=ghost; 5=knownobj;\\
                              &&&& 6=star; 7=trail; 8=sky; 9=notatype\\
petroMag\_u&F8&-&mag&Petrosian magnitude for $u'$-band in $r'$-band aperture\\
petroMagErr\_u&F8&-&mag&Error on petroMag\_u\\
modelMag\_u&F8&-&mag&Better of De Vaucouleurs/exponential $u'$-band magnitude fit in the $r'$-band\\
modelMagErr\_u&F8&-&mag&Error on modelMag\_u\\
cModelMag\_u&F8&-&mag&Linear combination of De Vaucouleurs+exponential magnitudes fit in the $u'$-band\\
cModelMagErr\_u&F8&-&mag&Error on cModelMag\_u\\
petroRad\_u&F8&arcsec&-&Petrosian radius in $u'$-band\\
petroRadErr\_u&F8&arcsec&-&Error on petroRad\_u\\
petroR50\_u&F8&arcsec&-&Radius containing 50\% of the Petrosian flux in $u'$-band\\
petroR50Err\_u&F8&arcsec&-&Error on petroR50\_u\\
petroR90\_u&F8&arcsec&-&Radius containing 90\% of the Petrosian flux in $u'$-band\\
petroR90Err\_u&F8&arcsec&-&Error on petroR90\_u\\
petroMag\_g&F8&-&mag&Petrosian magnitude for $g'$-band in $r'$-band Petrosian aperture\\
petroMagErr\_g&F8&-&mag&Error on petroMag\_g\\
modelMag\_g&F8&-&mag&Better of De Vaucouleurs/exponential $g'$-band magnitude fit in the $r'$-band\\
modelMagErr\_g&F8&-&mag&Error on modelMag\_g\\
cModelMag\_g&F8&-&mag&Linear combination of De Vaucouleurs+exponential magnitudes fit in the $g'$-band\\
cModelMagErr\_g&F8&-&mag&Error on cModelMag\_g\\
petroRad\_g&F8&arcsec&-&Petrosian radius in $g'$-band\\
petroRadErr\_g&F8&arcsec&-&Error on petroRad\_g\\
petroR50\_g&F8&arcsec&-&Radius containing 50\% of the Petrosian flux in $g'$-band\\
petroR50Err\_g&F8&arcsec&-&Error on petroR50\_g\\
petroR90\_g&F8&arcsec&-&Radius containing 90\% of the Petrosian flux in $g'$-band\\
petroR90Err\_g&F8&arcsec&-&Error on petroR90\_g\\
petroMag\_r&F8&-&-&Petrosian magnitude for $r'$-band in $r'$-band Petrosian aperture\\
petroMagErr\_r&F8&-&mag&Error on petroMag\_r\\
modelMag\_r&F8&-&mag&Better of De Vaucouleurs/exponential $r'$-band magnitude fit in the $r'$-band\\
modelMagErr\_r&F8&-&mag&Error on modelMag\_r\\
cModelMag\_r&F8&-&mag&Linear combination of De Vaucouleurs+exponential magnitudes fit in the $r'$-band\\
cModelMagErr\_r&F8&-&mag&Error on cModelMag\_r\\
petroRad\_r&F8&arcsec&dist&Petrosian radius in $r'$-band\\
petroRadErr\_r&F8&arcsec&-&Error on petroRad\_r\\
petroR50\_r&F8&arcsec&-&Radius containing 50\% of the Petrosian flux in $r'$-band\\
petroR50Err\_r&F8&arcsec&-&Error on petroR50\_r\\
petroR90\_r&F8&arcsec&-&Radius containing 90\% of the Petrosian flux in $r'$-band\\
petroR90Err\_r&F8&arcsec&-&Error on petroR90\_r\\
petroMag\_i&F8&-&mag&Petrosian magnitude for $i'$-band in $r'$-band Petrosian aperture\\
petroMagErr\_i&F8&-&mag&Error on petroMag\_i\\
modelMag\_i&F8&-&mag&Better of De Vaucouleurs/exponential $i'$-band magnitude fit in the $r'$-band\\
modelMagErr\_i&F8&-&mag&Error on modelMag\_i\\
cModelMag\_i&F8&-&mag&Linear combination of De Vaucouleurs+exponential magnitudes fit in the $i'$-band\\
cModelMagErr\_i&F8&-&mag&Error on cModelMag\_i\\
petroRad\_i&F8&arcsec&-&Petrosian radius in $i'$-band\\
petroRadErr\_i&F8&arcsec&-&Error on petroRad\_i\\
petroR50\_i&F8&arcsec&-&Radius containing 50\% of the Petrosian flux in $i'$-band\\
petroR50Err\_i&F8&arcsec&-&Error on petroR50\_i\\
petroR90\_i&F8&arcsec&-&Radius containing 90\% of the Petrosian flux in $i'$-band\\
petroR90Err\_i&F8&arcsec&-&Error on petroR90\_i\\
petroMag\_z&F8&-&-&Petrosian magnitude for $z'$-band in $r'$-band Petrosian aperture\\
petroMagErr\_z&F8&-&mag&Error on petroMag\_z\\
modelMag\_z&F8&-&mag&Better of De Vaucouleurs/exponential $z'$-band magnitude fit in the $r'$-band\\
modelMagErr\_z&F8&-&mag&Error on modelMag\_z\\
cModelMag\_z&F8&-&mag&Linear combination of De Vaucouleurs+exponential magnitudes fit in the $z'$-band\\
cModelMagErr\_z&F8&-&mag&Error on cModelMag\_z\\
petroRad\_z&F8&arcsec&-&Petrosian radius in $z'$-band\\
petroRadErr\_z&F8&arcsec&-&Error on petroRad\_z\\
petroR50\_z&F8&arcsec&-&Radius containing 50\% of the Petrosian flux in $z'$-band\\
petroR50Err\_z&F8&arcsec&-&Error on petroR50\_z\\
petroR90\_z&F8&arcsec&-&Radius containing 90\% of the Petrosian flux in $z'$-band\\
petroR90Err\_z&F8&arcsec&-&Error on petroR90\_z\\
SDSS\_survey&A30&-&name&SDSS survey name\\
SDSS\_specz&F8&$z$&redshift&Best SDSS spectroscopic redshift\\
SDSS\_specz\_Err&F8&$z$&redshift&Error on SDSS\_specz\\
SDSS\_specz\_class&A30&-&-&Classifcation of SDSS source as 'GALAXY', 'QSO', or 'STAR' based on features in the\\ 
                          &&&&objects spectrum\\
SDSS\_specz\_Warning&F8&value&flag&Bitmask of warning values where 0=good\\
SDSS\_specz\_noQSO&F8&$z$&redshift&Best SDSS spectroscopic redshift when excluding QSO fit\\
SDSS\_specz\_Err\_noQSO&F8&$z$&redshift&Error on SDSS\_specz\_noQSO\\
SDSS\_specz\_Warning\_noQSO&F8&value&flag&Same warning as SDSS\_specz\_Warning, but for SDSS\_specz\_noQSO\\
SDSS\_specz\_class\_noQSO&A30&-&-&Same classification as SDSS\_specz\_class, but for SDSS\_specz\_noQSO\\
SDSS\_photzNN&F8&$z$&redshift&SDSS photometric redshift from 'neural network' algorithm\\
SDSS\_photzNN\_Err&F8&$z$&redshift&Error on SDSS\_photzNN\\
SDSS\_photzRF&F8&$z$&redshift&SDSS photometric redshift from 'random forest' algorithm\\
SDSS\_photzRF\_Err&F8&$z$&redshift&Error on SDSS\_photzRF\\
     \hline
\multicolumn{5}{l}{
      \tablefoot{Many of the SDSS field descriptions are taken directly from the SDSS DR9 schema page here: http://skyserver.sdss3.org/dr9/en/help/browser/browser.asp}}
      \end{longtable}
\end{longtab}

\begin{sidewaystable*}
      \clearpage
       \caption{Sub-set of the GUViCS-SDSS catalog.}             
      \label{all_sdss_cat}      
      \centering     
      \tiny     
      \begin{tabular}{ccccccccccc}     
      \hline\hline     
       UV\_VPS\_ID & SDSS\_ID & SDSS\_RA & SDSS\_Dec & phot\_Clean& petroMag\_r& petorR50\_r&cModelMag\_r&modelMag\_r&$z_{spec}$&$z_{photRF}$\\
       \hline                    
1&1237648720700244214&194.570685056&-0.469382430018&1&19.86351&1.066034&19.87323&19.87323&-8888.0&0.0789564\\
2&1237648720700178676&194.5004379&-0.463474674025&1&19.05415&0.9225867&19.06404&19.06405&0.2451683&0.172791\\
3&1237648720700244003&194.597295792&-0.455630459169&1&17.06896&0.8109651&17.03068&17.03068&-8888.0&-8888.0\\
4&1237648720700244035&194.510200974&-0.454894912676&1&17.93353&0.779039&17.89274&17.89274&-8888.0&-8888.0\\
5&1237648720700309536&194.746662034&-0.452415921765&1&14.92753&0.8733402&14.89757&14.89757&-8888.0&-8888.0\\
6&1237648720700244142&194.586398592&-0.442746155224&1&18.16115&2.18277&18.11616&18.11616&-8888.0&0.195563\\
7&1237648720700309542&194.756667175&-0.438377969013&1&20.13555&0.9088713&20.14423&20.14426&-8888.0&-8888.0\\
8&1237648720700309513&194.677913463&-0.42809303877&1&16.83692&0.8310513&16.80174&16.80174&-8888.0&-8888.0\\
9&1237648720700178852&194.473059059&-0.422703589543&1&19.64554&1.304409&19.61404&19.61404&-8888.0&0.165244\\
10&1237648720700113080&194.313610924&-0.422684388298&1&18.16351&1.884779&18.08863&18.08863&-8888.0&0.191196\\
11&1237648720700178501&194.428818676&-0.420628586975&1&16.20097&0.7421287&16.1603&16.1603&-8888.0&-8888.0\\
11&1237648720700178502&194.429401131&-0.41918924813&1&20.0843&1.37854&20.40982&20.39767&-8888.0&0.14831\\
12&1237648704045121647&194.614942964&-0.405337074673&1&19.43814&1.105869&19.42088&19.42088&-8888.0&0.14715\\
12&1237648704045121648&194.615503031&-0.40663203932&1&20.2391&1.305289&20.14555&20.29172&-8888.0&0.0865192\\
13&1237648704045187093&194.750201902&-0.404410099976&1&14.43311&0.5863197&14.37228&14.37228&-8888.0&-8888.0\\
14&1237648704044990631&194.343503017&-0.395526996972&0&17.89945&2.842504&17.87007&17.87007&-8888.0&0.0463287\\
15&1237648704045056121&194.466746745&-0.393085190011&1&19.13187&0.9382241&19.10153&19.14655&-8888.0&0.125337\\
16&1237648704042304049&188.162644227&-0.384860113267&1&21.05508&1.047097&21.01303&21.01303&-8888.0&0.162085\\
18&1237648704042303707&188.212979348&-0.38304090509&1&19.60042&0.80029&19.5685&19.5685&-8888.0&-8888.0\\
19&1237648704044990536&194.341050177&-0.382847557805&0&13.44909&1.178099&14.97935&13.59838&-8888.0&-8888.0\\
20&1237648704042304117&188.196194543&-0.383096669973&1&21.69966&1.448838&21.6634&21.6634&-8888.0&0.415683\\
21&1237648704042304099&188.187575471&-0.380356371517&1&20.7378&1.063899&20.70365&20.748&-8888.0&0.198996\\
23&1237648704042369790&188.257982827&-0.380012386754&0&22.3031&0.9142082&22.2266&21.84088&-8888.0&0.342464\\
24&1237648704042303675&188.1800504&-0.378728375738&1&18.71882&0.8019442&18.68394&18.68405&-8888.0&-8888.0\\
25&1237648704042303898&188.107099202&-0.380197871207&0&20.77004&1.985193&20.99642&20.99642&-8888.0&-8888.0\\
26&1237648704042303826&188.192775827&-0.378245325478&1&20.56008&0.9576247&20.63448&20.63401&-8888.0&-8888.0\\
27&1237648704042370066&188.275799996&-0.378714343949&0&21.29313&3.762415&21.84264&21.86093&-8888.0&0.435027\\
28&1237648704044597280&193.384440771&-0.37779393787&0&14.16577&1.693035&14.61736&14.74322&-8888.0&-8888.0\\
29&1237648704042304454&188.225583678&-0.376638555902&0&23.6117&0.9046873&22.43729&23.80241&-8888.0&-8888.0\\
30&1237648704042303950&188.127495093&-0.37761584135&1&20.59755&1.751001&20.49957&20.53205&-8888.0&0.339657\\
31&1237648704042369450&188.263280895&-0.377826913289&1&21.65134&1.086265&21.23803&21.37473&-8888.0&0.392354\\
32&1237648704042303784&188.206375808&-0.377137995896&1&20.83275&0.9348387&20.78656&20.78661&-8888.0&0.253223\\
35&1237648704045056031&194.461259787&-0.375380895546&1&17.83416&0.617262&17.77811&17.77811&-8888.0&-8888.0\\
36&1237648704042369071&188.290235618&-0.375190667224&0&21.09913&1.281783&20.93743&20.93742&-8888.0&0.130005\\
36&1237648704042369072&188.289622129&-0.375997648686&1&21.63453&0.8854136&21.41649&21.41649&-8888.0&0.0984047\\
36&1237648704042369076&188.289064923&-0.375377792797&0&24.09786&0.5908738&23.91677&23.87782&-8888.0&0.676087\\
36&1237648704042369077&188.289680512&-0.374501855036&0&23.38869&0.5280144&23.02468&23.02467&-8888.0&-8888.0\\
39&1237648704042238492&188.076385929&-0.373243019609&1&24.4804&0.2705168&24.11954&24.11954&-8888.0&-8888.0\\
41&1237648704042303639&188.144225366&-0.37393654897&1&19.24933&0.7845541&19.19751&19.19751&-8888.0&-8888.0\\
42&1237648704042303738&188.11778755&-0.372790420495&1&19.55833&0.8050651&19.53878&19.53878&-8888.0&-8888.0\\
42&1237648704042303739&188.115843754&-0.37316966269&1&20.84318&1.230178&20.78664&20.86171&-8888.0&0.0887103\\
43&1237648704045056132&194.49838628&-0.372234387989&1&19.41976&0.8397874&19.3998&19.38776&-8888.0&0.280437\\
44&1237648704042238731&188.1054057&-0.372172492487&1&22.25185&0.7856979&22.09014&22.09014&-8888.0&0.367416\\
45&1237648704042304295&188.151587834&-0.372551405353&1&22.75891&0.6515285&22.53342&22.53342&-8888.0&-8888.0\\
47&1237648704042369167&188.280957964&-0.371717788558&1&18.50293&0.7366127&18.45677&18.45677&-8888.0&-8888.0\\
48&1237648704042304170&188.216358795&-0.371422131258&1&21.63396&0.9046321&21.48607&21.48607&-8888.0&0.344373\\
50&1237648704044597315&193.478164871&-0.370592819405&0&14.67558&0.9659479&15.99006&15.99006&-8888.0&-8888.0\\
       \hline
      \end{tabular}
       \tablefoot{This is a sub-sample of the entire table that shows a variety of fields as a preview of this catalog. Please see the online version of this paper for access to the entire catalog content.\\}
\end{sidewaystable*}

\end{appendix}
\end{document}